\documentclass[fleqn,10pt]{wlscirep}
\usepackage[utf8]{inputenc}
\usepackage[T1]{fontenc}
\usepackage{xspace}
\usepackage{xcolor}
\usepackage{scalerel}

\usepackage{placeins}

\usepackage{pdflscape} % this is just for supplementary material!

\newcommand{\apjl}{Astrophys. J. Lett.}   % Astrophysical Journal, Letters
\newcommand{\apjs}{Astrophys. J. Supplementary Ser.}   % Astrophysical Journal, Supplement
\newcommand{\ao}{Appl. Opt.}   % Applied Optics
\newcommand{\aap}{Astron. Astrophys.}   % Astronomy and Astrophysics
\renewcommand{\citet}{\cite}

\title{
Cosmic accretion shocks as a tool to measure the dark matter mass of galaxy clusters
}

\author[1,*]{David Vallés-Pérez}
\author[1,2,$\dagger$]{Vicent Quilis}
\author[1,2,$\dagger$]{Susana Planelles}
\affil[1]{Departament d'Astronomia i Astrofísica, Universitat de València, C/Doctor Moliner, 50, Burjassot, 46100, València, Spain}
\affil[2]{Observatori Astronòmic, Universitat de València, C/Doctor Moliner, 50, Burjassot, 46100, València, Spain}

\affil[*]{Corresponding author: David Vallés-Pérez (david.valles-perez@uv.es), }
\affil[$\dagger$]{Vicent Quilis (vicent.quilis@uv.es), Susana Planelles (susana.planelles@uv.es)} 

\begin{abstract}
Cosmological accretion shocks created during the formation of galaxy clusters are a ubiquitous phenomenon all around the Universe. These shocks, and their features, are intimately related with the gravitational energy put into play during galaxy cluster formation. Studying a sample of simulated galaxy clusters and their associated accretion shocks, we show that objects in our sample sit in a plane within the three dimensional-space of cluster total mass, shock radius, and Mach number (a measure of shock intensity). Using this relation, and considering that forthcoming new observations will be able to measure shock radii and intensities, we put forward the idea that the dark matter content of galaxy clusters could be indirectly measured with an error up to around $30$ per cent at the $1\sigma$ confidence level. This procedure would be a new and independent method to measure the dark matter mass in cosmic structures, and a novel constraint to the accepted $\Lambda$CDM paradigm.
\end{abstract}

\begin{document}

\flushbottom
\maketitle
\thispagestyle{empty}

%\section*{Introduction}
Galaxy clusters and galaxy groups sit at the top of the hierarchy of gravitationally-bound cosmological structures, as the result of a hierarchical formation history which has lasted for around 14 Gyr \citep{Zeldovich_1970, Press_1974, Gott_1975}. While their matter content is dominated by dark matter (DM), which accounts for roughly $85\%$ of the gravitational mass, baryons (most of them, in the form of a diffuse, multiphase plasma; see, e.g., \cite{Bohringer_2010, Kravtsov_2012, Planelles_2015, Walker_2019}, for reviews) also play a pivotal role in their evolution and, especially, in the modelling of their observational properties \citep[e.g.,][]{Tozzi_2001, Nagai_2007, Bykov_2015}.

The singular place of galaxy clusters makes them valuable tools for precision cosmology \citep{Allen_2011, Weinberg_2013, Clerc_2022}, yet the determination of their masses is not straightforward and can only be done by different indirect approaches, often relying on rather strong assumptions. For instance, radial profiles of some thermodynamic properties of the intracluster medium (ICM; the bulk of baryonic mass in galaxy clusters) that can be obtained from X-ray or Sunyaev-Zeldovich (SZ) observations can be used to constrain the total mass, under the assumption of sphericity and hydrostatic equilibrium \citep[e.g.,][]{Biffi_2016, Ettori_2019}. Related to this, masses can also be obtained from integrated properties invoking self-similarity \citep{Lovisari_2022}. Both these approaches depend crucially on baryonic physics and feedback mechanisms, are affected by biases in the order of $(10-30)\%$ \citep{Biffi_2016}, and usually break at the group scale \citep[e.g.,][]{Giodini_2013}. Alternatively, masses can be determined by weak-lensing measurements \citep{Umetsu_2020}, or looking at the kinematics of galaxies in the cluster (e.g., through the caustic technique, \cite{Diaferio_1999}), which are ultimately probes of the underlying gravitational potential. A more in-depth, recent review on current mass estimation techniques for galaxy clusters can be found in \cite{Pratt_2019}.

As a direct consequence of the collisional nature of the baryonic component, shock waves appear as pervading phenomena in cosmological structure formation. They play a central role in the evolution of galaxy groups and clusters, by providing the necessary non-adiabatic heating of the diffuse plasma up to the temperatures observed within cosmic structures \citep{Quilis_1998, Miniati_2000, Ryu_2003} at the expense of removing kinetic energy from bulk flows, so that the infalling baryons can virialise within the gravitational potential wells of DM haloes. In particular, both analytic models \citep{Bertschinger_1983, Shi_2016} and simulations \citep{Zhang_2021, Aung_2021} have studied the location of the outermost accretion shocks of galaxy groups and clusters, whose position and evolution are ultimately determined by the gravitational collapse of the given structure.

The detection of these elusive features in observations has been devoted considerable attention recently, being the target of numerous surveys using new telescopic facilities. Although very preliminary, first detections of large-scale shocks have been reported by using very different observational approaches: pressure jumps using Sunyaev-Zeldovich effect observed with the Planck satellite and the South Pole Telescope (SPT) \citep{Anbajagane_2022}, {$\gamma$}-ray observations with the VERITAS Cherenkov array \cite{Keshet_2017} or the Fermi Large Area Telescope \citep{Reiss_2018}, using UV absorption spectroscopy \citep{Holguin_2022}, or by measuring polarisation in radio observations \citep{Vernstrom_2023}.

In this work, we propose that total masses of clusters can be inferred from the physical size of the accretion shock shell and its intensity, and calibrate these relations across a broad redshift interval using high-resolution cosmological simulations. We refer the reader to the Methods section and the Supplementary Material for technical details, while we summarise our main results below.

\section*{Results}
%%%%%%%%%%%%%%%%%%%%%%%%%%%

We exemplify and further discuss our procedure for detecting the accretion shock shell and its equivalent radius, $R_\mathrm{sh}$, and strength, quantified through the average Mach number, $\mathcal{M}_\mathrm{sh}$, in the Methods section and in Supplementary Section \ref{s:suppl.visual}. Below, we discuss our results regarding the calibration of the multi-dimensional scaling relation, its evolution, and its scatter. 

\subsection*{Multivariate relation between total mass, shock intensity and shock radii}

Following the approach described in more detail in the Methods section, we use our simulation data to fit a linearised relation $\log_{10} M(<2R_\mathrm{vir})=f(\log_{10} R_\mathrm{sh}, \mathcal{M}_\mathrm{sh})$. The choice of measuring the mass in spherical apertures of $2 R_\mathrm{vir}$ is motivated by the fact that accretion shocks are located most of the times between $2R_\mathrm{vir}$ and $3 R_\mathrm{vir}$ (see Supplementary Section \ref{s:suppl.statistics} for a statistical summary of our sample), and hence this radius is a reasonable proxy for the mass producing the collapse of the overdensity and driving the shock evolution. However, we note that using sensibly smaller aperture radii leads to similarly accurate results (see Supplementary Section \ref{s:suppl.1rvir} for the equivalent results using $M(<R_\mathrm{vir})$).

\begin{figure}
    \centering
    \includegraphics[width=.8\textwidth]{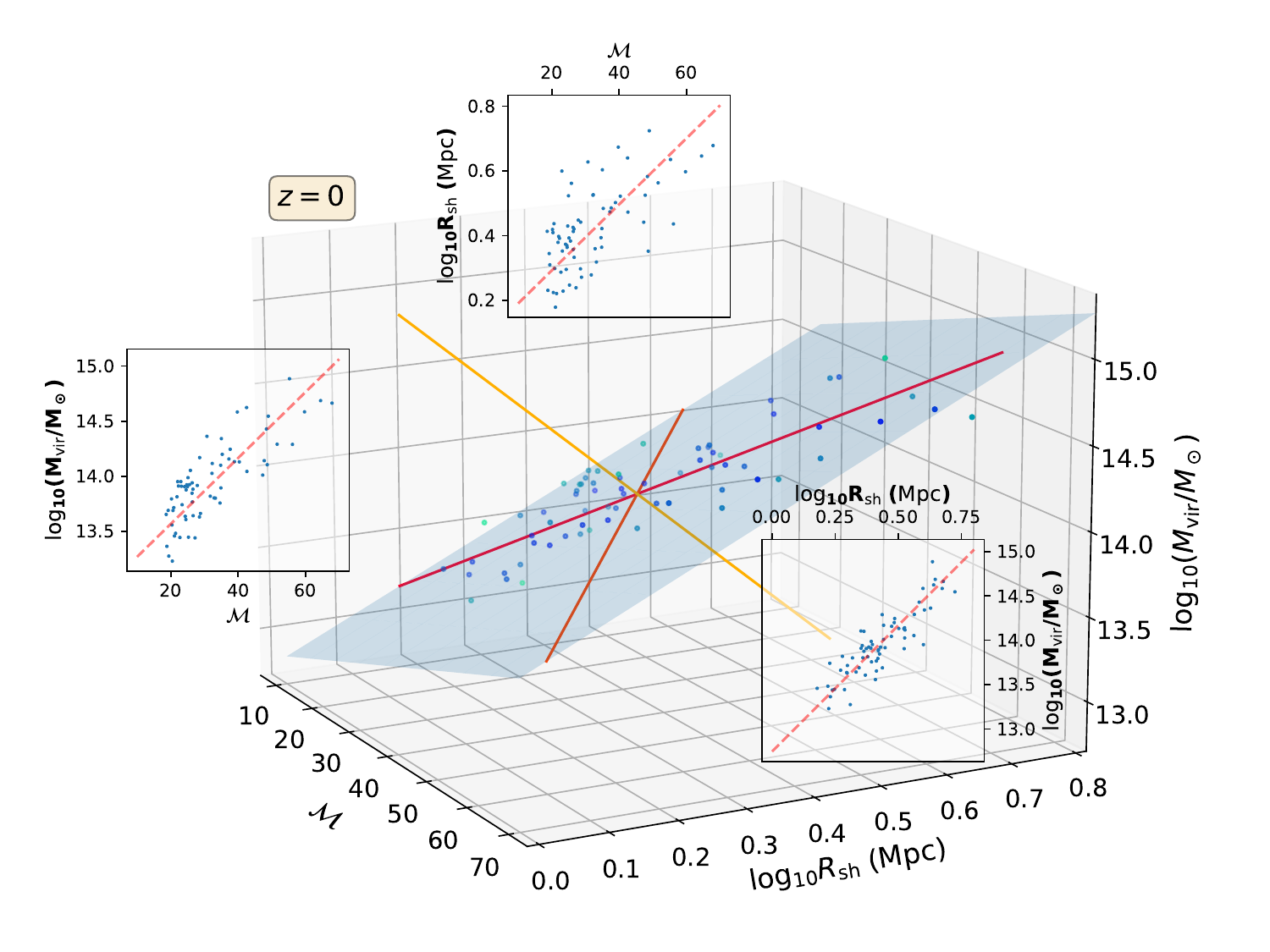}
    \includegraphics[width=.8\textwidth]{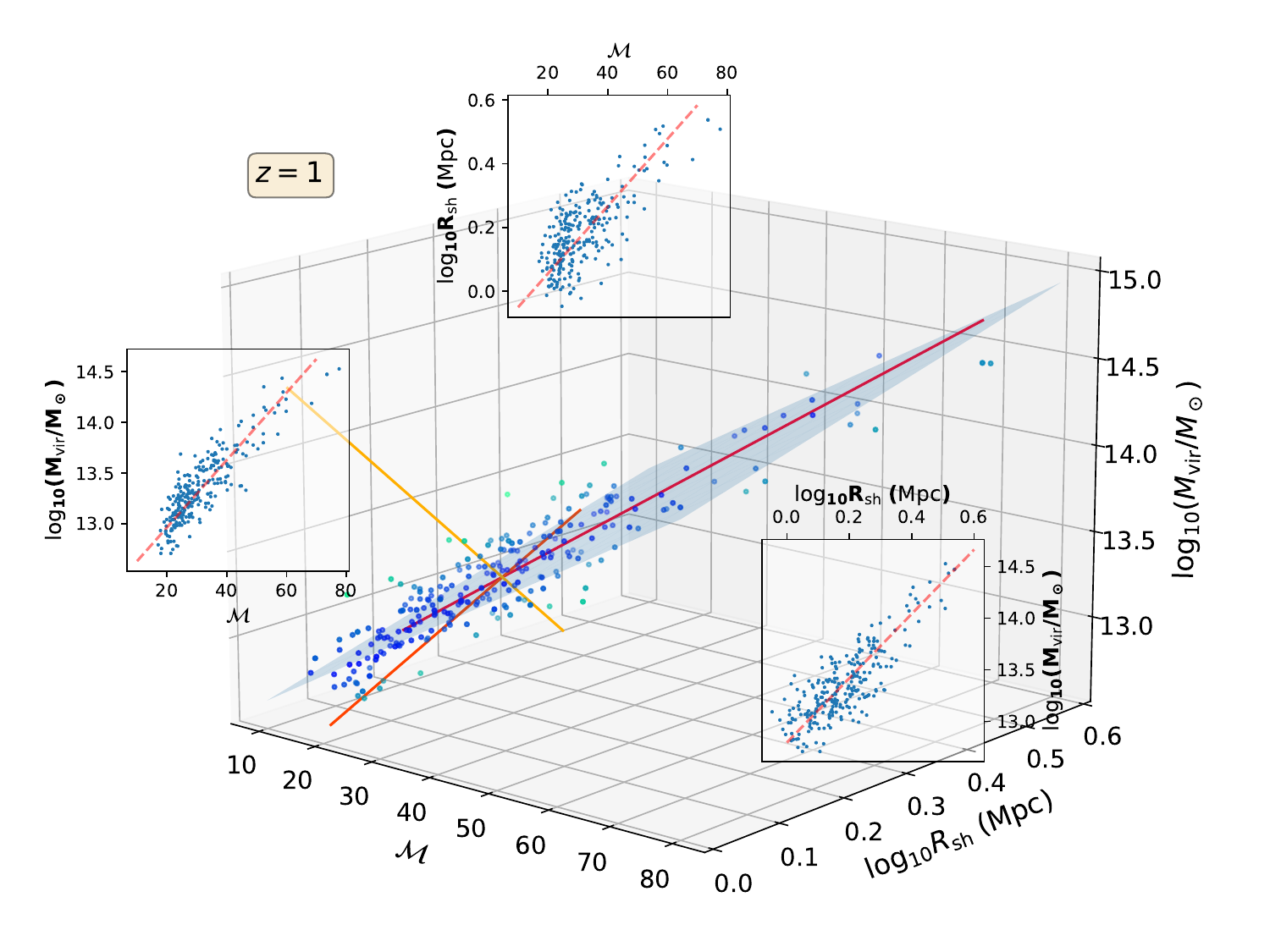}
    \caption{Best-fit relations $\log_{10} M(<2R_\mathrm{vir})=f(\log_{10} R_\mathrm{sh}, \mathcal{M}_\mathrm{sh})$, for $z=0$ (upper panel) and for $z=1$ (lower panel). In both panels, the information is presented as follows. Dots represent each individual cluster or group used for the fit, with their colour encoding the distance to the plane, darker being closer. The blue plane represents the best-fitting relation, which is spanned by the red and orange directions (representing the first and second principal components of the standardised data, respectively). The yellow line represents the third principal component. The plane is defined by setting this component to 0 (its mean value). The insets present the three marginal distributions, with the dashed, red line representing the best-fit linear relation to the marginal distribution.
    }
    \label{fig:3dscatter}
\end{figure}

Our best-fit relation for the sample at $z=0$ can be represented by the functional form

\begin{equation}
    \log_{10} \frac{M(<2R_\mathrm{vir})}{M_\odot} = 
    12.760 + 
    1.910 \log_{10} \frac{R_\mathrm{sh}}{\mathrm{Mpc}}  +
    0.0117 \mathcal{M}_\mathrm{sh}, \qquad \text{(at $z=0$)}
    \label{eq:fit_3d_z0}
\end{equation}

\noindent and is presented graphically in the upper panel of Fig. \ref{fig:3dscatter}. The mean scatter around this relation is $\sigma_{\log M} = 0.134$, implying $1 \sigma$ errors of $~^{+36\%}_{-27\%}$ on the total mass measured within $2 R_\mathrm{vir}$ apertures. The relation marginalised over shock intensities (the mass-shock radius relation, which we will hereon refer to as \textit{2d relation}, in contrast to the complete \text{3d relation}), has an intrinsic scatter of $\sigma_{2d,\,\log M} = 0.177$ ($~^{+50\%}_{-34\%}$ errors) at $z=0$, suggesting that the inclusion of information about shock intensity is crucial to recover more precise mass estimates. The slope of the 2d relation is marginally consistent with a self-similar scaling ($M(<2R_\mathrm{vir}) \propto R_\mathrm{sh}^3$) within a $1 \sigma$ interval (see Supplementary Section \ref{s:suppl.2d_relation} for more details on the 2d relation).

The lower panel of Fig. \ref{fig:3dscatter} contains the same information at $z=1$, which is best-fitted by the relation,

\begin{equation}
    \log_{10} \frac{M(<2R_\mathrm{vir})}{M_\odot} = 
    12.440 + 
    1.122 \log_{10} \frac{R_\mathrm{sh}}{\mathrm{cMpc}}  +
    0.0225 \mathcal{M}_\mathrm{sh}, \qquad \text{(at $z=1$)}
    \label{eq:fit_3d_z1}
\end{equation}

\noindent whose intrinsic scatter, $\sigma_{\log M} = 0.144$, implies typical errors of $~^{+39\%}_{-28\%}$ on the estimation of the mass enclosed by two virial radii. Also in this case, the addition of information about shock intensity decreases the error figure importantly, from a 2d value of $\sigma_{2d,\log M} = 0.209$ ($~^{+62\%}_{-38\%}$ errors).

Additional information about the fits in equations (\ref{eq:fit_3d_z0}) and (\ref{eq:fit_3d_z1}) can be found in Supplementary Section \ref{s:suppl.scatter_relations}. We have also checked that the mass accretion rate, when determined at the virial radius, does not correlate significantly with the position of the shock (see Supplementary Section \ref{s:suppl.accretion_rate}), and hence including it does not improve our results.

\subsection*{Evolution of the relation and fitting formulae}

We have studied the evolution of our three-dimensional scaling relation with cosmic time, for the redshift interval $0 \leq z \lesssim 1.5$. The resulting evolution can be fitted by a generalisation of the relation in equations (\ref{eq:fit_3d_z0}) and (\ref{eq:fit_3d_z1}), as

\begin{equation}
    \log_{10} \frac{M(<2R_\mathrm{vir})}{M_\odot} = \gamma(z) + \alpha(z) \log_{10} \frac{R_\mathrm{sh}}{\mathrm{cMpc}} + \beta(z) \mathcal{M}_\mathrm{sh}, \qquad \text{(at $0 \leq z \lesssim 1.5$)}
    \label{eq:fit_3d_evolution}
\end{equation}

\noindent where radii are expressed in comoving coordinates and the functions $\alpha(z)$, $\beta(z)$ and $\gamma(z)$ encapsulate the redshift evolution, which can be well-fitted by the polynomial forms below:

\begin{equation}
    \alpha(z) = 1.523 + 0.208 z - 0.386 z^2
    \label{eq:evolution_alpha}
\end{equation}

\begin{equation}
    \beta(z) = 0.01721 - 0.00374 z +0.00704 z^2
    \label{eq:evolution_beta}
\end{equation}

\begin{equation}
    \gamma(z) = 12.75 - 0.143 z - 0.144 z^2
    \label{eq:evolution_gamma}
\end{equation}

\begin{figure}
    \centering
    \includegraphics[width=.5\textwidth]{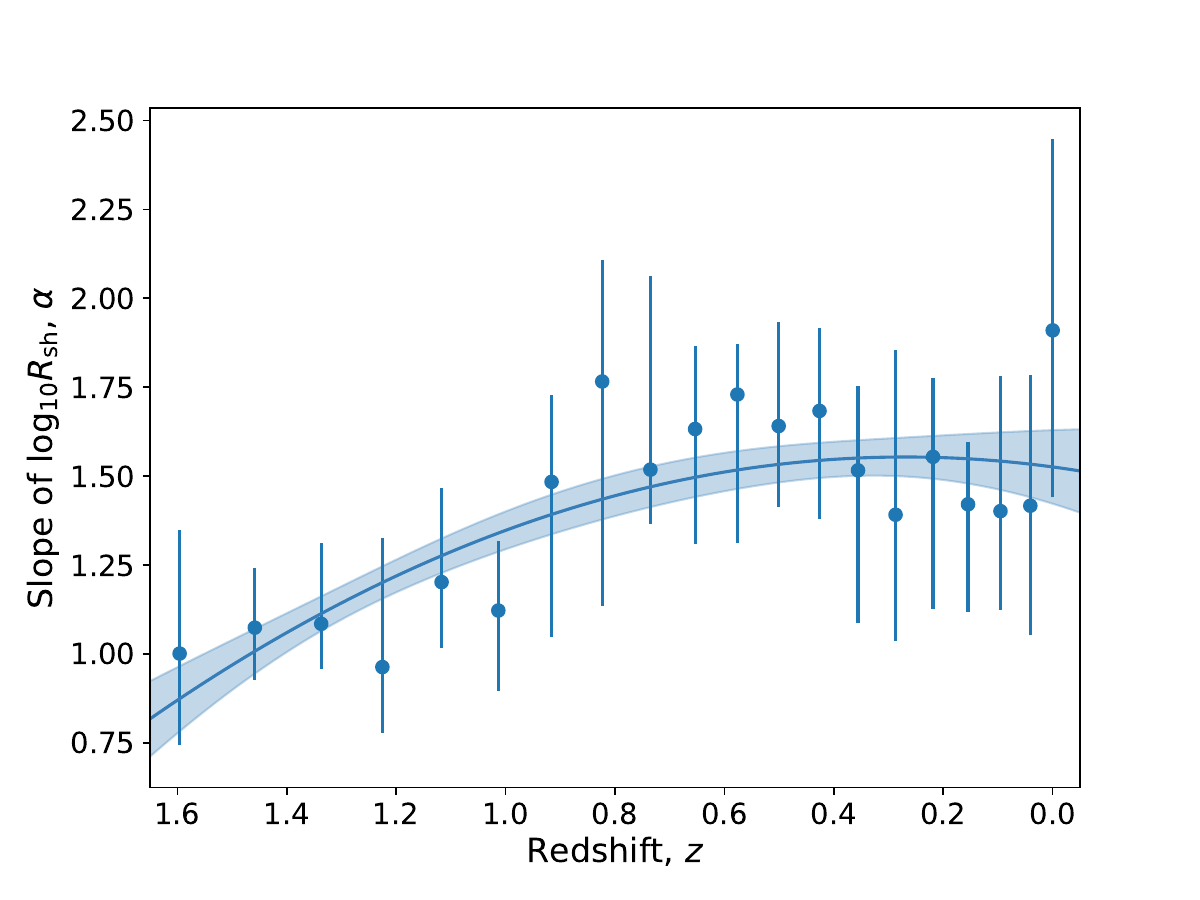}~
    \includegraphics[width=.5\textwidth]{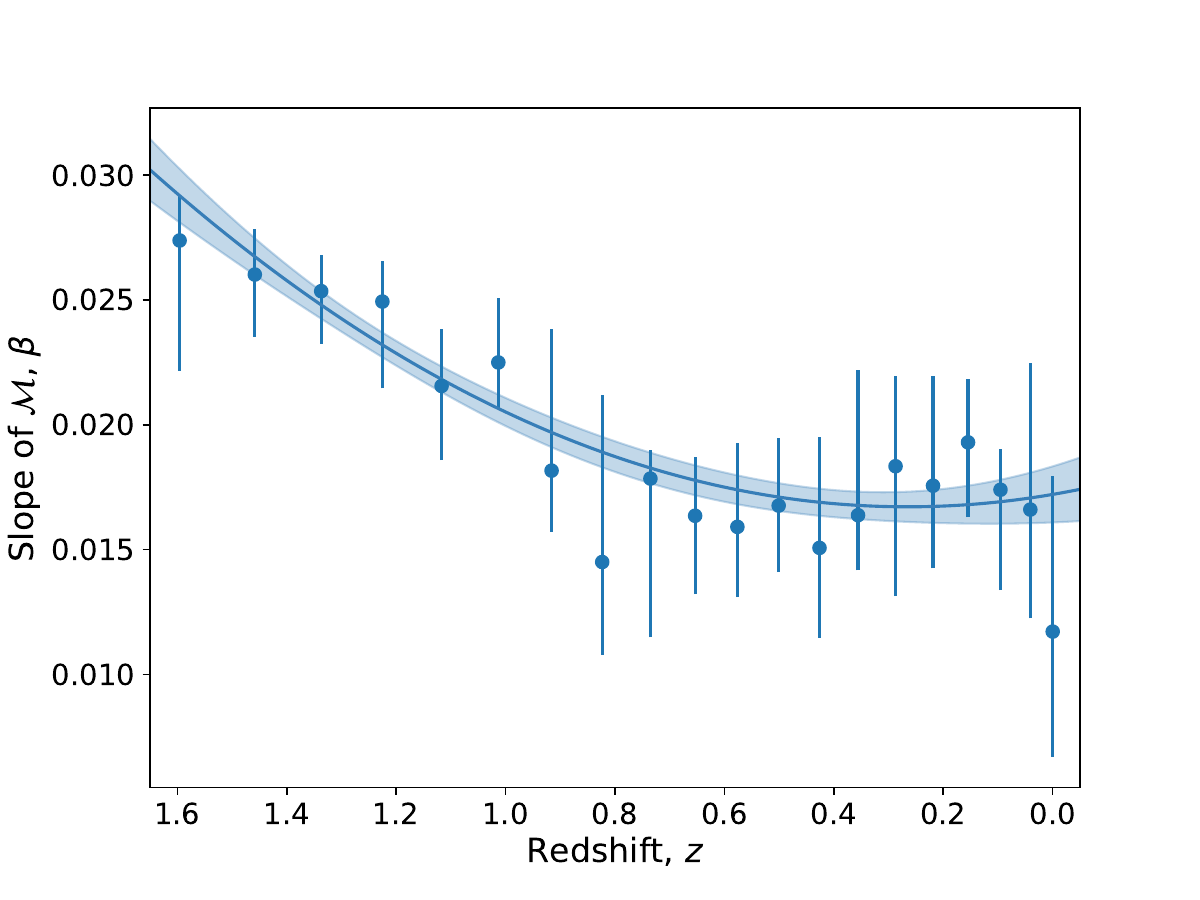}
    \includegraphics[width=.5\textwidth]{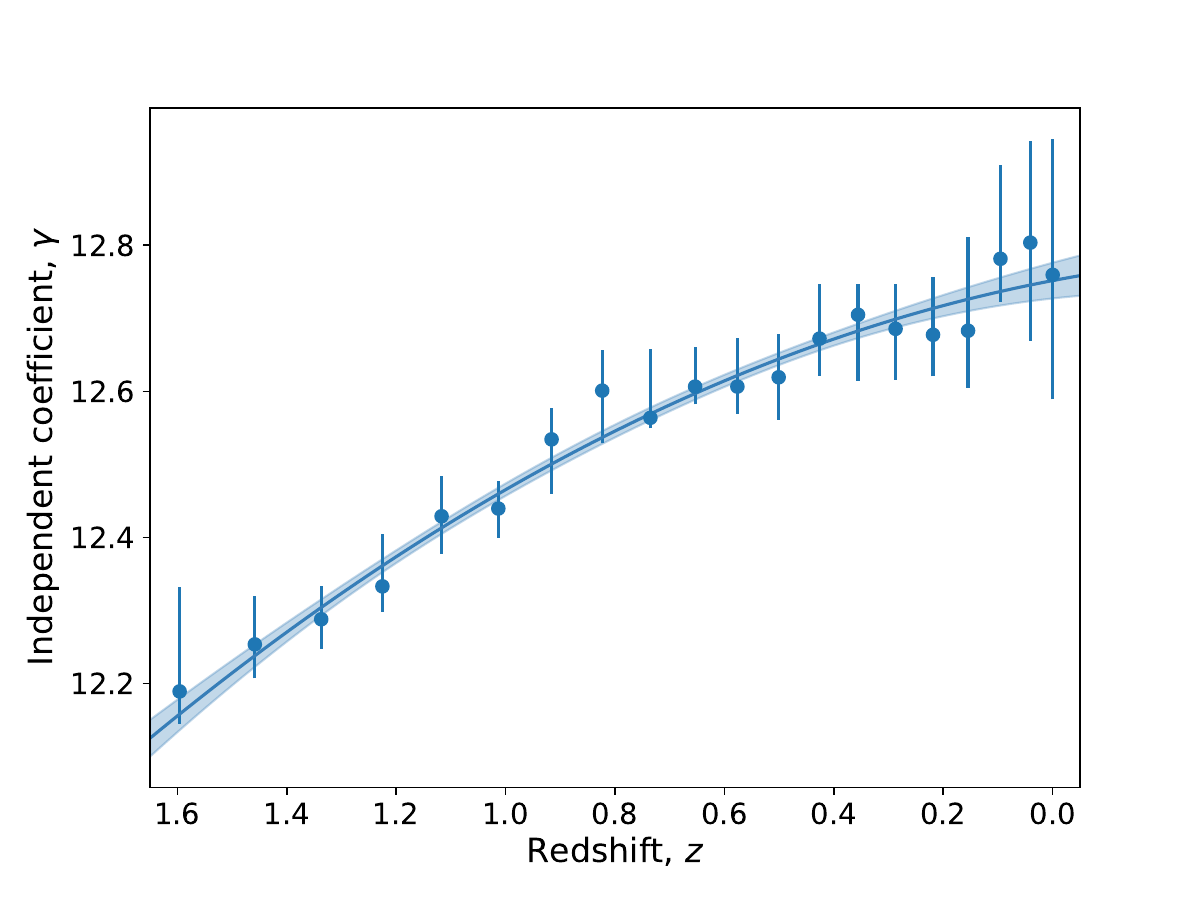}~
    \includegraphics[width=.5\textwidth]{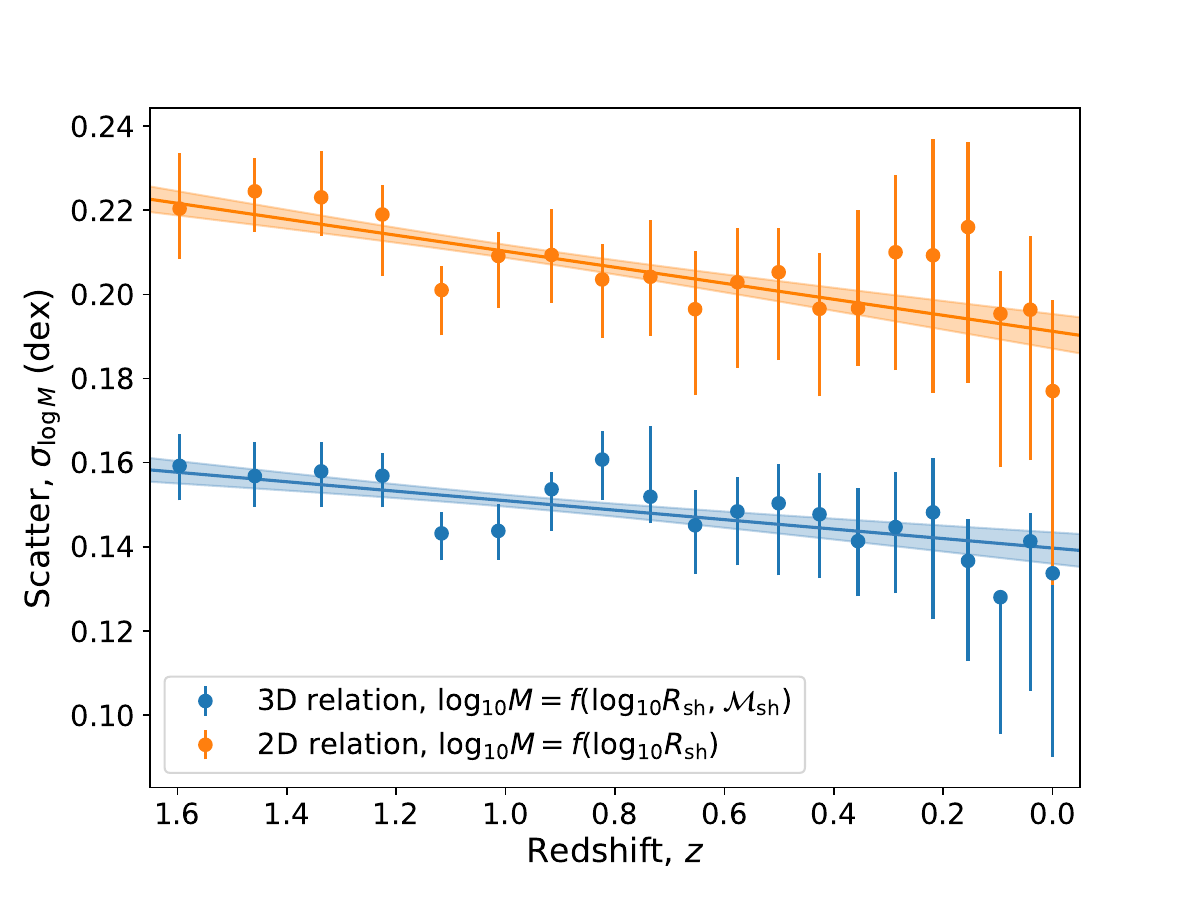}
    \caption{Scaling relation evolution summary. In all panels, dots represent the determinations at each redshift, with the bars accounting for $(16-84)\%$ percentiles obtained through bootstrap resampling. Solid lines with their shaded regions represent the best polynomial fits accounting for the evolution, with their $1\sigma$ confidence region. \textit{Top left:} evolution of the coefficient of the shock radius. \textit{Top right:} evolution of the coefficient of Mach number. \textit{Bottom left:} evolution of the independent term. \textit{Bottom right:} evolution of the scatter (in dex) around the three-dimensional scaling relation (blue line), and around the bidimensional scaling relation (orange line).}
    \label{fig:evolution}
\end{figure}

The results for $\alpha(z)$, $\beta(z)$ and $\gamma(z)$ are presented graphically in the upper left, upper right, and lower left panels of Fig. \ref{fig:evolution}, respectively. The lower right panel contains instead the evolution of the scatter around the fitted relation, evaluated over the same sample, $\sigma_{\log M}$. Here the blue line refers to the scatter on the three-dimensional relation, which obeys a slightly decreasing trend (from $\sigma_{\log M} \lesssim 0.16 \; \mathrm{dex}$ at $z\simeq 1.5$ to $\sigma_{\log M} \lesssim 0.14 \; \mathrm{dex}$ at $z\simeq 0$). As a comparison, the orange line presents the corresponding evolution for the bidimensional relation, i.e., the one not using information on shock intensity, confirming the behaviour observed in the examples in the previous section: adding information about shock intensity decreases the scatter across the whole redshift interval considered, and appears to be crucial in recovering precise mass estimates. We refer the reader to Supplementary Section \ref{s:suppl.covariances_evolution} for more details on the evolution of the fit results. The corresponding evolution, for $M(<R_\mathrm{vir})$, is also presented in Supplementary Section \ref{s:suppl.1rvir}.

\subsection*{Assessment of the relation and its scatter}

The study of the residuals and the scatter around our best-fit relations can be used as a calibration of the goodness of the fit and a prediction on the error intrinsic to the relation. At the same time, it is useful to assess the validity of the underlying assumptions and discussing the limitations of the model. 

\begin{figure}
    \centering
    \includegraphics[width=.5\textwidth]{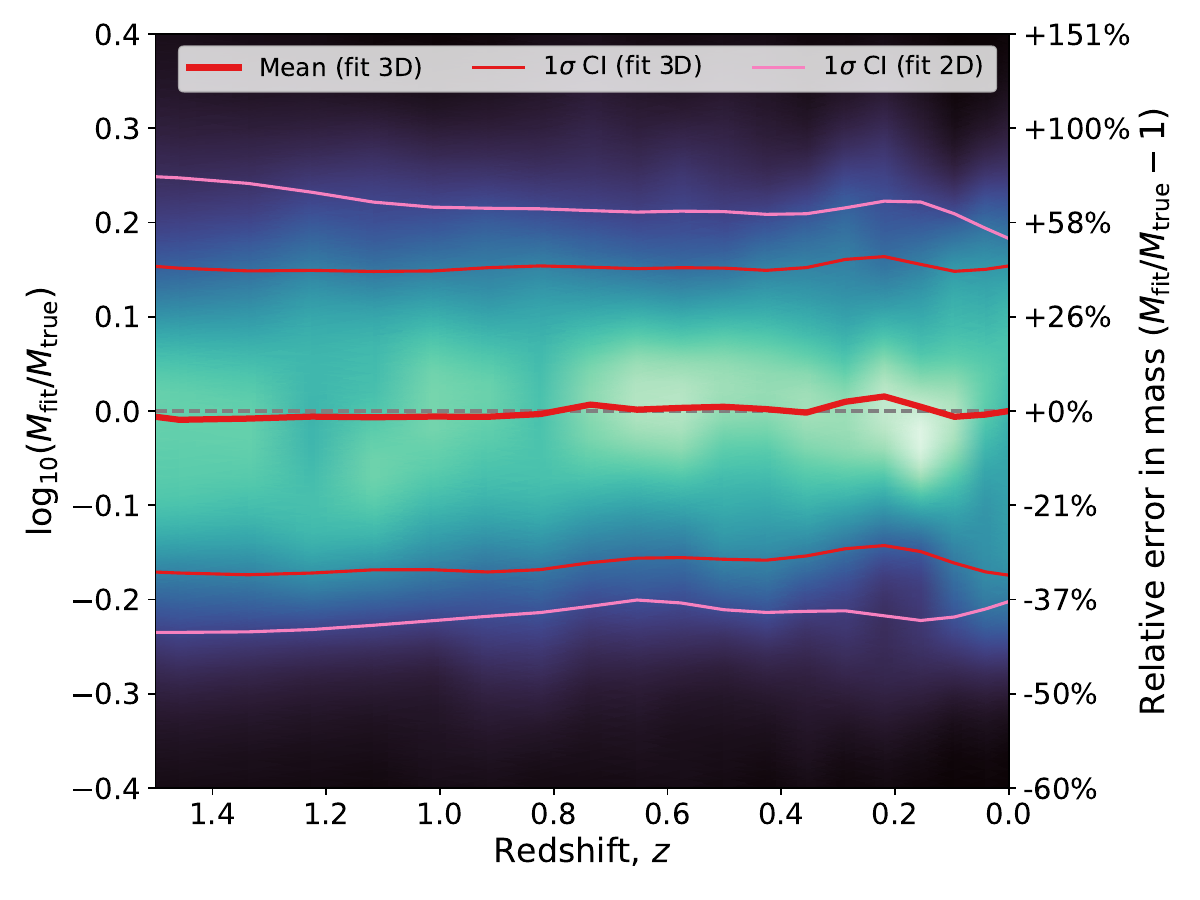}~
    \includegraphics[width=.5\textwidth]{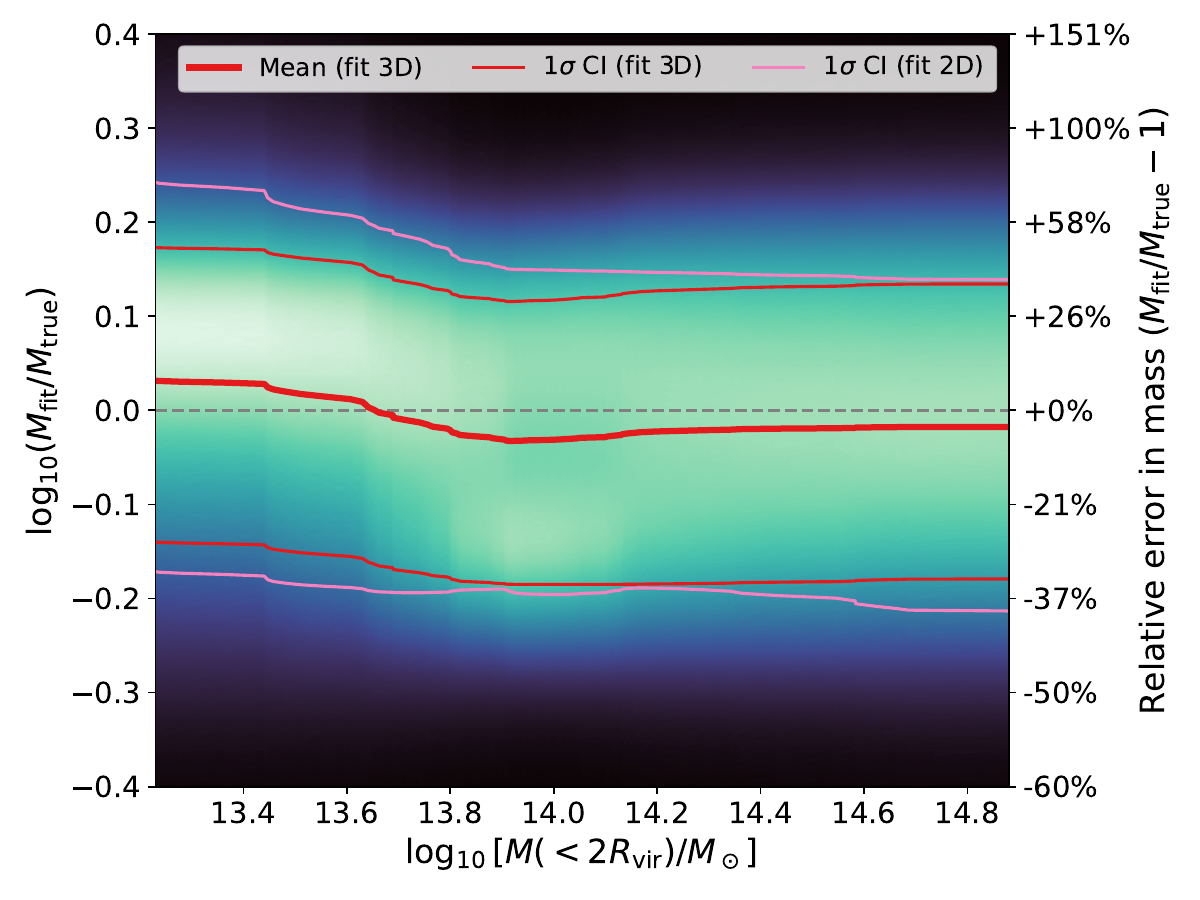}
    \caption{Distribution of the masses recovered by our best-fit relation (equation \ref{eq:fit_3d_evolution}, with the fits given in equations \ref{eq:evolution_alpha}-\ref{eq:evolution_gamma}) to assess its intrinsic scatter. In both cases, the vertical axis presents a conditional probability density estimation of the logarithmic error (in dex; shown as a percent error on the right of each panel), for a given redshift (left-hand panel) or for a given mass (right-hand panel; at $z=0$). That is to say, each vertical line depicts, colour-coded (the brightest, the highest), the probability distribution of the residuals, for a given redshift or mass. The thick, red line presents the mean value, while thin, red lines enclose the $(16-84)\%$ confidence region. Similarly, the thin, pink line contains the $(16-84)\%$ confidence region around the bidimensional, $\log_{10} M(<2R_\mathrm{vir}) = f(\log_{10} R_\mathrm{sh})$ relation.}
    \label{fig:residuals}
\end{figure}

In the left panel of Fig. \ref{fig:residuals}, we represent the evolution with cosmic time (decreasing redshift towards the right) of the scatter around our relation. For each value on the $x$-axis (i.e., for each redshift), the colour along the $y$-axis represents the distribution of the residuals with respect to our relation, the brightest being the most probable. Our relation is hence unbiased with redshift, in the sense that the mean value of $\log_{10} M_\mathrm{fit} / M_\mathrm{true}$ averages to 0 at all times. Here, red [pink] thin lines enclose the $(16-84)\%$ confidence region for the complete $\log_{10} M(<2R_\mathrm{vir})=f(\log_{10} R_\mathrm{sh}, \mathcal{M}_\mathrm{sh})$ [bidimensional, $\log_{10} M(<2R_\mathrm{vir})=f(\log_{10} R_\mathrm{sh})$] relation, showing again how the inclusion of shock intensity information makes the predictions considerably more precise.

At $z=0$, the residuals with respect to our scaling relation are presented as a function of the true mass in the right-hand side panel of Fig. \ref{fig:residuals}. The mean of the probability distribution of $\log_{10} M_\mathrm{fit}/M_\mathrm{true}$ (thick, red line) presents a noticeable, although rather small ($\lesssim 0.03 \, \mathrm{dex}$, corresponding to a $\lesssim 7\%$ bias), trend with mass. Note, however, that the mode of the distribution of $\log_{10} M_\mathrm{fit}/M_\mathrm{true}$ peaks at around $0.07 \, \mathrm{dex}$ for $\log_{10} M(<2R_\mathrm{vir}) \lesssim 13.8$, meaning that the most likely bias on the mass determination of a single low-mass group would be $\approx +17\%$. This behaviour, which is consistent with redshift (see Supplementary Section \ref{s:supplementary_scattermass}), is a natural consequence of having assumed a linearised model (equation \ref{eq:fit_3d_evolution}). While this model captures to a reasonably high extent the trends of the data, the fact that this behaviour with mass is consistent accross cosmic times, and not just a result of a statistical fluctuation, would imply that it can be easily corrected. We defer this possibility for future works using simulations with more statistics.

\section*{Discussion}

By using cosmological simulations, which track the coupled evolution of baryons and DM through the formation history of galaxy groups and clusters, we have studied the locations, intensities, and evolution of the outermost accretion shocks linked to these structures, where the collisional nature of baryonic matter enters into play and produces distinctive features on their structural, dynamical, thermodynamical and, potentially, observable profiles. 

Our main results show that location and intensity of these shocks, together with the total mass of the hosting structures, lie on a two-dimensional plane. We have calibrated this relation to provide an empirical, fitting formula for obtaining the total (dark and baryonic) mass of galaxy groups and clusters within large aperture radii of $2R_\mathrm{vir}$, valid in the broad redshift interval $ 0 \leq z \lesssim 1.5$.

From the numerical standpoint, future simulation works with enhanced statistics should go in the direction of confirming this relation with varying physical models and simulation codes, as well as deepening our understanding on possible residual biases. From the physical perspective, while the impact of star formation physics and associated feedback mechanisms are unlikely to make any noticeable difference on the fate of external shocks (see, for example, \cite{Planelles_2015}), other physical processes, such as physical viscosity or thermal conduction, might be worth considering when describing cluster outskirts \cite{Walker_2019, Walker_2022}. 

The determination of galaxy cluster masses is a long-standing problem that has been studied extensively and intensively in the literature. Several methods and observational strategies have been designed to tackle this issue, each of them having its own limitations and intrinsic biases. For instance, using galaxy kinematics, \citet{Andreon_2017} quote a $35\%$ scatter between caustic and true masses. Studies based on the hydrostatic equilibrium assumption estimate masses that are affected from a bias from the breach of this condition, which usually lies in the range $(10-30)\%$, as reported by different works \citep{Lau_2009, Biffi_2016, Angelinelli_2020}, while other factors (e.g., gas clumping, temperature inhomogeneities, etc.) may contribute to the scatter with similar magnitude \citep{Rasia_2014}. Regarding weak-lensing masses, the uncertainties introduced by the contamination by background, unlensed galaxies may reach for errors up to $40\%$ \citep{Okabe_2016}, to be added to uncertainties due to the mass modelling in the order of $\sim 10\%$ \citep{Meneghetti_2010}.

In this context, we put forward the idea that, provided that new strategies and observational facilities will detect and characterise these very large scale shocks, the results in the present work would allow to measure galaxy clusters masses within large apertures with  $1\sigma$ errors in the order of $\sim 30\%$.
Given these error figures, this novel method would become a new and independent manner to measure galaxy cluster masses that would be fully complementary, and with similar degree of uncertainty, to the previously mentioned procedures. 
 
As a result, our work would lead to an independent method to indirectly constrain the dark matter content in cosmological structures, and hence the currently accepted $\Lambda$CDM paradigm.

\FloatBarrier
\clearpage

\section*{Methods}

In this section, we describe the simulation where the results have been extracted from, the algorithm employed to detect and characterise shocks in the simulation outputs, the sample of galaxy groups and clusters extracted from the simulation, the method devised to identify the outer accretion shock of each object from our sample and, finally, the procedure we follow to perform the fits shown in this article.

%---------------------
\subsection*{The simulation}
\label{s:methods.simulation}
%---------------------

In this work, we analyse the outcomes of a $\Lambda$CDM simulation of a $L^3=(100 \, h^{-1} \, \mathrm{Mpc})^3$ cubic volume, run with the adaptive-mesh refinement (AMR; \citealp{Berger_1989}) code \texttt{MASCLET} \citep{Quilis_2004}. The evolution of the collisionless dynamics of dark matter is addressed by means of a multilevel and multispecies particle-mesh (PM; \citealp{Hockney_Eastwood_1988}), which takes advantage of the AMR grid to increase the force resolution of a monolithic PM as long as the mesh is refined. For the baryonic component, we make use of Eulerian, high-resolution shock-capturing techniques (higher-order versions of the method by \citep{Godunov_1959}), based on the piecewise parabolic method (PPM; \citealp{Colella_1984, Marti_1996}). These methods provide a faithful description of flow discontinuities and, therefore, are especially well-suited to handle shock waves without the need of introducing explicit artificial viscosity, and accurately describing the energy conversion processes (see, e.g., the discussion in \citep{Vazza_2009}).

The simulation assumes periodic boundary conditions and a flat cosmology with fiducial values for the cosmological parameters, which are consistent with the latest Planck Collaboration reported values \cite{Planck_2020}: matter density parameter $\Omega_\mathrm{m} = 0.31$, baryonic density parameter $ \Omega_\mathrm{b}=0.048$, Hubble dimensionless parameter $h=0.678$. The initial conditions correspond to a realisation of a Gaussian random field with spectrum $P(k) = A k^{n_s} T(k)$, whose spectral index is $n_s=0.96$ and the amplitude yields a normalisation $\sigma_8 = 0.82$ on $8 \, h^{-1} \, \mathrm{Mpc}$ scales. $T(k)$ is a CDM transfer function at $z=1000$ \cite{Eisenstein_1998}. These conditions are evolved, using the Zeldovich approximation \cite{Zeldovich_1970}, up to $z_\mathrm{ini}=100$, where the evolution with \texttt{MASCLET} starts. In order to refine the initial conditions taking advantage of the AMR scheme, we perform a first evolution on a $N_x^3 = 256^3$ cubic grid, from whose results at $z_\mathrm{fin}=0$ we choose the Lagrangian regions which will be mapped with enhanced resolution (using three nested levels of initial conditions) back at $z_\mathrm{ini}$, with a final best DM mass resolution of $1.48 \times 10^7 \, M_\odot$ and a total of around 392 million DM particles.

During the evolution, we dynamically and recursively refine regions based on a pseudo-Lagrangian criterion (local DM or baryonic density), as well as other criteria based on converging flows, Jeans length, or the presence of DM particles from refined regions in the initial conditions. With $n_\ell = 6$ refinement levels and $\Delta x_{\ell+1} / \Delta x_\ell = 1/2$, we achieve a peak resolution of $\Delta x_6 \simeq 9 \, \mathrm{kpc}$. Besides gravity and hydrodynamics, the simulation includes cooling \citep{Sutherland_1993} and a parametrization of heating from an extragalactic UV background \citep{Haardt_1996}, but does not include star formation nor other feedback mechanism. While feedback from supernovae and active galactic nuclei may have an important impact on the distribution of weak, internal shocks, they are most likely irrelevant for the study of accretion shocks \citep[see, e.g.,][]{Planelles_2021}.

%---------------------
\subsection*{The shock finder}
\label{s:methods.shock_finder}
%---------------------

In post-processing, shock waves in each snapshot of the simulation are detected and characterised making use of the shock finder presented by \cite{Planelles_2013}, which exploits the whole multi-resolution information of the outputs of our AMR code. The basic steps can be summarised as follows:

\begin{enumerate}
    \item Tentative shocked cells are flagged as those with a converging gas flow ($\nabla \cdot \vec{v} < 0$) and aligned temperature and entropy gradients ($\nabla T \cdot \nabla S > 0$).
    \item For each tentative shocked cell, we move to its left and to its right along the $x$ axis (equivalently, we repeat the process with the $y$ and $z$ axes), until we reach a non-shocked cell. We shall refer to these cells with the subindex `pre' and `post' (standing for preshock [or upstream of the shock] and postshock [or downstream of the shock], respectively), where $T_{\mathrm{post},x} > T_{\mathrm{pre},x}$. For a cell to be shocked, we also require that the gas densities verify $\rho_{\mathrm{post},x} > \rho_{\mathrm{pre},x}$ in consistency with the Rankine-Hugoniot jump conditions \citep[e.g.,][]{Landau_1959}. 
    \item The one-dimensional Mach number, $\mathcal{M}_x$, is computed from the Rankine-Hugoniot relation,

    \begin{equation}
        \frac{T_{\mathrm{post},x}}{T_{\mathrm{pre},x}} = \frac{(5 \mathcal{M}_x^2 - 1)(\mathcal{M}_x^2 + 3)}{16 \mathcal{M}_x^2},
        \label{eq:rankine-hugoniot}
    \end{equation}

    \noindent whose inversion yields

    \begin{equation}
        \mathcal{M}_x^2 = \frac{8q - 7 + 4 \sqrt{4q^2-7q+4}}{5}, \quad \text{with } q\equiv \frac{T_{\mathrm{post},x}}{T_{\mathrm{pre},x}} > 1.
        \label{eq:rankine-hugoniot_mach}
    \end{equation}

    \item The three-dimensional Mach number is obtained when combining the three directions as $\mathcal{M} = \left(\mathcal{M}_x^2 + \mathcal{M}_y^2 + \mathcal{M}_z^2 \right)^{1/2}$.
\end{enumerate}

Thus, the shock finder corresponds to a coordinate-splitting strategy \citep{Ryu_2003, Vazza_2009}, in which projection effects and numerical artefacts due to the Cartesian geometry are minimised by the averaging over the three spatial directions. This artificially imposed geometry is overcome by other family of shock finders, often dubbed \textit{coordinate-unsplit}, which use the direction of the local temperature gradient to identify the pre- and post-shock states \citep{Skillman_2008, Schaal_2015}. While splitting and unsplit algorithms differ significantly when dealing with weak, internal shocks, remarkable convergence is found for strong, external shocks \citep[their figure 4]{Skillman_2008}. Finally, while there are other possibilities to solve for the Mach number than the temperature jump, such as the velocity jump, \cite{Vazza_2009, Vazza_2011} report generally comparable results for both methods.

%---------------------
\subsection*{Halo catalogues}
\label{s:methods.haloes}
%---------------------

For each snapshot of the simulation, we have extracted halo catalogues using the public halo finder \texttt{ASOHF} \citep{Planelles_2010, Knebe_2011, Valles-Perez_2022}, which uses the spherical overdensity definition \citep{Cole_1996} to delimit the extent of haloes around density peaks, together with a number of particularities regarding unbinding mechanisms and handling of substructures. The extent of haloes is defined in our case by the virial radius, $R_\mathrm{vir}$, which is the radius enclosing an overdensity $\Delta_c = \langle \rho \rangle / \rho_\mathrm{c} = 18 \pi^2 + 82 x - 39 x^2$, where $x = \Omega_m(z) - 1$ and $\Omega_m(z)$ being the matter density parameter at a redshift $z$ \citep{Bryan_1998}.

At $z=0$, the simulated volume contains 31 well-resolved clusters with masses above $10^{14} \, M_\odot$ (the largest of them having a mass of $5.7\times 10^{14} \, M_\odot$), as well as 358 groups with DM masses exceeding $10^{13} \, M_\odot$.

%---------------------
\subsubsection*{Merger trees}
\label{s:methods.haloes.mtree}
%---------------------

Haloes amongst different snapshots are connected using the auxiliary \texttt{mtree.py} code of the \texttt{ASOHF} package, which identifies all objects contributing to a particular halo in a posterior snapshot. In particular, this code also determines the main progenitor of each halo at the immediately previous snapshot, using a criterion based on the most-bound particles. For more details, we refer the interested reader to the original paper of the halo finder \citep{Valles-Perez_2022}. Using the main branch we can compute the accretion rate proxy $\Gamma_\Delta$ \citep{Diemer_2014},

\begin{equation}
    \Gamma_\Delta = \frac{\Delta \log M_\Delta}{\Delta \log a},
    \label{eq:gamma_delta}
\end{equation}

\noindent where $a = (1+z)^{-1}$ is the scale factor of the FLRW metric, and $M_\Delta$ is any spherical-overdensity mass. While in \citet{Valles-Perez_2020} we characterised the instantaneous accretion rates using Savitzky-Golay \citet{Savitzky_1964} filters, here we have chosen to compute the accretion rate over the last dynamical time, given that the main effect we are studying in this work, i.e. the propagation of the external accretion shock, is the result of the continued accretion history over the object's history, rather than a quantity linked to the instantaneous growth of the halo at a particular overdensity.

%---------------------
\subsubsection*{Dynamical state classification}
\label{s:methods.haloes.dynstate}
%---------------------
At any given time, we split our sample of galaxy groups and clusters in a totally relaxed, a marginally relaxed and a disturbed class, following the methodology introduced by \citet{Valles-Perez_2023}, who calibrated redshift-dependent thresholds and weights on several parameters, such as centre offset ($\Delta_r$), virial ratio ($\eta$), mean radial velocity ($\langle \tilde v_r\rangle$), sparsity ($s_{200c,500c}$) and three-dimensional ellipticity $\epsilon$ of the dark matter halo, so as to correlate with the presence of mergers and/or strong accretion periods.

By $z=0$, out of the 31 clusters, 5 ($16\%$) are classified as totally relaxed, 15 ($58\%$) as marginally relaxed, and the rest (11, a $35\%$) are disturbed.

%---------------------
\subsection*{Accretion shock characterisation}
\label{s:methods.accretion_shock}
%---------------------

While above we have described the procedure to identify shock waves throughout the computational domain, here we describe the procedure by which we identify the outermost accretion shock of each group or cluster, and how we characterise it. 

We start by computing directional profiles of the Mach number, using $N_\theta \times N_\phi = 50 \times 50$ bins in solid angle, equally spaced in $\cos \theta$ and $\phi$. Alongside each angular direction, we take the directional profile of the Mach number, $\mathcal{M}$, from $r_\mathrm{min}=R_\mathrm{vir}$ to $r_\mathrm{max}=5 R_\mathrm{vir}$, using logarithmically spaced bins with $\Delta r = 0.01 \, \mathrm{dex}$. For these regards, unshocked cells are considered to have $\mathcal{M}=0$. The directional profiles are taken using linear interpolation from the data at the highest resolution (coarser than $\Delta r$) available at each point.

For each directional profile, $\mathcal{M}_{\theta,\phi}(r)$, we consider it crosses the accretion shock surface at a radial distance $r=R_\mathrm{sh}(\theta,\phi)$ if this distance corresponds to the largest local maximum of $\mathcal{M}_{\theta,\phi}(r)$ in the interval $[ r_\mathrm{min}, r_\mathrm{max} ]$, and its value exceeds a given threshold on the Mach number to be regarded as a strong shock, which we have set at $\mathcal{M}^\mathrm{strong}_\mathrm{thr} = 10$. Naturally, it may happen that these conditions are not met for a particular angular bin, in which case we consider that there is no crossing with the accretion shock shell in this direction. Objects where $R_\mathrm{sh}(\theta,\phi)$ is not identified in at least $75\%$ of the directions are purged from the sample.

In Suppl. Fig. \ref{figS1}, we show several visual impressions of the accretion shock shells recovered by this process, in order to demonstrate the robustness of the procedure. Besides the characterisation of $R_\mathrm{sh}(\theta,\phi)$, we also extract additional information, which we describe below:

\paragraph{Accretion shock equivalent radius.} In order to characterise the size of the accretion shock shell, we aim to obtain a single equivalent, or \textit{effective} radius, $R_\mathrm{sh}$. Near the interface between the accretion shock of the halo (where matter is infalling smoothly) and a filament, the radius may get increased (see Suppl. Fig. \ref{figS1}, e.g. its left panel), causing the distribution of values of $R_\mathrm{sh}(\theta,\phi)$ to be right-skewed. Therefore, the mean may be strongly biased high by these directions, even if they correspond to a small fraction of the solid angle around the halo. While the median is a more robust indicator, it is also sensitive to the presence of extreme values, especially when the number of directions ($\leq N_\theta N_\phi = 2500$) is not very high. The mode, or the most probable value of the distribution of $R_\mathrm{sh}(\theta,\phi)$, however, is insensitive to this skewness.

In order to determine the mode of the values of $R_\mathrm{sh}(\theta,\phi)$, which is a continuous variable, we estimate its probability density function (PDF), $f(R)$, by means of a kernel density estimation using Gaussian kernels (see Suppl. Fig. \ref{figS2}), whose bandwidth is fixed according to Scott's rule \citet{Scott_1992}. The shock radius is then set to the absolute maximum of the resulting PDF, $R_\mathrm{sh} = \mathrm{argmax} \, f(R)$. 

\paragraph{Accretion shock Mach number.} From the accretion shock shell, we can obtain the distribution of Mach number across it in a straightforward way, by setting $\mathcal{M}_\mathrm{sh}(\theta,\phi) = \mathcal{M}_{\theta, \phi} \left(r=R_\mathrm{sh}(\theta,\phi)\right)$. In order to obtain the characteristic strength of the shock, we define the effective Mach number as the average of $\mathcal{M}_\mathrm{sh}(\theta,\phi)$ over the shock shell.

%---------------------
\subsection*{Fits for the multi-dimensional scaling relations}
\label{s:methods.fits}
%---------------------

We describe here the procedure we follow to fit the multi-dimensional (linear) scaling relations that we show in Fig. \ref{fig:3dscatter}. We may exemplify it here with a bivariate relation, $Z(X,Y)$, although the procedure is totally analogous for relations of higher or lower number of independent variables.

\subsubsection*{Outlier removal}

Even though the variable $Z$ may be strongly correlated to $X$ and $Y$, the presence of outliers can significantly bias the fitting of linear relations and, thus, hinder our ability to recover the underlying relation. These outliers can correspond to a variety of effects, ranging from underresolved objects, to complex configurations where the outer shock detection algorithm has not reached a satisfactory solution, or either it is strongly disrupted. Therefore, before proceeding with the fits, we clean the sample from data points which depart significantly from the underlying trend.

Since $Z$ will most often be the mass, and this variable is not uniformly distributed (but instead there is an important preponderance of low-mass objects with respect to high-mass ones) care must be taken in this procedure so as not to get rid of the objects in the high-mass range.

To this end, we compute the conditional probability distribution of $Z$ given $(X,Y)$, by

\begin{equation}
    \rho(Z | X, Y) = \frac{\rho(X,Y,Z)}{\rho(X,Y)},
    \label{eq:conditional_probability}
\end{equation}

\noindent where $\rho(X,Y)$ is the two-dimensional probability density function of $X$ and $Y$ (marginalised over $Z$), and $\rho(X,Y,Z)$ is the complete, three-dimensional probability density function of $X$, $Y$ and $Z$. Both of them are estimated via a Gaussian kernel density estimation procedure, in a similar manner to what we have described before for the accretion shock location procedure. Then, we choose to remove the $5\%$ most unlikely data points (the ones with smallest value of $\rho(Z|X,Y)$) from our sample to fit our scaling relations. While this threshold is arbitrary, in our experiments it seems sufficient to prune the values that are visually far away from the general trend.

\paragraph{Filtering out of objects with a high uncertainty in $R_\mathrm{sh}$.} In some cases, as discussed in Supplementary Section \ref{s:suppl.visual} (see the right panel of Suppl. Fig. \ref{figS1}), the distribution of values of $R_\mathrm{sh}(\theta,\phi)$ is not monomodal, thus difficulting the determination of the equivalent radius, $R_\mathrm{sh}$. Aiming to filter out these cases to prevent these uncertainties from propagating to our results, we have constructed an ad-hoc criterion to flag them. Given a kernel-density estimate (KDE) distribution $f(R)$ of the values of $R_\mathrm{sh}(\theta,\phi)$, we consider it to be multimodal (and thus we discard it) if there is at least one local minimum of $f$ that simultaneously fulfils:

\begin{itemize}
    \item Its value is below $0.8$ times the minimum of the relative maxima that surround it.
    \item Both these maxima are above $0.5$ times the absolute maximum of $f$.
\end{itemize}

Strictly, $f$ would be multimodal if there it has at least one relative minimum. The first condition prevents that objects are discarded due to a relative minimum between two maxima that is not significant, and may be present due to sampling noise. The second condition prevents that we discard objects that exhibit fluctuations in the tails of the KDE, which do not introduce confusion in our determination of $R_\mathrm{sh}$.

\subsubsection*{Fit using principal component analysis}

We describe here the process we follow in order to find the best plane (or, equivalently, the line or the hyperplane, for fewer or higher dimensions) fitting our data. Although a straightforward possibility would be to use ordinary least squares (OLS) to fit $Z(X,Y)$ to the desired functional form, we do not pursue this procedure here. The main reason behind this is the fact that, by minimising the residuals between $Z_i$ and $Z(X_i, Y_i)$, the symmetry amongst variables is broken, i.e., the result of a fit $Z(X,Y)$ would be different from a fit $X(Y,Z)$ or $Y(X,Z)$. Aiming to find a general relation between these variables, we find no reason to break this symmetry, and perform instead a total least squares fit (TLS), where the distance from the points to the plane is minimised (instead of just the $Z$-axis distance, as it is done with OLS). Naturally, in doing these fits, we work with standardised variables, since otherwise the distances we made reference to, and hence our resulting fit, would depend on the unit system,

\begin{equation}
    \tilde{x} = \frac{X - \mu_x}{\sigma_x}, \quad
    \tilde{y} = \frac{Y - \mu_y}{\sigma_y}, \quad
    \tilde{z} = \frac{Z - \mu_z}{\sigma_z},
\end{equation}

\noindent where $\mu_{x(,y,z)}$ and $\sigma_{x(,y,z)}$ are, respectively, the mean and standard deviation of $X$(, $Y$, $Z$). 

To find the plane that these data best fit to, we apply standard principal component analysis (PCA; \cite{pca}). We compute the covariance matrix,

\begin{equation}
    \Sigma_{ij} = \sum_{k=1}^N \tilde x^i_{(k)} \tilde x^j_{(k)}, \qquad \qquad \text{where } \{ \tilde x^i_{(k)} \}_{i=1}^3 = \{\tilde{x}_{(k)}, \tilde{y}_{(k)}, \tilde{z}_{(k)} \}
    \label{eq:correlation}
\end{equation}

\noindent which is symmetric and hence diagonalisable with real eigenvalues and orthogonal eigenvectors. Therefore, there exists a basis in which the correlation matrix is diagonal, implying that a change of basis would make the components uncorrelated. These are the so-called principal components ($\{ PC_i \}_{i=1}^3$). The eigenvalue associated to each principal component represents, if properly normalised, the fraction of the variance explained by this component ($\lambda_i$). Therefore, if we sort the principal components in non-increasing order of their eigenvalues ($\lambda_1 \geq \lambda_2 \geq \lambda_3$), then the best-fitting plane will be the one spanned by $PC_1$ and $PC_2$, i.e., it will be orthogonal to the third eigenvector. It must be borne in mind that this orthogonality is not preserved when going back to the original variables from the standardised ones. Therefore, if we call $\vec{v_3}$ the eigenvector associated to $PC_3$, the best-fitting plane would then be

\begin{equation}
    v_3^{(1)} \tilde x + v_3^{(2)} \tilde y + v_3^{(3)} \tilde z = 0,
    \label{eq:methods_best_fit_plane_std}
\end{equation}

\noindent which, converted back to the original variables would read:

\begin{equation}
    Z = \frac{\sigma_z}{v_3^{(3)}} \left[ \left(\frac{v_3^{(1)} \mu_x}{\sigma_x} + \frac{v_3^{(2)} \mu_y}{\sigma_y} + \frac{v_3^{(3)} \mu_z}{\sigma_z} \right) - \frac{v_3^{(1)}}{\sigma_x} X - \frac{v_3^{(2)}}{\sigma_y} Y  \right].
    \label{eq:methods_best_fit_plane}
\end{equation}

This methodology can be extended in a straightforward way to fit a $(N-1)$-dimensional hyperplane to $N$-dimensional data.

%---------------------
\subsection*{Fits for the evolution of the parameters}
\label{s:methods.fits_evolution}
%---------------------

In order to produce the fits shown in Fig. \ref{fig:evolution}, we have followed a similar methodology to the one described in \citet{Valles-Perez_2023}. The evolution with redshift of each of the quantities shown is fitted, by ordinary least squares weighted to the inverse variance (which had been estimated through bootstrap resampling), by polynomials of increasing degree, until the highest degree coefficient is insignificant ($p$-value above 0.046 or reduced chi-squared below 0.25).

\newpage

\subsection*{Acknowledgements}

This work has been supported by the Agencia Estatal de Investigación Española (AEI; grant PID2022-138855NB-C33), by the Ministerio de Ciencia e Innovación (MCIN) within the Plan de Recuperación, Transformación y Resiliencia del Gobierno de España through the project ASFAE/2022/001, with funding from European Union NextGenerationEU (PRTR-C17.I1), and by the Generalitat Valenciana (grant CIPROM/2022/49). DVP acknowledges support from Universitat de València through an Atracció de Talent fellowship. Simulations have been carried out using the supercomputer Lluís Vives at the Servei d'Informàtica of the Universitat de València.

\subsection*{Author contributions statement}

VQ initiated the project. VQ and DVP ran the cosmological simulation. SP and DVP developed the shock and halo finders. DVP performed the data analysis and produced the figures. DVP, SP, and VQ discussed the results and wrote this paper.  

\subsection*{Additional information}
The authors declare no competing interest.

\subsection*{Data availability}
The data underlying this article will be shared upon reasonable request to the corresponding author.

\subsection*{Code availability}
The halo finder \texttt{ASOHF} is publicly available in \url{https://github.com/dvallesp/ASOHF}. The shock finder, the simulation code (\texttt{MASCLET}) and the codes for analysing the output data and producing the figures will be shared upon reasonable request to the corresponding author.

\bibliography{main}

\begin{thebibliography}{10}
\urlstyle{rm}
\expandafter\ifx\csname url\endcsname\relax
  \def\url#1{\texttt{#1}}\fi
\expandafter\ifx\csname urlprefix\endcsname\relax\def\urlprefix{URL }\fi
\expandafter\ifx\csname doiprefix\endcsname\relax\def\doiprefix{DOI: }\fi
\providecommand{\bibinfo}[2]{#2}
\providecommand{\eprint}[2][]{\url{#2}}

\bibitem{Zeldovich_1970}
\bibinfo{author}{{Zel'dovich}, Y.~B.}
\newblock \bibinfo{journal}{\bibinfo{title}{{Gravitational instability: An approximate theory for large density perturbations.}}}
\newblock {\emph{\JournalTitle{\aap}}} \textbf{\bibinfo{volume}{5}}, \bibinfo{pages}{84--89} (\bibinfo{year}{1970}).

\bibitem{Press_1974}
\bibinfo{author}{{Press}, W.~H.} \& \bibinfo{author}{{Schechter}, P.}
\newblock \bibinfo{journal}{\bibinfo{title}{{Formation of Galaxies and Clusters of Galaxies by Self-Similar Gravitational Condensation}}}.
\newblock {\emph{\JournalTitle{\apj}}} \textbf{\bibinfo{volume}{187}}, \bibinfo{pages}{425--438}, \doiprefix\url{10.1086/152650} (\bibinfo{year}{1974}).

\bibitem{Gott_1975}
\bibinfo{author}{{Gott}, I., J.~R.} \& \bibinfo{author}{{Rees}, M.~J.}
\newblock \bibinfo{journal}{\bibinfo{title}{{A theory of galaxy formation and clustering.}}}
\newblock {\emph{\JournalTitle{\aap}}} \textbf{\bibinfo{volume}{45}}, \bibinfo{pages}{365--376} (\bibinfo{year}{1975}).

\bibitem{Bohringer_2010}
\bibinfo{author}{{B{\"o}hringer}, H.} \& \bibinfo{author}{{Werner}, N.}
\newblock \bibinfo{journal}{\bibinfo{title}{{X-ray spectroscopy of galaxy clusters: studying astrophysical processes in the largest celestial laboratories}}}.
\newblock {\emph{\JournalTitle{\aapr}}} \textbf{\bibinfo{volume}{18}}, \bibinfo{pages}{127--196}, \doiprefix\url{10.1007/s00159-009-0023-3} (\bibinfo{year}{2010}).

\bibitem{Kravtsov_2012}
\bibinfo{author}{{Kravtsov}, A.~V.} \& \bibinfo{author}{{Borgani}, S.}
\newblock \bibinfo{journal}{\bibinfo{title}{{Formation of Galaxy Clusters}}}.
\newblock {\emph{\JournalTitle{\araa}}} \textbf{\bibinfo{volume}{50}}, \bibinfo{pages}{353--409}, \doiprefix\url{10.1146/annurev-astro-081811-125502} (\bibinfo{year}{2012}).
\newblock \eprint{1205.5556}.

\bibitem{Planelles_2015}
\bibinfo{author}{{Planelles}, S.}, \bibinfo{author}{{Schleicher}, D.~R.~G.} \& \bibinfo{author}{{Bykov}, A.~M.}
\newblock \bibinfo{journal}{\bibinfo{title}{{Large-Scale Structure Formation: From the First Non-linear Objects to Massive Galaxy Clusters}}}.
\newblock {\emph{\JournalTitle{\ssr}}} \textbf{\bibinfo{volume}{188}}, \bibinfo{pages}{93--139}, \doiprefix\url{10.1007/s11214-014-0045-7} (\bibinfo{year}{2015}).
\newblock \eprint{1404.3956}.

\bibitem{Walker_2019}
\bibinfo{author}{{Walker}, S.} \emph{et~al.}
\newblock \bibinfo{journal}{\bibinfo{title}{{The Physics of Galaxy Cluster Outskirts}}}.
\newblock {\emph{\JournalTitle{\ssr}}} \textbf{\bibinfo{volume}{215}}, \bibinfo{pages}{7}, \doiprefix\url{10.1007/s11214-018-0572-8} (\bibinfo{year}{2019}).
\newblock \eprint{1810.00890}.

\bibitem{Tozzi_2001}
\bibinfo{author}{{Tozzi}, P.} \& \bibinfo{author}{{Norman}, C.}
\newblock \bibinfo{journal}{\bibinfo{title}{{The Evolution of X-Ray Clusters and the Entropy of the Intracluster Medium}}}.
\newblock {\emph{\JournalTitle{\apj}}} \textbf{\bibinfo{volume}{546}}, \bibinfo{pages}{63--84}, \doiprefix\url{10.1086/318237} (\bibinfo{year}{2001}).
\newblock \eprint{astro-ph/0003289}.

\bibitem{Nagai_2007}
\bibinfo{author}{{Nagai}, D.}, \bibinfo{author}{{Kravtsov}, A.~V.} \& \bibinfo{author}{{Vikhlinin}, A.}
\newblock \bibinfo{journal}{\bibinfo{title}{{Effects of Galaxy Formation on Thermodynamics of the Intracluster Medium}}}.
\newblock {\emph{\JournalTitle{\apj}}} \textbf{\bibinfo{volume}{668}}, \bibinfo{pages}{1--14}, \doiprefix\url{10.1086/521328} (\bibinfo{year}{2007}).
\newblock \eprint{astro-ph/0703661}.

\bibitem{Bykov_2015}
\bibinfo{author}{{Bykov}, A.~M.} \emph{et~al.}
\newblock \bibinfo{journal}{\bibinfo{title}{{Structures and Components in Galaxy Clusters: Observations and Models}}}.
\newblock {\emph{\JournalTitle{\ssr}}} \textbf{\bibinfo{volume}{188}}, \bibinfo{pages}{141--185}, \doiprefix\url{10.1007/s11214-014-0129-4} (\bibinfo{year}{2015}).
\newblock \eprint{1512.01456}.

\bibitem{Allen_2011}
\bibinfo{author}{{Allen}, S.~W.}, \bibinfo{author}{{Evrard}, A.~E.} \& \bibinfo{author}{{Mantz}, A.~B.}
\newblock \bibinfo{journal}{\bibinfo{title}{{Cosmological Parameters from Observations of Galaxy Clusters}}}.
\newblock {\emph{\JournalTitle{\araa}}} \textbf{\bibinfo{volume}{49}}, \bibinfo{pages}{409--470}, \doiprefix\url{10.1146/annurev-astro-081710-102514} (\bibinfo{year}{2011}).
\newblock \eprint{1103.4829}.

\bibitem{Weinberg_2013}
\bibinfo{author}{{Weinberg}, D.~H.} \emph{et~al.}
\newblock \bibinfo{journal}{\bibinfo{title}{{Observational probes of cosmic acceleration}}}.
\newblock {\emph{\JournalTitle{\physrep}}} \textbf{\bibinfo{volume}{530}}, \bibinfo{pages}{87--255}, \doiprefix\url{10.1016/j.physrep.2013.05.001} (\bibinfo{year}{2013}).
\newblock \eprint{1201.2434}.

\bibitem{Clerc_2022}
\bibinfo{author}{Clerc, N.} \& \bibinfo{author}{Finoguenov, A.}
\newblock \emph{\bibinfo{title}{X-Ray Cluster Cosmology}}, \bibinfo{pages}{1--52} (\bibinfo{publisher}{Springer Nature Singapore}, \bibinfo{address}{Singapore}, \bibinfo{year}{2022}).

\bibitem{Biffi_2016}
\bibinfo{author}{{Biffi}, V.} \emph{et~al.}
\newblock \bibinfo{journal}{\bibinfo{title}{{On the Nature of Hydrostatic Equilibrium in Galaxy Clusters}}}.
\newblock {\emph{\JournalTitle{\apj}}} \textbf{\bibinfo{volume}{827}}, \bibinfo{pages}{112}, \doiprefix\url{10.3847/0004-637X/827/2/112} (\bibinfo{year}{2016}).
\newblock \eprint{1606.02293}.

\bibitem{Ettori_2019}
\bibinfo{author}{{Ettori}, S.} \emph{et~al.}
\newblock \bibinfo{journal}{\bibinfo{title}{{Hydrostatic mass profiles in X-COP galaxy clusters}}}.
\newblock {\emph{\JournalTitle{\aap}}} \textbf{\bibinfo{volume}{621}}, \bibinfo{pages}{A39}, \doiprefix\url{10.1051/0004-6361/201833323} (\bibinfo{year}{2019}).
\newblock \eprint{1805.00035}.

\bibitem{Lovisari_2022}
\bibinfo{author}{Lovisari, L.} \& \bibinfo{author}{Maughan, B.~J.}
\newblock \emph{\bibinfo{title}{Scaling Relations of Clusters and Groups and Their Evolution}}, \bibinfo{pages}{1--50} (\bibinfo{publisher}{Springer Nature Singapore}, \bibinfo{address}{Singapore}, \bibinfo{year}{2022}).

\bibitem{Giodini_2013}
\bibinfo{author}{{Giodini}, S.} \emph{et~al.}
\newblock \bibinfo{journal}{\bibinfo{title}{{Scaling Relations for Galaxy Clusters: Properties and Evolution}}}.
\newblock {\emph{\JournalTitle{\ssr}}} \textbf{\bibinfo{volume}{177}}, \bibinfo{pages}{247--282}, \doiprefix\url{10.1007/s11214-013-9994-5} (\bibinfo{year}{2013}).
\newblock \eprint{1305.3286}.

\bibitem{Umetsu_2020}
\bibinfo{author}{{Umetsu}, K.}
\newblock \bibinfo{journal}{\bibinfo{title}{{Cluster-galaxy weak lensing}}}.
\newblock {\emph{\JournalTitle{\aapr}}} \textbf{\bibinfo{volume}{28}}, \bibinfo{pages}{7}, \doiprefix\url{10.1007/s00159-020-00129-w} (\bibinfo{year}{2020}).
\newblock \eprint{2007.00506}.

\bibitem{Diaferio_1999}
\bibinfo{author}{{Diaferio}, A.}
\newblock \bibinfo{journal}{\bibinfo{title}{{Mass estimation in the outer regions of galaxy clusters}}}.
\newblock {\emph{\JournalTitle{\mnras}}} \textbf{\bibinfo{volume}{309}}, \bibinfo{pages}{610--622}, \doiprefix\url{10.1046/j.1365-8711.1999.02864.x} (\bibinfo{year}{1999}).
\newblock \eprint{astro-ph/9906331}.

\bibitem{Pratt_2019}
\bibinfo{author}{{Pratt}, G.~W.} \emph{et~al.}
\newblock \bibinfo{journal}{\bibinfo{title}{{The Galaxy Cluster Mass Scale and Its Impact on Cosmological Constraints from the Cluster Population}}}.
\newblock {\emph{\JournalTitle{\ssr}}} \textbf{\bibinfo{volume}{215}}, \bibinfo{pages}{25}, \doiprefix\url{10.1007/s11214-019-0591-0} (\bibinfo{year}{2019}).
\newblock \eprint{1902.10837}.

\bibitem{Quilis_1998}
\bibinfo{author}{Quilis, V.}, \bibinfo{author}{Ibáñez, J.~M.} \& \bibinfo{author}{Sáez, D.}
\newblock \bibinfo{journal}{\bibinfo{title}{On the role of shock waves in galaxy cluster evolution}}.
\newblock {\emph{\JournalTitle{The Astrophysical Journal}}} \textbf{\bibinfo{volume}{502}}, \bibinfo{pages}{518}, \doiprefix\url{10.1086/305932} (\bibinfo{year}{1998}).

\bibitem{Miniati_2000}
\bibinfo{author}{Miniati, F.} \emph{et~al.}
\newblock \bibinfo{journal}{\bibinfo{title}{Properties of cosmic shock waves in large-scale structure formation}}.
\newblock {\emph{\JournalTitle{The Astrophysical Journal}}} \textbf{\bibinfo{volume}{542}}, \bibinfo{pages}{608}, \doiprefix\url{10.1086/317027} (\bibinfo{year}{2000}).

\bibitem{Ryu_2003}
\bibinfo{author}{{Ryu}, D.}, \bibinfo{author}{{Kang}, H.}, \bibinfo{author}{{Hallman}, E.} \& \bibinfo{author}{{Jones}, T.~W.}
\newblock \bibinfo{journal}{\bibinfo{title}{{Cosmological Shock Waves and Their Role in the Large-Scale Structure of the Universe}}}.
\newblock {\emph{\JournalTitle{\apj}}} \textbf{\bibinfo{volume}{593}}, \bibinfo{pages}{599--610}, \doiprefix\url{10.1086/376723} (\bibinfo{year}{2003}).
\newblock \eprint{astro-ph/0305164}.

\bibitem{Bertschinger_1983}
\bibinfo{author}{{Bertschinger}, E.}
\newblock \bibinfo{journal}{\bibinfo{title}{{Cosmological self-similar shock waves and galaxy formation}}}.
\newblock {\emph{\JournalTitle{\apj}}} \textbf{\bibinfo{volume}{268}}, \bibinfo{pages}{17--29}, \doiprefix\url{10.1086/160925} (\bibinfo{year}{1983}).

\bibitem{Shi_2016}
\bibinfo{author}{{Shi}, X.}
\newblock \bibinfo{journal}{\bibinfo{title}{{Locations of accretion shocks around galaxy clusters and the ICM properties: insights from self-similar spherical collapse with arbitrary mass accretion rates}}}.
\newblock {\emph{\JournalTitle{\mnras}}} \textbf{\bibinfo{volume}{461}}, \bibinfo{pages}{1804--1815}, \doiprefix\url{10.1093/mnras/stw1418} (\bibinfo{year}{2016}).
\newblock \eprint{1603.07183}.

\bibitem{Zhang_2021}
\bibinfo{author}{{Zhang}, C.}, \bibinfo{author}{{Zhuravleva}, I.}, \bibinfo{author}{{Kravtsov}, A.} \& \bibinfo{author}{{Churazov}, E.}
\newblock \bibinfo{journal}{\bibinfo{title}{{Evolution of splashback boundaries and gaseous outskirts: insights from mergers of self-similar galaxy clusters}}}.
\newblock {\emph{\JournalTitle{\mnras}}} \textbf{\bibinfo{volume}{506}}, \bibinfo{pages}{839--863}, \doiprefix\url{10.1093/mnras/stab1546} (\bibinfo{year}{2021}).
\newblock \eprint{2103.03850}.

\bibitem{Aung_2021}
\bibinfo{author}{{Aung}, H.}, \bibinfo{author}{{Nagai}, D.} \& \bibinfo{author}{{Lau}, E.~T.}
\newblock \bibinfo{journal}{\bibinfo{title}{{Shock and splash: gas and dark matter halo boundaries around {\ensuremath{\Lambda}}CDM galaxy clusters}}}.
\newblock {\emph{\JournalTitle{\mnras}}} \textbf{\bibinfo{volume}{508}}, \bibinfo{pages}{2071--2078}, \doiprefix\url{10.1093/mnras/stab2598} (\bibinfo{year}{2021}).
\newblock \eprint{2012.00977}.

\bibitem{Anbajagane_2022}
\bibinfo{author}{{Anbajagane}, D.} \emph{et~al.}
\newblock \bibinfo{journal}{\bibinfo{title}{{Shocks in the stacked Sunyaev-Zel'dovich profiles of clusters II: Measurements from SPT-SZ + Planck Compton-y map}}}.
\newblock {\emph{\JournalTitle{\mnras}}} \textbf{\bibinfo{volume}{514}}, \bibinfo{pages}{1645--1663}, \doiprefix\url{10.1093/mnras/stac1376} (\bibinfo{year}{2022}).
\newblock \eprint{2111.04778}.

\bibitem{Keshet_2017}
\bibinfo{author}{{Keshet}, U.}, \bibinfo{author}{{Kushnir}, D.}, \bibinfo{author}{{Loeb}, A.} \& \bibinfo{author}{{Waxman}, E.}
\newblock \bibinfo{journal}{\bibinfo{title}{{Preliminary Evidence for a Virial Shock around the Coma Galaxy Cluster}}}.
\newblock {\emph{\JournalTitle{\apj}}} \textbf{\bibinfo{volume}{845}}, \bibinfo{pages}{24}, \doiprefix\url{10.3847/1538-4357/aa794b} (\bibinfo{year}{2017}).
\newblock \eprint{1210.1574}.

\bibitem{Reiss_2018}
\bibinfo{author}{{Reiss}, I.} \& \bibinfo{author}{{Keshet}, U.}
\newblock \bibinfo{journal}{\bibinfo{title}{{Detection of virial shocks in stacked Fermi-LAT galaxy clusters}}}.
\newblock {\emph{\JournalTitle{\jcap}}} \textbf{\bibinfo{volume}{2018}}, \bibinfo{pages}{010}, \doiprefix\url{10.1088/1475-7516/2018/10/010} (\bibinfo{year}{2018}).
\newblock \eprint{1705.05376}.

\bibitem{Holguin_2022}
\bibinfo{author}{{Holguin Luna}, P.} \& \bibinfo{author}{{Burchett}, J.}
\newblock \bibinfo{title}{{Localizing the accretion shock and constraining gaseous conditions in galaxy cluster outskirts with UV absorption spectroscopy}}.
\newblock In \emph{\bibinfo{booktitle}{American Astronomical Society Meeting Abstracts}}, vol.~\bibinfo{volume}{54} of \emph{\bibinfo{series}{American Astronomical Society Meeting Abstracts}}, \bibinfo{pages}{427.02} (\bibinfo{year}{2022}).

\bibitem{Vernstrom_2023}
\bibinfo{author}{{Vernstrom}, T.} \emph{et~al.}
\newblock \bibinfo{journal}{\bibinfo{title}{{Polarized accretion shocks from the cosmic web}}}.
\newblock {\emph{\JournalTitle{Science Advances}}} \textbf{\bibinfo{volume}{9}}, \bibinfo{pages}{eade7233}, \doiprefix\url{10.1126/sciadv.ade7233} (\bibinfo{year}{2023}).
\newblock \eprint{2302.08072}.

\bibitem{Walker_2022}
\bibinfo{author}{{Walker}, S.} \& \bibinfo{author}{{Lau}, E.}
\newblock \bibinfo{title}{{Cluster Outskirts and Their Connection to the Cosmic Web}}.
\newblock In \emph{\bibinfo{booktitle}{Handbook of X-ray and Gamma-ray Astrophysics}}, \bibinfo{pages}{13}, \doiprefix\url{10.1007/978-981-16-4544-0_120-1} (\bibinfo{publisher}{Springer Living Reference Work}, \bibinfo{year}{2022}).

\bibitem{Andreon_2017}
\bibinfo{author}{{Andreon}, S.}, \bibinfo{author}{{Trinchieri}, G.}, \bibinfo{author}{{Moretti}, A.} \& \bibinfo{author}{{Wang}, J.}
\newblock \bibinfo{journal}{\bibinfo{title}{{Intrinsic scatter of caustic masses and hydrostatic bias: An observational study}}}.
\newblock {\emph{\JournalTitle{\aap}}} \textbf{\bibinfo{volume}{606}}, \bibinfo{pages}{A25}, \doiprefix\url{10.1051/0004-6361/201730920} (\bibinfo{year}{2017}).
\newblock \eprint{1706.08353}.

\bibitem{Lau_2009}
\bibinfo{author}{{Lau}, E.~T.}, \bibinfo{author}{{Kravtsov}, A.~V.} \& \bibinfo{author}{{Nagai}, D.}
\newblock \bibinfo{journal}{\bibinfo{title}{{Residual Gas Motions in the Intracluster Medium and Bias in Hydrostatic Measurements of Mass Profiles of Clusters}}}.
\newblock {\emph{\JournalTitle{\apj}}} \textbf{\bibinfo{volume}{705}}, \bibinfo{pages}{1129--1138}, \doiprefix\url{10.1088/0004-637X/705/2/1129} (\bibinfo{year}{2009}).
\newblock \eprint{0903.4895}.

\bibitem{Angelinelli_2020}
\bibinfo{author}{{Angelinelli}, M.} \emph{et~al.}
\newblock \bibinfo{journal}{\bibinfo{title}{{Turbulent pressure support and hydrostatic mass bias in the intracluster medium}}}.
\newblock {\emph{\JournalTitle{\mnras}}} \textbf{\bibinfo{volume}{495}}, \bibinfo{pages}{864--885}, \doiprefix\url{10.1093/mnras/staa975} (\bibinfo{year}{2020}).
\newblock \eprint{1905.04896}.

\bibitem{Rasia_2014}
\bibinfo{author}{{Rasia}, E.} \emph{et~al.}
\newblock \bibinfo{journal}{\bibinfo{title}{{Temperature Structure of the Intracluster Medium from Smoothed-particle Hydrodynamics and Adaptive-mesh Refinement Simulations}}}.
\newblock {\emph{\JournalTitle{\apj}}} \textbf{\bibinfo{volume}{791}}, \bibinfo{pages}{96}, \doiprefix\url{10.1088/0004-637X/791/2/96} (\bibinfo{year}{2014}).
\newblock \eprint{1406.4410}.

\bibitem{Okabe_2016}
\bibinfo{author}{{Okabe}, N.} \& \bibinfo{author}{{Smith}, G.~P.}
\newblock \bibinfo{journal}{\bibinfo{title}{{LoCuSS: weak-lensing mass calibration of galaxy clusters}}}.
\newblock {\emph{\JournalTitle{\mnras}}} \textbf{\bibinfo{volume}{461}}, \bibinfo{pages}{3794--3821}, \doiprefix\url{10.1093/mnras/stw1539} (\bibinfo{year}{2016}).
\newblock \eprint{1507.04493}.

\bibitem{Meneghetti_2010}
\bibinfo{author}{{Meneghetti}, M.} \emph{et~al.}
\newblock \bibinfo{journal}{\bibinfo{title}{{Weighing simulated galaxy clusters using lensing and X-ray}}}.
\newblock {\emph{\JournalTitle{\aap}}} \textbf{\bibinfo{volume}{514}}, \bibinfo{pages}{A93}, \doiprefix\url{10.1051/0004-6361/200913222} (\bibinfo{year}{2010}).
\newblock \eprint{0912.1343}.

\bibitem{Berger_1989}
\bibinfo{author}{{Berger}, M.~J.} \& \bibinfo{author}{{Colella}, P.}
\newblock \bibinfo{journal}{\bibinfo{title}{{Local Adaptive Mesh Refinement for Shock Hydrodynamics}}}.
\newblock {\emph{\JournalTitle{Journal of Computational Physics}}} \textbf{\bibinfo{volume}{82}}, \bibinfo{pages}{64--84}, \doiprefix\url{10.1016/0021-9991(89)90035-1} (\bibinfo{year}{1989}).

\bibitem{Quilis_2004}
\bibinfo{author}{{Quilis}, V.}
\newblock \bibinfo{journal}{\bibinfo{title}{{A new multidimensional adaptive mesh refinement hydro + gravity cosmological code}}}.
\newblock {\emph{\JournalTitle{\mnras}}} \textbf{\bibinfo{volume}{352}}, \bibinfo{pages}{1426--1438}, \doiprefix\url{10.1111/j.1365-2966.2004.08040.x} (\bibinfo{year}{2004}).
\newblock \eprint{astro-ph/0405389}.

\bibitem{Hockney_Eastwood_1988}
\bibinfo{author}{{Hockney}, R.~W.} \& \bibinfo{author}{{Eastwood}, J.~W.}
\newblock \emph{\bibinfo{title}{{Computer simulation using particles}}} (\bibinfo{publisher}{Institute of Physics Publishing}, \bibinfo{year}{1988}).

\bibitem{Godunov_1959}
\bibinfo{author}{Godunov, S.~K.} \& \bibinfo{author}{Bohachevsky, I.}
\newblock \bibinfo{journal}{\bibinfo{title}{{Finite difference method for numerical computation of discontinuous solutions of the equations of fluid dynamics}}}.
\newblock {\emph{\JournalTitle{{Matemati{\v c}eskij sbornik}}}} \textbf{\bibinfo{volume}{47(89)}}, \bibinfo{pages}{271--306} (\bibinfo{year}{1959}).

\bibitem{Colella_1984}
\bibinfo{author}{{Colella}, P.} \& \bibinfo{author}{{Woodward}, P.~R.}
\newblock \bibinfo{journal}{\bibinfo{title}{{The Piecewise Parabolic Method (PPM) for Gas-Dynamical Simulations}}}.
\newblock {\emph{\JournalTitle{Journal of Computational Physics}}} \textbf{\bibinfo{volume}{54}}, \bibinfo{pages}{174--201}, \doiprefix\url{10.1016/0021-9991(84)90143-8} (\bibinfo{year}{1984}).

\bibitem{Marti_1996}
\bibinfo{author}{{Mart{\'\i}}, J. M. S.~S.} \& \bibinfo{author}{{M{\"u}ller}, E.}
\newblock \bibinfo{journal}{\bibinfo{title}{{Extension of the Piecewise Parabolic Method to One-Dimensional Relativistic Hydrodynamics}}}.
\newblock {\emph{\JournalTitle{Journal of Computational Physics}}} \textbf{\bibinfo{volume}{123}}, \bibinfo{pages}{1--14}, \doiprefix\url{10.1006/jcph.1996.0001} (\bibinfo{year}{1996}).

\bibitem{Vazza_2009}
\bibinfo{author}{{Vazza}, F.}, \bibinfo{author}{{Brunetti}, G.} \& \bibinfo{author}{{Gheller}, C.}
\newblock \bibinfo{journal}{\bibinfo{title}{{Shock waves in Eulerian cosmological simulations: main properties and acceleration of cosmic rays}}}.
\newblock {\emph{\JournalTitle{\mnras}}} \textbf{\bibinfo{volume}{395}}, \bibinfo{pages}{1333--1354}, \doiprefix\url{10.1111/j.1365-2966.2009.14691.x} (\bibinfo{year}{2009}).
\newblock \eprint{0808.0609}.

\bibitem{Planck_2020}
\bibinfo{author}{{Planck Collaboration}} \emph{et~al.}
\newblock \bibinfo{journal}{\bibinfo{title}{{Planck 2018 results. VI. Cosmological parameters}}}.
\newblock {\emph{\JournalTitle{\aap}}} \textbf{\bibinfo{volume}{641}}, \bibinfo{pages}{A6}, \doiprefix\url{10.1051/0004-6361/201833910} (\bibinfo{year}{2020}).
\newblock \eprint{1807.06209}.

\bibitem{Eisenstein_1998}
\bibinfo{author}{{Eisenstein}, D.~J.} \& \bibinfo{author}{{Hu}, W.}
\newblock \bibinfo{journal}{\bibinfo{title}{{Baryonic Features in the Matter Transfer Function}}}.
\newblock {\emph{\JournalTitle{\apj}}} \textbf{\bibinfo{volume}{496}}, \bibinfo{pages}{605--614}, \doiprefix\url{10.1086/305424} (\bibinfo{year}{1998}).
\newblock \eprint{astro-ph/9709112}.

\bibitem{Sutherland_1993}
\bibinfo{author}{{Sutherland}, R.~S.} \& \bibinfo{author}{{Dopita}, M.~A.}
\newblock \bibinfo{journal}{\bibinfo{title}{{Cooling Functions for Low-Density Astrophysical Plasmas}}}.
\newblock {\emph{\JournalTitle{\apjs}}} \textbf{\bibinfo{volume}{88}}, \bibinfo{pages}{253}, \doiprefix\url{10.1086/191823} (\bibinfo{year}{1993}).

\bibitem{Haardt_1996}
\bibinfo{author}{{Haardt}, F.} \& \bibinfo{author}{{Madau}, P.}
\newblock \bibinfo{journal}{\bibinfo{title}{{Radiative Transfer in a Clumpy Universe. II. The Ultraviolet Extragalactic Background}}}.
\newblock {\emph{\JournalTitle{\apj}}} \textbf{\bibinfo{volume}{461}}, \bibinfo{pages}{20}, \doiprefix\url{10.1086/177035} (\bibinfo{year}{1996}).
\newblock \eprint{astro-ph/9509093}.

\bibitem{Planelles_2021}
\bibinfo{author}{{Planelles}, S.} \emph{et~al.}
\newblock \bibinfo{journal}{\bibinfo{title}{{Exploring the role of cosmological shock waves in the Dianoga simulations of galaxy clusters}}}.
\newblock {\emph{\JournalTitle{\mnras}}} \textbf{\bibinfo{volume}{507}}, \bibinfo{pages}{5703--5719}, \doiprefix\url{10.1093/mnras/stab2436} (\bibinfo{year}{2021}).
\newblock \eprint{2108.09670}.

\bibitem{Planelles_2013}
\bibinfo{author}{{Planelles}, S.} \& \bibinfo{author}{{Quilis}, V.}
\newblock \bibinfo{journal}{\bibinfo{title}{{Cosmological shock waves: clues to the formation history of haloes}}}.
\newblock {\emph{\JournalTitle{\mnras}}} \textbf{\bibinfo{volume}{428}}, \bibinfo{pages}{1643--1655}, \doiprefix\url{10.1093/mnras/sts142} (\bibinfo{year}{2013}).
\newblock \eprint{1210.1369}.

\bibitem{Landau_1959}
\bibinfo{author}{{Landau}, L.~D.} \& \bibinfo{author}{{Lifshitz}, E.~M.}
\newblock \emph{\bibinfo{title}{{Fluid mechanics}}} (\bibinfo{publisher}{Pergamon}, \bibinfo{year}{1959}).

\bibitem{Skillman_2008}
\bibinfo{author}{{Skillman}, S.~W.}, \bibinfo{author}{{O'Shea}, B.~W.}, \bibinfo{author}{{Hallman}, E.~J.}, \bibinfo{author}{{Burns}, J.~O.} \& \bibinfo{author}{{Norman}, M.~L.}
\newblock \bibinfo{journal}{\bibinfo{title}{{Cosmological Shocks in Adaptive Mesh Refinement Simulations and the Acceleration of Cosmic Rays}}}.
\newblock {\emph{\JournalTitle{\apj}}} \textbf{\bibinfo{volume}{689}}, \bibinfo{pages}{1063--1077}, \doiprefix\url{10.1086/592496} (\bibinfo{year}{2008}).
\newblock \eprint{0806.1522}.

\bibitem{Schaal_2015}
\bibinfo{author}{{Schaal}, K.} \& \bibinfo{author}{{Springel}, V.}
\newblock \bibinfo{journal}{\bibinfo{title}{{Shock finding on a moving mesh - I. Shock statistics in non-radiative cosmological simulations}}}.
\newblock {\emph{\JournalTitle{\mnras}}} \textbf{\bibinfo{volume}{446}}, \bibinfo{pages}{3992--4007}, \doiprefix\url{10.1093/mnras/stu2386} (\bibinfo{year}{2015}).
\newblock \eprint{1407.4117}.

\bibitem{Vazza_2011}
\bibinfo{author}{{Vazza}, F.} \emph{et~al.}
\newblock \bibinfo{journal}{\bibinfo{title}{{A comparison of cosmological codes: properties of thermal gas and shock waves in large-scale structures}}}.
\newblock {\emph{\JournalTitle{\mnras}}} \textbf{\bibinfo{volume}{418}}, \bibinfo{pages}{960--985}, \doiprefix\url{10.1111/j.1365-2966.2011.19546.x} (\bibinfo{year}{2011}).
\newblock \eprint{1106.2159}.

\bibitem{Planelles_2010}
\bibinfo{author}{{Planelles}, S.} \& \bibinfo{author}{{Quilis}, V.}
\newblock \bibinfo{journal}{\bibinfo{title}{{ASOHF: a new adaptive spherical overdensity halo finder}}}.
\newblock {\emph{\JournalTitle{\aap}}} \textbf{\bibinfo{volume}{519}}, \bibinfo{pages}{A94}, \doiprefix\url{10.1051/0004-6361/201014214} (\bibinfo{year}{2010}).
\newblock \eprint{1006.3205}.

\bibitem{Knebe_2011}
\bibinfo{author}{{Knebe}, A.} \emph{et~al.}
\newblock \bibinfo{journal}{\bibinfo{title}{{Haloes gone MAD: The Halo-Finder Comparison Project}}}.
\newblock {\emph{\JournalTitle{\mnras}}} \textbf{\bibinfo{volume}{415}}, \bibinfo{pages}{2293--2318}, \doiprefix\url{10.1111/j.1365-2966.2011.18858.x} (\bibinfo{year}{2011}).
\newblock \eprint{1104.0949}.

\bibitem{Valles-Perez_2022}
\bibinfo{author}{{Vall{\'e}s-P{\'e}rez}, D.}, \bibinfo{author}{{Planelles}, S.} \& \bibinfo{author}{{Quilis}, V.}
\newblock \bibinfo{journal}{\bibinfo{title}{{The halo-finding problem revisited: a deep revision of the ASOHF code}}}.
\newblock {\emph{\JournalTitle{\aap}}} \textbf{\bibinfo{volume}{664}}, \bibinfo{pages}{A42}, \doiprefix\url{10.1051/0004-6361/202243712} (\bibinfo{year}{2022}).
\newblock \eprint{2205.02245}.

\bibitem{Cole_1996}
\bibinfo{author}{{Cole}, S.} \& \bibinfo{author}{{Lacey}, C.}
\newblock \bibinfo{journal}{\bibinfo{title}{{The structure of dark matter haloes in hierarchical clustering models}}}.
\newblock {\emph{\JournalTitle{\mnras}}} \textbf{\bibinfo{volume}{281}}, \bibinfo{pages}{716}, \doiprefix\url{10.1093/mnras/281.2.716} (\bibinfo{year}{1996}).
\newblock \eprint{astro-ph/9510147}.

\bibitem{Bryan_1998}
\bibinfo{author}{{Bryan}, G.~L.} \& \bibinfo{author}{{Norman}, M.~L.}
\newblock \bibinfo{journal}{\bibinfo{title}{{Statistical Properties of X-Ray Clusters: Analytic and Numerical Comparisons}}}.
\newblock {\emph{\JournalTitle{\apj}}} \textbf{\bibinfo{volume}{495}}, \bibinfo{pages}{80--99}, \doiprefix\url{10.1086/305262} (\bibinfo{year}{1998}).
\newblock \eprint{astro-ph/9710107}.

\bibitem{Diemer_2014}
\bibinfo{author}{{Diemer}, B.} \& \bibinfo{author}{{Kravtsov}, A.~V.}
\newblock \bibinfo{journal}{\bibinfo{title}{{Dependence of the Outer Density Profiles of Halos on Their Mass Accretion Rate}}}.
\newblock {\emph{\JournalTitle{\apj}}} \textbf{\bibinfo{volume}{789}}, \bibinfo{pages}{1}, \doiprefix\url{10.1088/0004-637X/789/1/1} (\bibinfo{year}{2014}).
\newblock \eprint{1401.1216}.

\bibitem{Valles-Perez_2020}
\bibinfo{author}{{Vall{\'e}s-P{\'e}rez}, D.}, \bibinfo{author}{{Planelles}, S.} \& \bibinfo{author}{{Quilis}, V.}
\newblock \bibinfo{journal}{\bibinfo{title}{{On the accretion history of galaxy clusters: temporal and spatial distribution}}}.
\newblock {\emph{\JournalTitle{\mnras}}} \textbf{\bibinfo{volume}{499}}, \bibinfo{pages}{2303--2318}, \doiprefix\url{10.1093/mnras/staa3035} (\bibinfo{year}{2020}).
\newblock \eprint{2009.13882}.

\bibitem{Savitzky_1964}
\bibinfo{author}{{Savitzky}, A.} \& \bibinfo{author}{{Golay}, M.~J.~E.}
\newblock \bibinfo{journal}{\bibinfo{title}{{Smoothing and differentiation of data by simplified least squares procedures}}}.
\newblock {\emph{\JournalTitle{Analytical Chemistry}}} \textbf{\bibinfo{volume}{36}}, \bibinfo{pages}{1627--1639}, \doiprefix\url{10.1021/ac60214a047} (\bibinfo{year}{1964}).

\bibitem{Valles-Perez_2023}
\bibinfo{author}{{Vall{\'e}s-P{\'e}rez}, D.}, \bibinfo{author}{{Planelles}, S.}, \bibinfo{author}{{Monllor-Berbegal}, {\'O}.} \& \bibinfo{author}{{Quilis}, V.}
\newblock \bibinfo{journal}{\bibinfo{title}{{On the choice of the most suitable indicator for the assembly state of dark matter haloes through cosmic time}}}.
\newblock {\emph{\JournalTitle{\mnras}}} \textbf{\bibinfo{volume}{519}}, \bibinfo{pages}{6111--6125}, \doiprefix\url{10.1093/mnras/stad059} (\bibinfo{year}{2023}).
\newblock \eprint{2301.02253}.

\bibitem{Scott_1992}
\bibinfo{author}{{Scott}, D.~W.}
\newblock \emph{\bibinfo{title}{{Multivariate Density Estimation}}} (\bibinfo{publisher}{John Wiley \& Sons}, \bibinfo{year}{1992}).

\bibitem{pca}
\bibinfo{author}{Tipping, M.} \& \bibinfo{author}{Bishop, C.}
\newblock \bibinfo{journal}{\bibinfo{title}{Mixtures of probabilistic principal component analysers}}.
\newblock {\emph{\JournalTitle{Neural Computation}}} \textbf{\bibinfo{volume}{11}}, \bibinfo{pages}{443--482}, \doiprefix\url{10.1162/089976699300016728} (\bibinfo{year}{1999}).
\newblock \bibinfo{note}{Copyright of the Massachusetts Institute of Technology Press (MIT Press)}.

\end{thebibliography}

\label{LastPage}

\FloatBarrier
\newpage
\section*{Supplementary material}
\renewcommand{\thesubsection}{\Alph{subsection}}
\renewcommand{\figurename}{Suppl. Fig.}
\setcounter{figure}{0}

%---------------------
\subsection{Visual impression of the accretion shock identification procedure}
\label{s:suppl.visual}
%---------------------

\begin{figure}[h]
    \centering
    \includegraphics[height=0.3\textwidth]{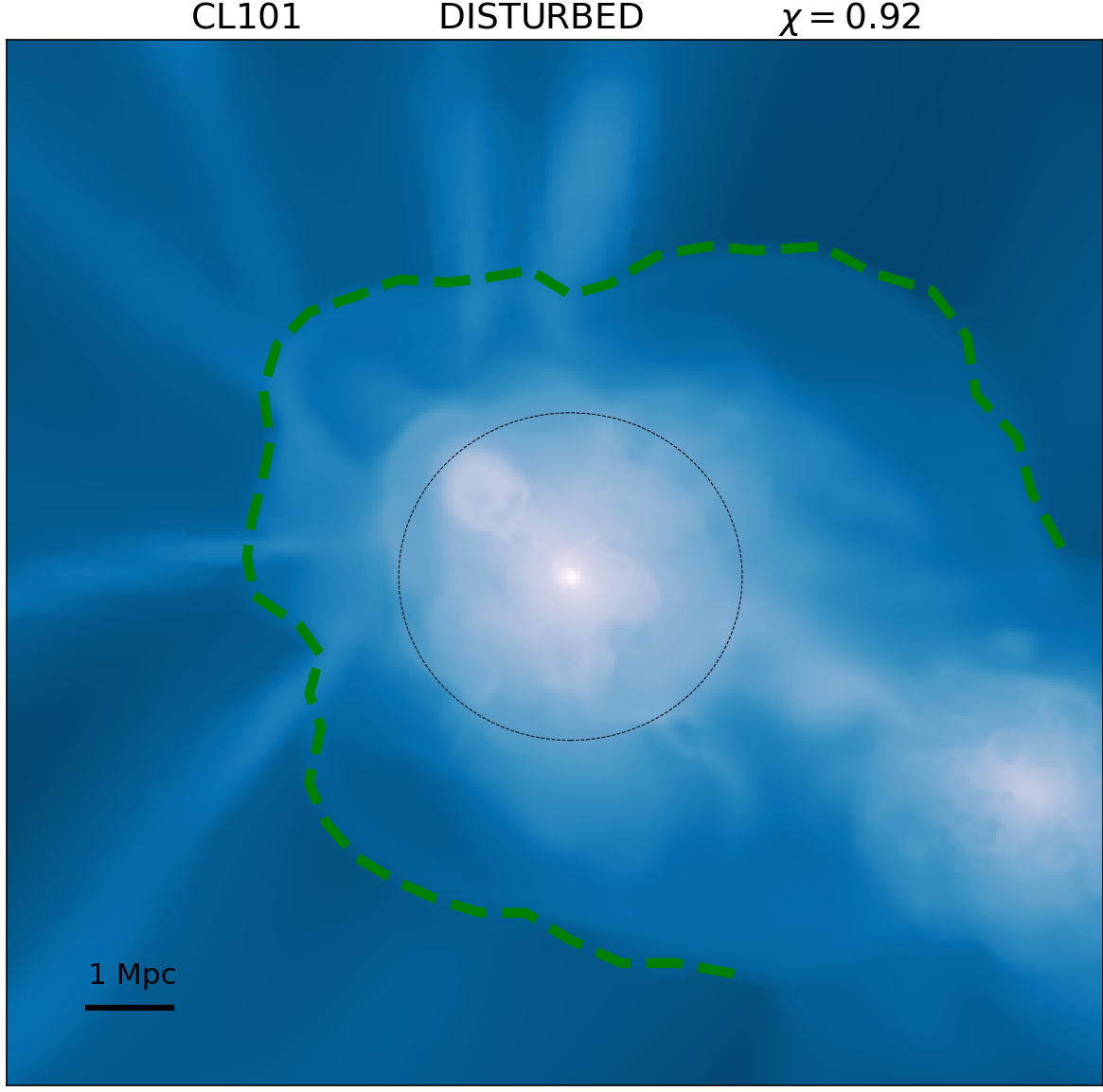}~
    \includegraphics[height=0.3\textwidth]{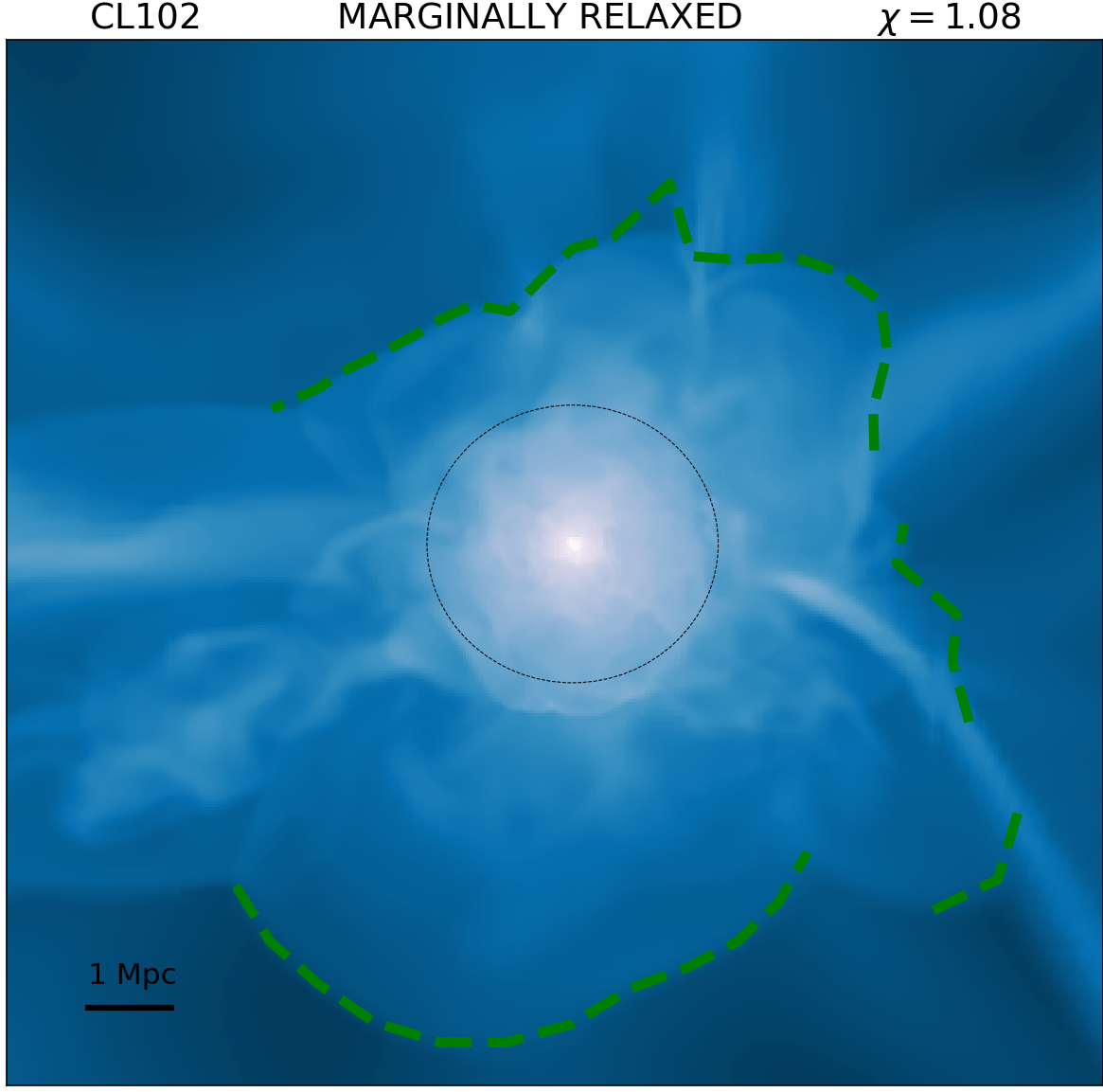}~
    \includegraphics[height=0.3\textwidth]{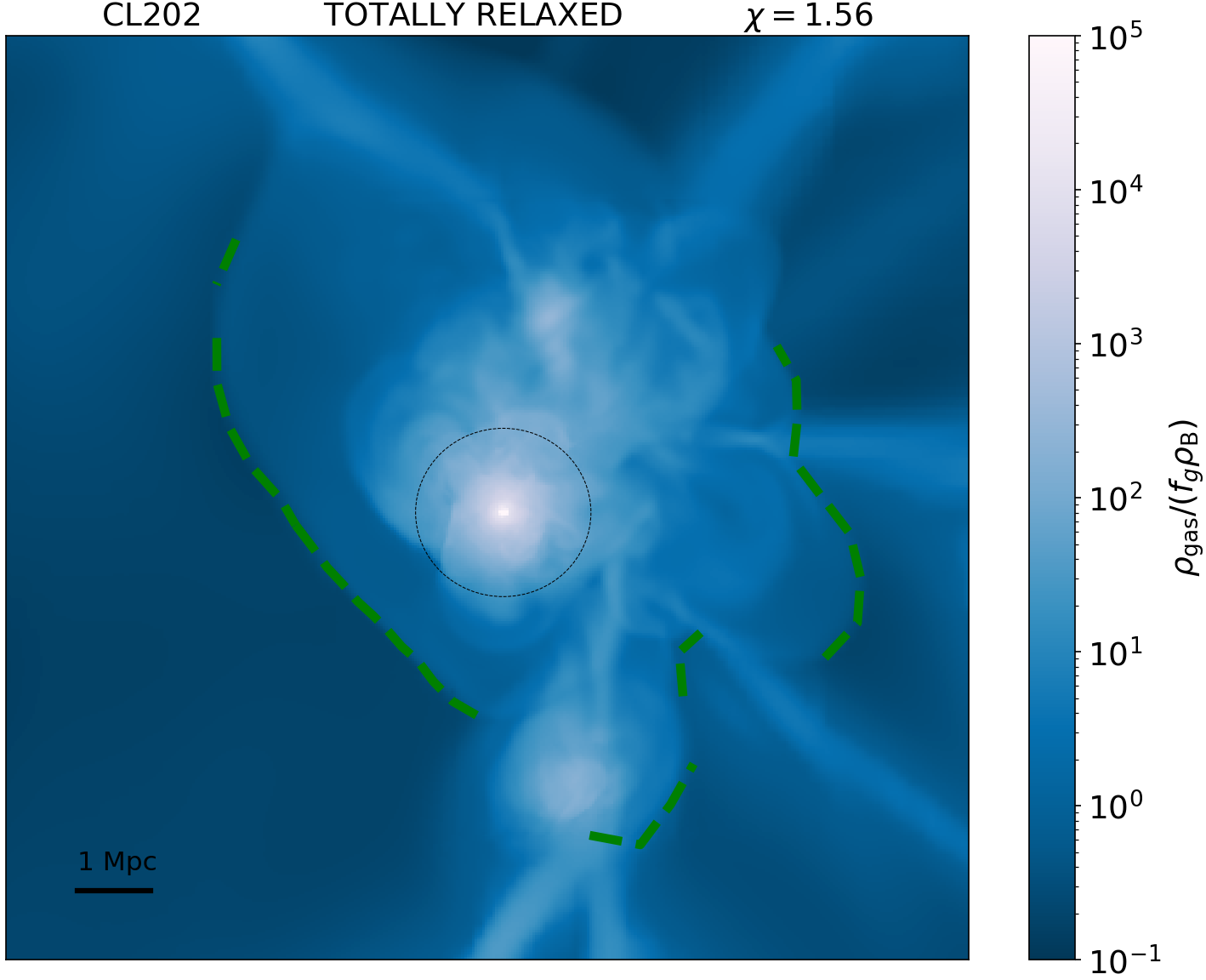}
    \caption{Visual impression of the shock shell identification results. The three panels correspond to three different galaxy clusters at $z=0$, one of each dynamical state category. The colourmaps show thin ($\sim 36 \, \mathrm{kpc}$) slices of the gas densities (in units of the cosmic mean gas density) through the cluster centres. The virial radius is indicated by the black, dotted circle. The green, dashed lines indicate the shock shell, in the directions where it can be located. The line in the bottom left corner shows a $1 \, \mathrm{Mpc}$ ruler for visual reference.}
    \label{figS1}
\end{figure}

With the aim of showcasing the typical appearance of our detected accretion shocks, as well as presenting a visual test of the trustworthiness of our detection process, we present in Suppl. Fig. \ref{figS1} three panels corresponding to three galaxy clusters at $z=0$, each of them corresponding to a different dynamical state at the present time according to our classification. In these panels we overplot the the virial sphere (black, dotted line) and the shell corresponding to the accretion shock in the directions it has been detected (green, dashed line).

The left panel corresponds to our most massive galaxy cluster, \texttt{CL101}, which is disturbed by $z \simeq 0$ due to a recent merger with mass ratio 1:5 at $z \simeq 0.19$, that has proceeded from the bottom-right of the slice. Also in this direction, another massive structure is approaching the cluster. Thus, the accretion shock shell is disrupted in this direction, and correspondingly not detected by our procedure. In the remaining directions, the algorithm delimits correctly the location of the outer accretion shock, which can be seen as a jump in the underlying density map.

The middle panel shows a slice through \texttt{CL102}, which is a marginally relaxed cluster which suffered its last merger, with a mass ratio of $\sim$1:9, at $z \simeq 0.3$. In this case, the shock shell is rather asymmetric, mainly due to its anisotropic environment, dominated by several filaments, and to several minor mergers at $z \simeq 0.5$. In this case, it can be seen how our shock shell identification procedure naturally excludes the directions of the filaments, where the accretion shock is not present.

Finally, the right panel of Suppl. Fig. \ref{figS1} corresponds to a totally relaxed cluster, \texttt{CL202}. This object suffered a minor merger (mass ratio of 1:8) at $z \simeq 0.3$, but unlike the example in the middle panel, where the merger was almost head-on, in this case the merger had a larger impact parameter. This may justify why in this case the cluster has had time to fall back to relaxation, when assessed at the virial volume, while outside the virial radius the gas density shows a more disturbed morphology, with a very anisotropic accretion shell. This cluster is on the verge of a merger with another structure, proceeding from the lower part of the slice. It is worth mentioning how, in this situation, our algorithm is picking the accretion shock boundary of the infalling structure as part of the accretion shell of the main cluster. Generally, when two haloes are merging and their accretion shocks join, it is not trivial to establish since when they must be regarded as a single object for the purpose of identifying its joint outer shock. However, our accretion shock equivalent radius definition, based on the mode of the PDF of the angular distribution of distances to the shock shell, unlike other statistics such as the arithmetic mean, gets minimally biased by these events (since the mode is robust to the presence of tails in the distribution).

\begin{figure}
    \centering
    \includegraphics[width=0.333\textwidth]{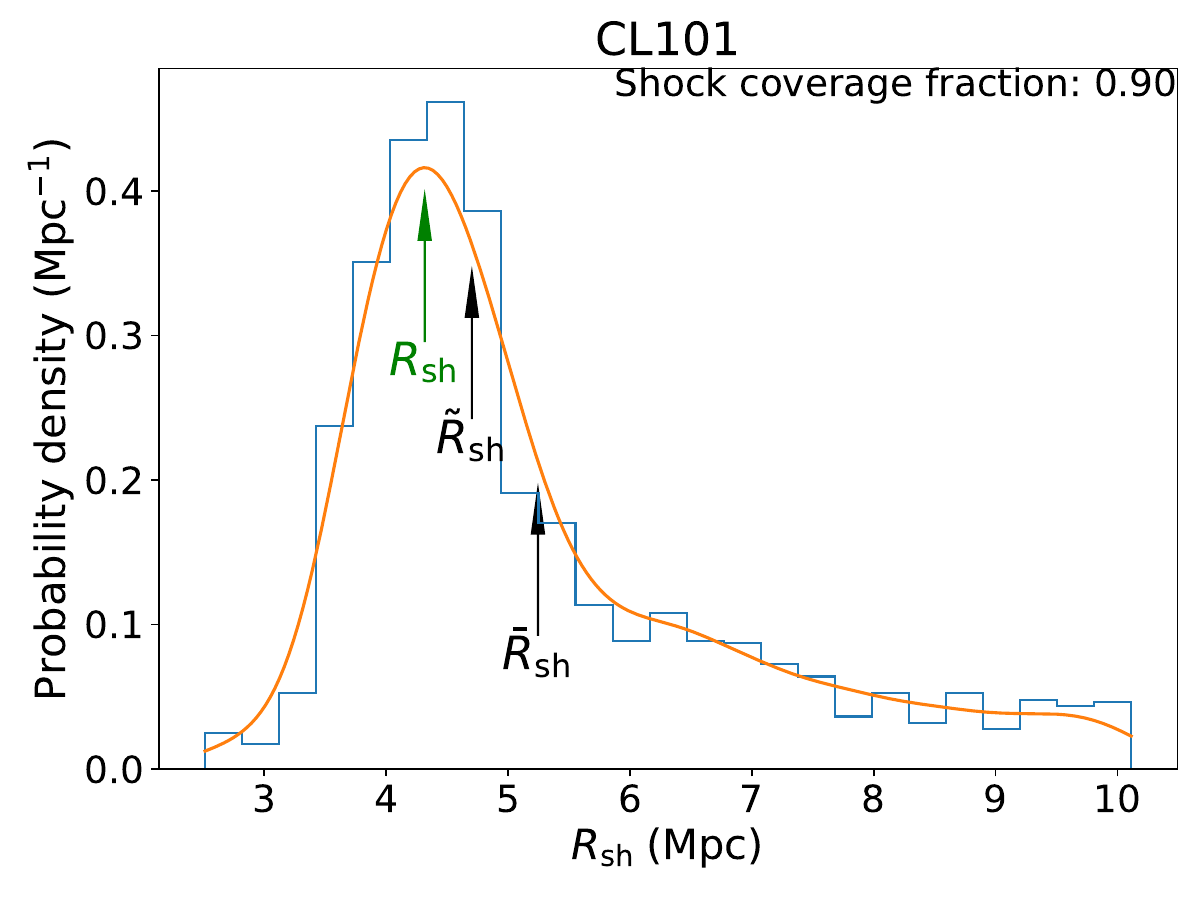}~
    \includegraphics[width=0.333\textwidth]{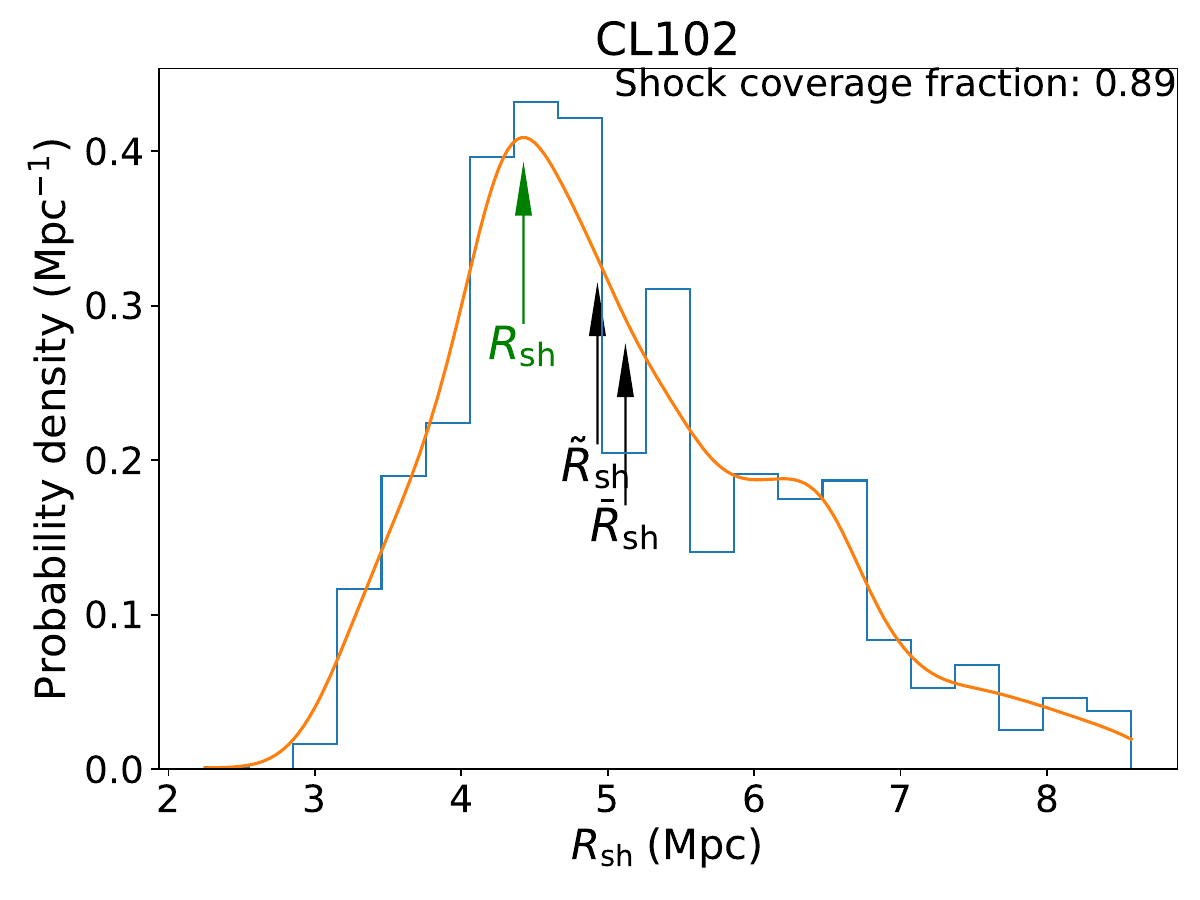}~
    \includegraphics[width=0.333\textwidth]{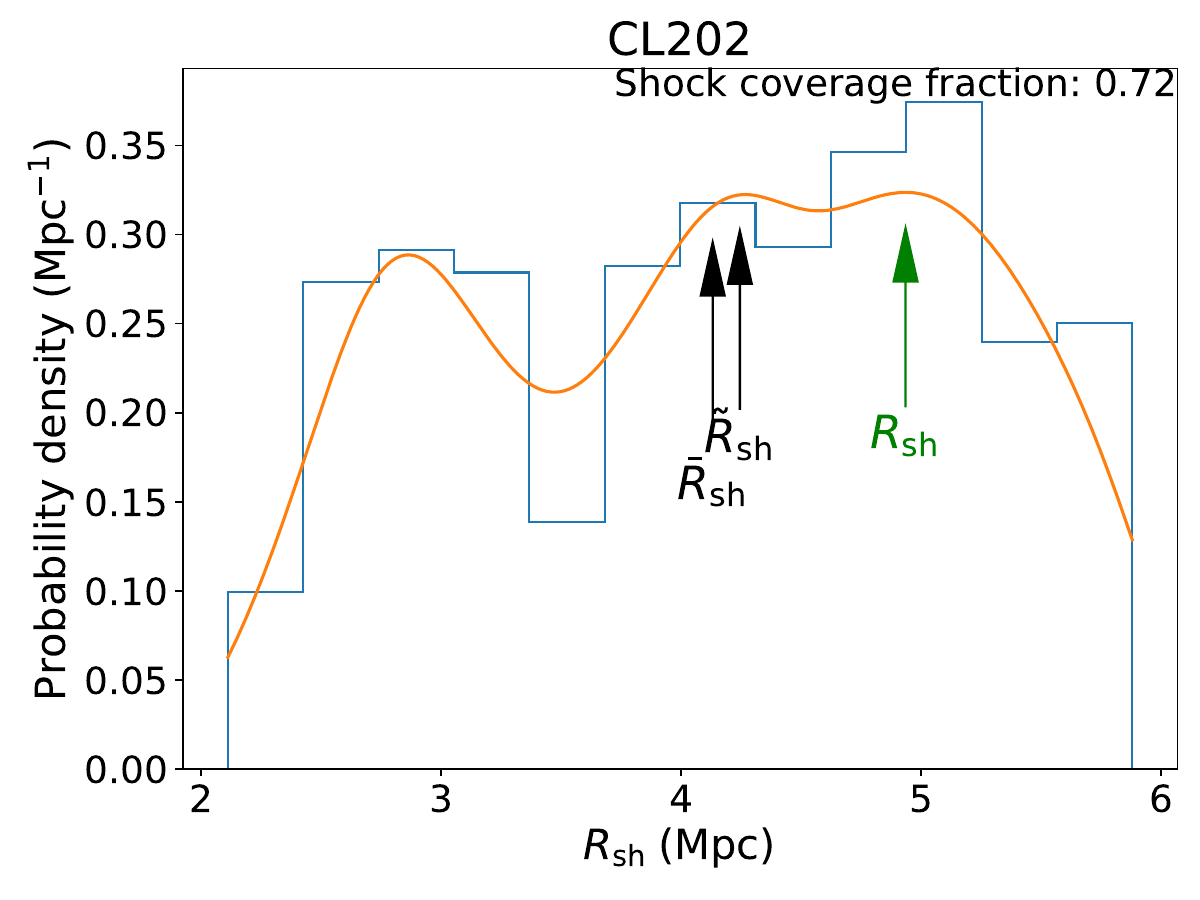}
    \includegraphics[width=0.333\textwidth]{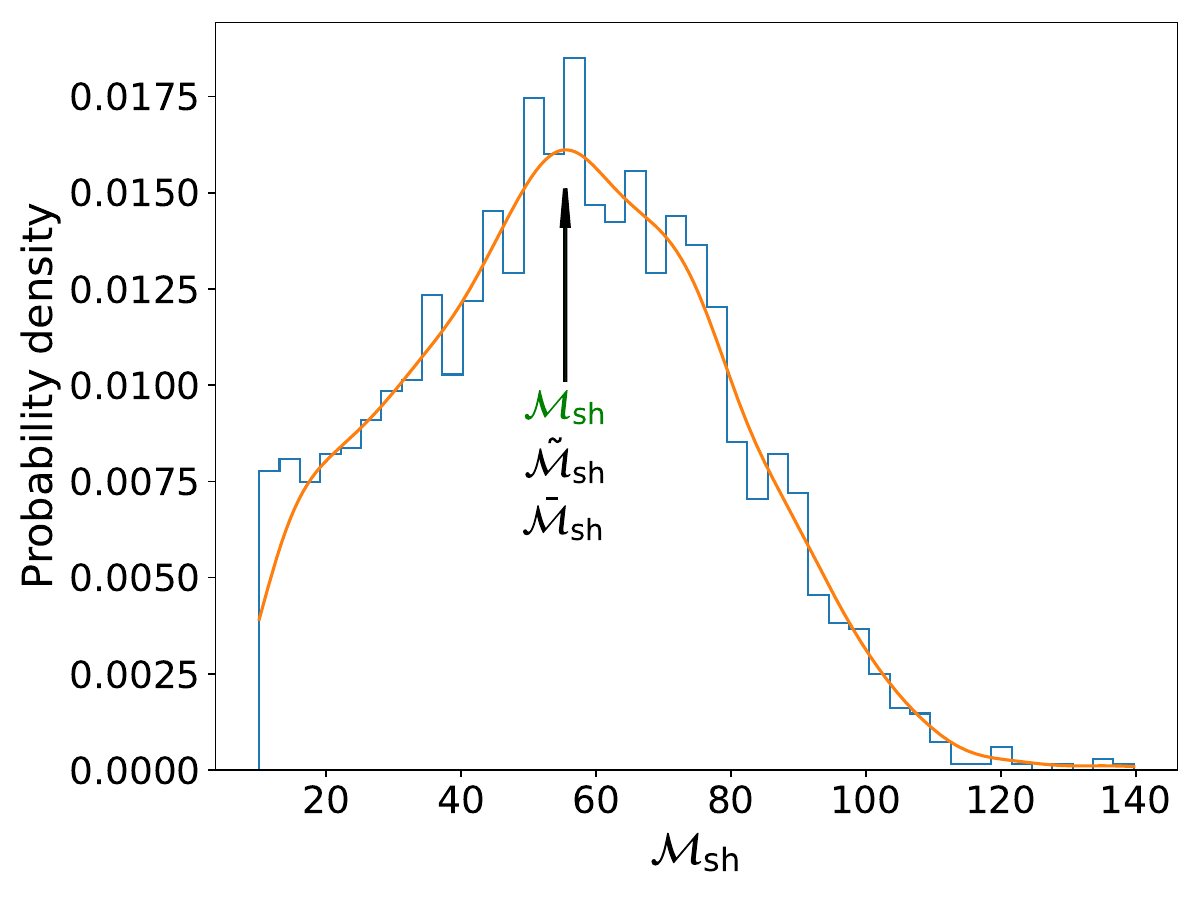}~
    \includegraphics[width=0.333\textwidth]{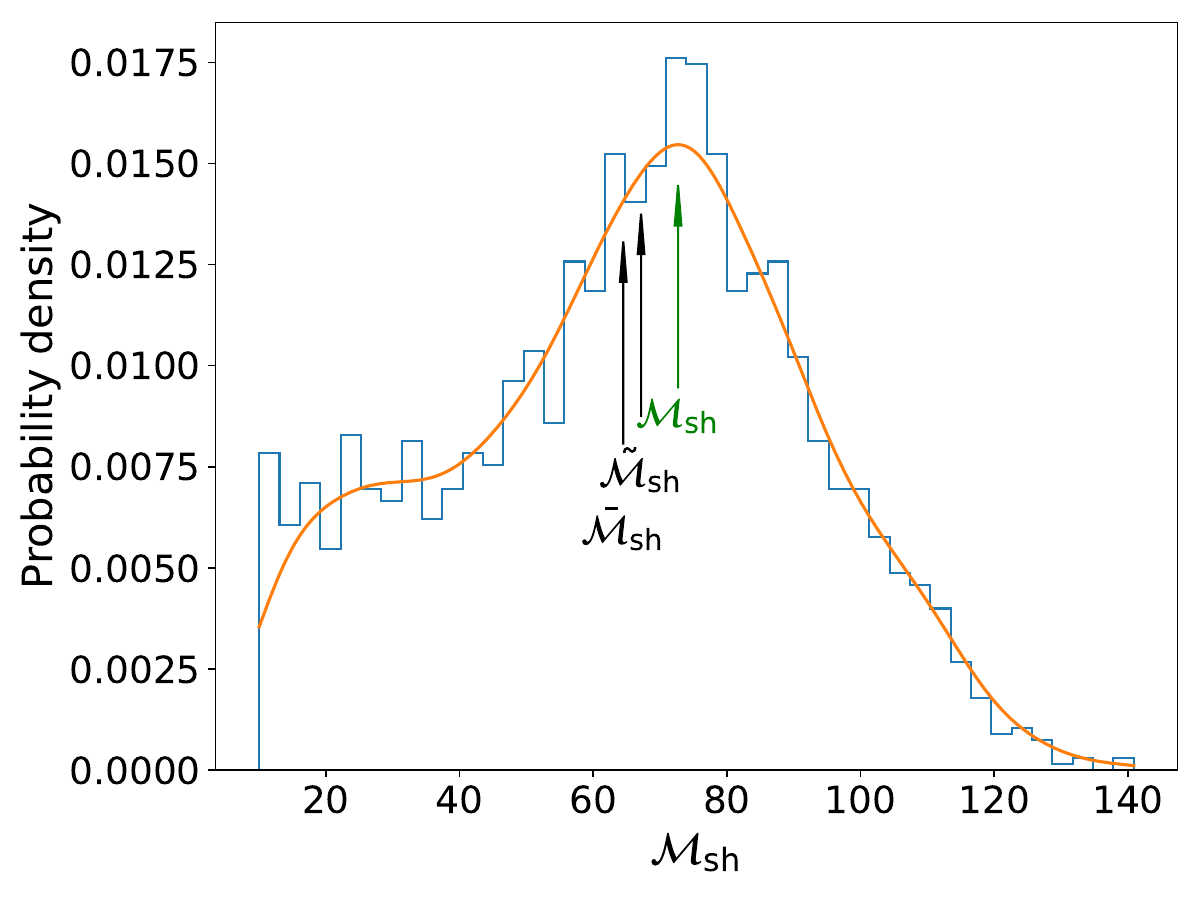}~
    \includegraphics[width=0.333\textwidth]{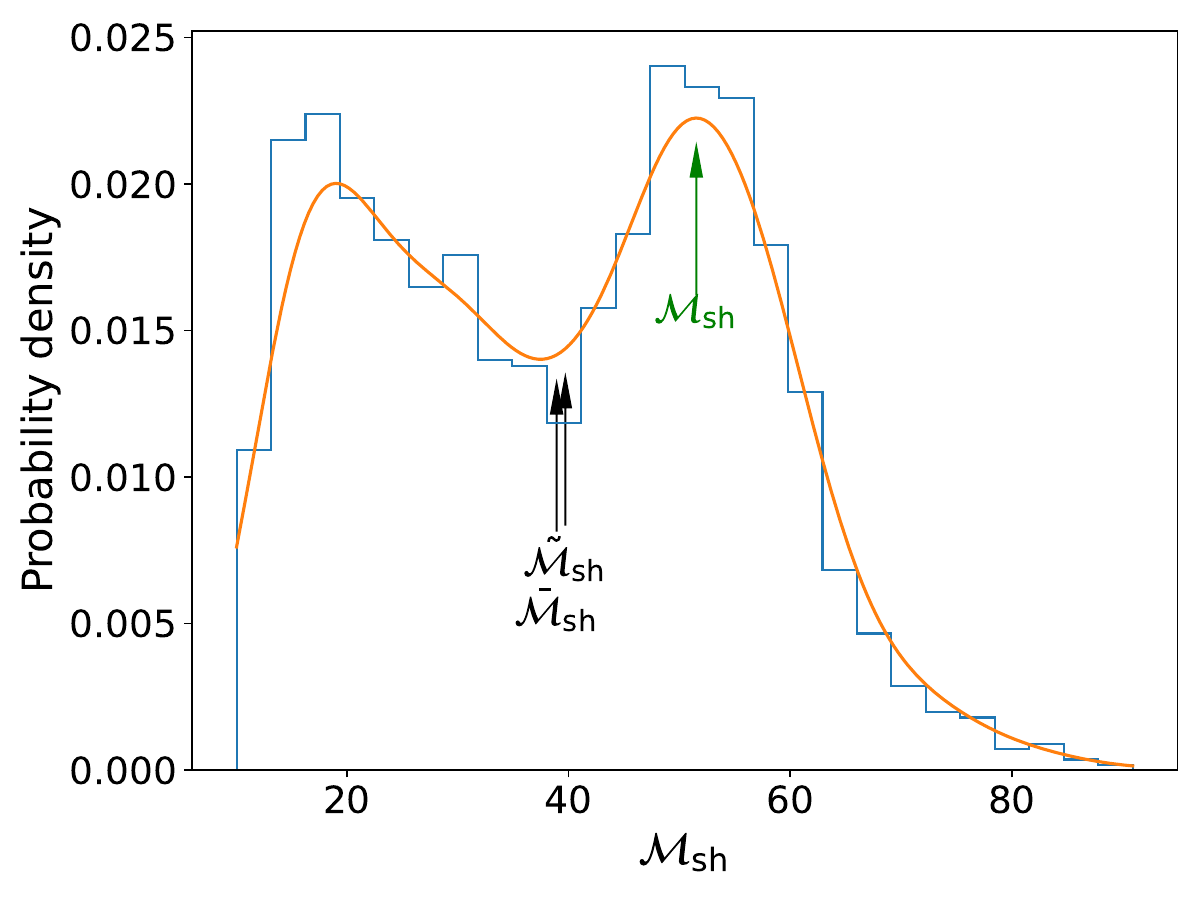}
    \caption{Examples on the determination of $R_\mathrm{sh}$ and $\mathcal{M}_\mathrm{sh}$, on the same clusters as Suppl. Fig. \ref{figS1}. The top rows present the distribution of values of $R_\mathrm{sh}(\theta,\phi)$, with a histogram (blue line) and with its corresponding Gaussian kernel density estimation (orange line). The green arrow indicates the most frequent value (i.e., the peak of the probability density function, or mode), which we associate to the shock radius ($R_\mathrm{sh}$), while the two black arrows show the mean ($\bar R_\mathrm{sh}$) and median ($\tilde R_\mathrm{sh}$) values, for comparison. The bottom row contains the same information regarding the distribution of $\mathcal{M}_\mathrm{sh}(\theta,\phi)$. The upper panels also inform about the shock coverage fraction, i.e., the fraction of the solid angle around the object in which we identify the shock cell.}
    \label{figS2}
\end{figure}

Given that, in what follows, we aim to represent these shock shells with a single value of a so-called \textit{equivalent} radius, in Suppl. Fig. \ref{figS2} we comment on this issue by presenting the distribution of radial distances to the accretion shock shell, $R_\mathrm{sh}(\theta,\phi)$ (top row), and the distribution of Mach numbers through the accretion shock shell, $\mathcal{M}_\mathrm{sh}(\theta,\phi)$ (bottom row), for the same three clusters displayed in Suppl. Fig. \ref{figS1}.

In most cases (such as \texttt{CL101} or \texttt{CL102}; left and central figures), the distribution of shock radii is monomodal and right-skewed, implying that the mean and, to a lesser extent, the median get dragged towards larger values by the tails of the distribution. In these cases, however, the peak of the PDF is well-defined and provides a robust measurement of the shock radius, in the sense that it does not depend on the particular value of $r_\mathrm{max}$ (which has a large influence on the right tail of the distribution). In the vast majority of these cases, also the distribution of $\mathcal{M}_\mathrm{sh}(\theta,\phi)$ is monomodal, with typically small differences between the mean, median and mode of the distribution.

On the other hand, in some cases the distribution of values of $R_\mathrm{sh}(\theta,\phi)$ might be multimodal. Such is the case of \texttt{CL202} (right column), where three peaks are captured by the kernel density estimate. Additionally, in this case, a smaller fraction of the solid angle around the cluster is found to contain the shock shell, indicating a disrupted morphology, most likely associated to the presence of the infalling object in the lower direction (see right panel of Suppl. Fig. \ref{figS1}). Also the Mach number distribution appears to be bimodal in this case. In cases like this, though infrequent, the characterisation is not so straightforward. This is why we have cleaned our sample from multimodal distributions of $R_\mathrm{sh}(\theta,\phi)$ (see the Methods section).

It is interesting to highlight how shock morphologies, at least from the qualitative point of view, do not seem to be significantly correlated to our usual measures of dynamical state. This must not be surprising, since these measures are associated to a particular aperture (i.e., $R_\mathrm{vir}$) much smaller than the typical volume enclosed by the shock shells (a factor $2-3$ in radius). Indeed, we may find clusters that are dynamically disturbed, when looking at the virial volume, that exhibit rather spherical accretion shocks (e.g., \texttt{CL101}; see Suppl. Fig. \ref{figS1}); or highly relaxed clusters with more complex-shaped outer boundaries (such as \texttt{CL202}).

%---------------------
\subsection{Statistical summary of the sample}
\label{s:suppl.statistics}
%---------------------

\begin{landscape}
\begin{figure}
    \centering
    \includegraphics[width=.71\textwidth]{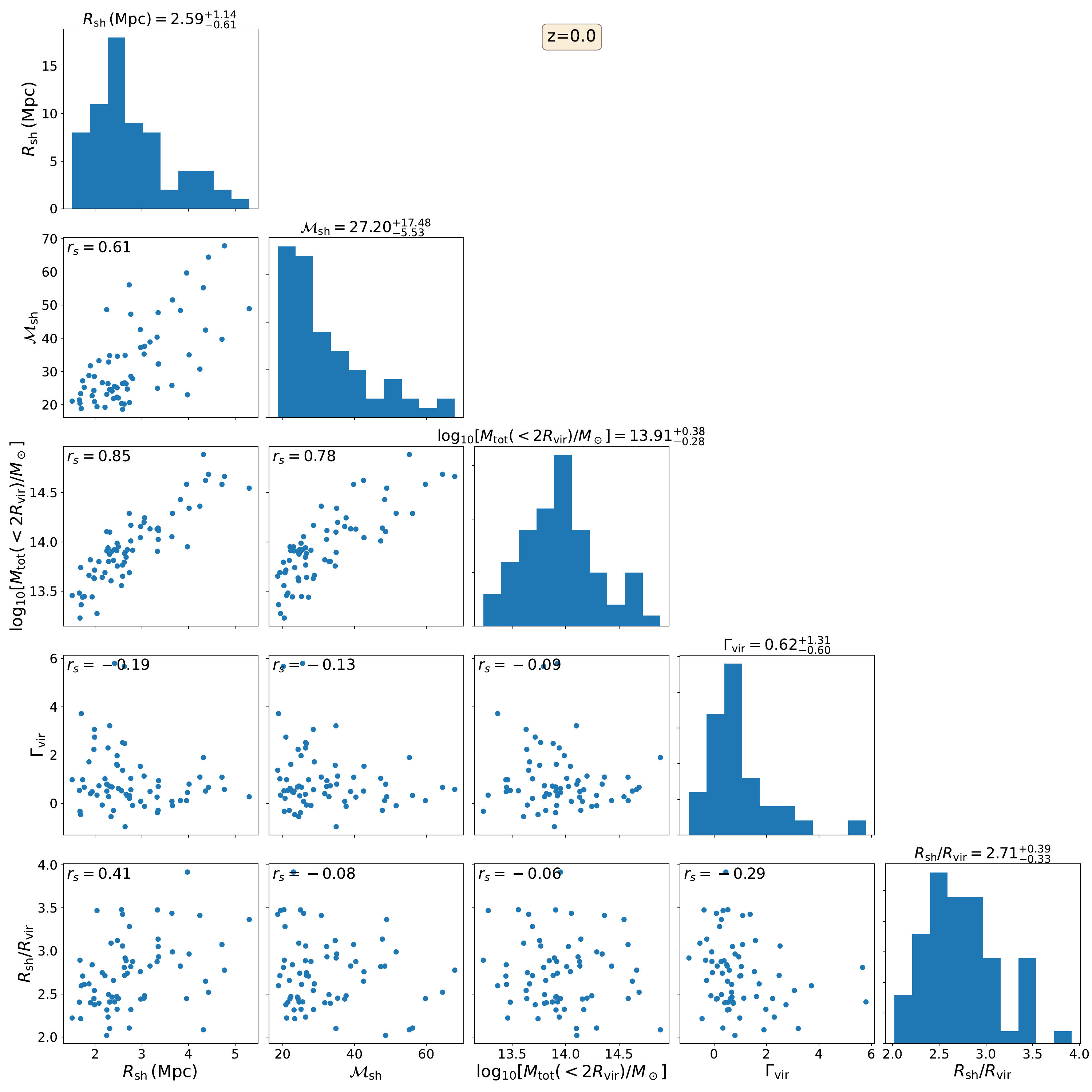}~
    \includegraphics[width=.71\textwidth]{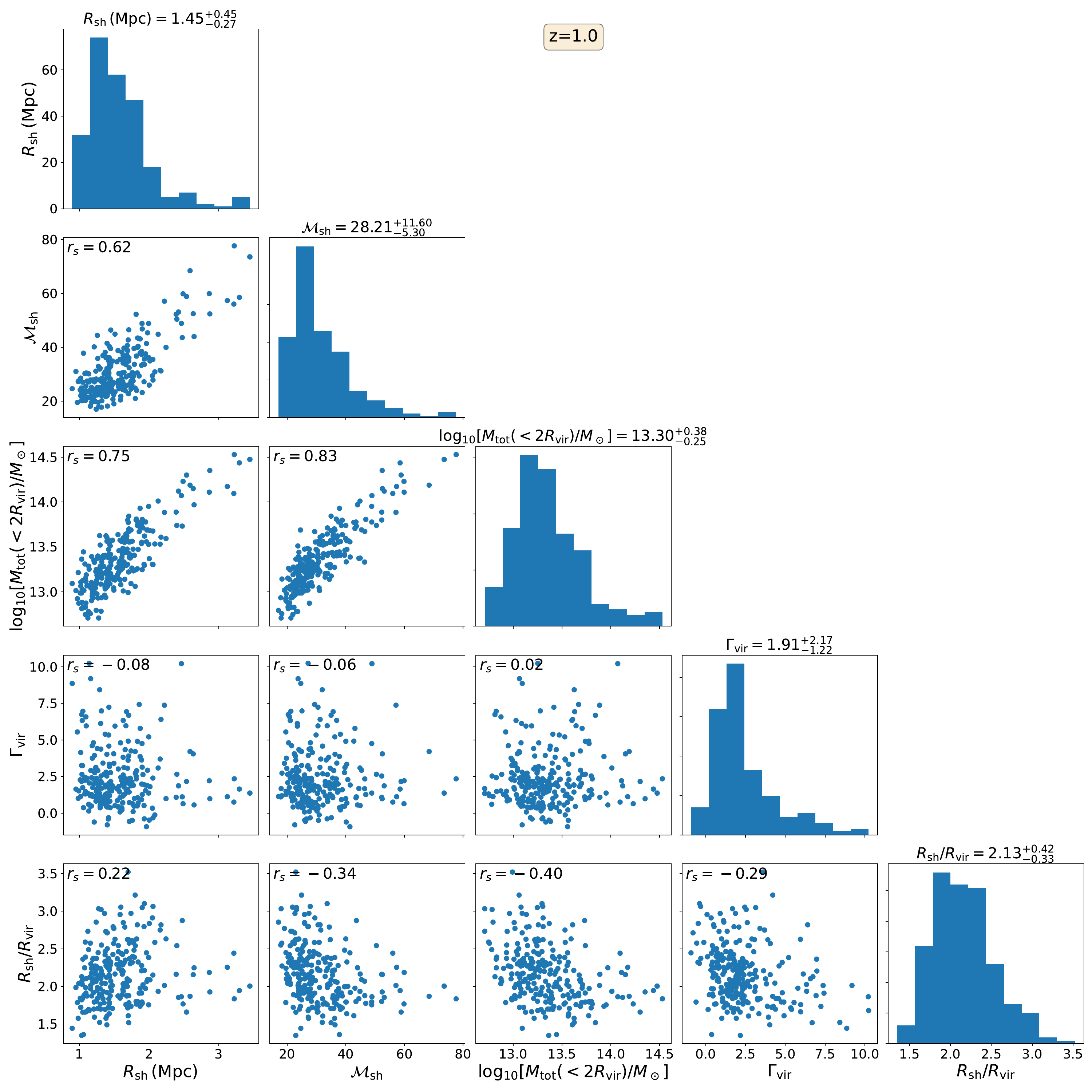}
    \caption{Statistical summary of our galaxy clusters and galaxy groups sample. The left-hand side panel corresponds to the sample at $z=0$, comprising 67 objects. The right-hand side panel contains the same information for $z=1$, when the sample is composed of 249 objects. The plots over the principal diagonal contain histograms of the corresponding variable (shock equivalent radius, mean Mach number, total mass within $2R_\mathrm{vir}$, accretion rate $\Gamma_\mathrm{vir}$, and shock radius in units of the virial radius). Above each of these panels we inform about the median and the scatter (quantified through the distance to the first and third quartiles) of the variable. The off-diagonal panels contain scatter plots for each pair of variables. Written on each of these plots is the Spearman correlation coefficient ($r_s$) between the corresponding pair of variables.}
    \label{figS3}
\end{figure}
\end{landscape}

Suppl. Fig. \ref{figS3} presents a summary, in statistical terms, of our sample, both at $z=0$ and at $z=1$ (left-hand side and right-hand side panels, respectively). For each cosmic epoch, we present a corner-like plot, including the univariate distributions of shock radius $R_\mathrm{sh}$, mean Mach number $\mathcal{M}_\mathrm{sh}$, total mass within $2 R_\mathrm{vir}$, accretion rate $\Gamma_\mathrm{vir}$, and $R_\mathrm{sh} / R_\mathrm{vir}$; and scatter plots between each pair of variables to illustrate their correlations or lack thereof.

%---------------------
\subsection{Fits for the mass measured in an aperture of $R_\mathrm{vir}$}
\label{s:suppl.1rvir}
%---------------------

While in the results that we report in the main article we consistently use $2R_\mathrm{vir}$ as our mass definition, and this is the mass that we have proposed to infer using shock sizes and intensities through our fitting relation, there is no particular reason to choose this specific aperture. In principle, the properties of the shock shell would be best correlated to the gravitational mass driving the collapse of the structure and, hence, the generation and propagation of the accretion shock. However, in a fully three-dimensional picture, it is not straightforward to agree on the definition of this mass. Therefore, the mass aperture could be varied in a sensibly large interval, as long as it is still a good estimate with the mass driving the collapse of the galaxy cluster or group.

\begin{figure}[h]
    \centering
    \includegraphics[width=.5\textwidth]{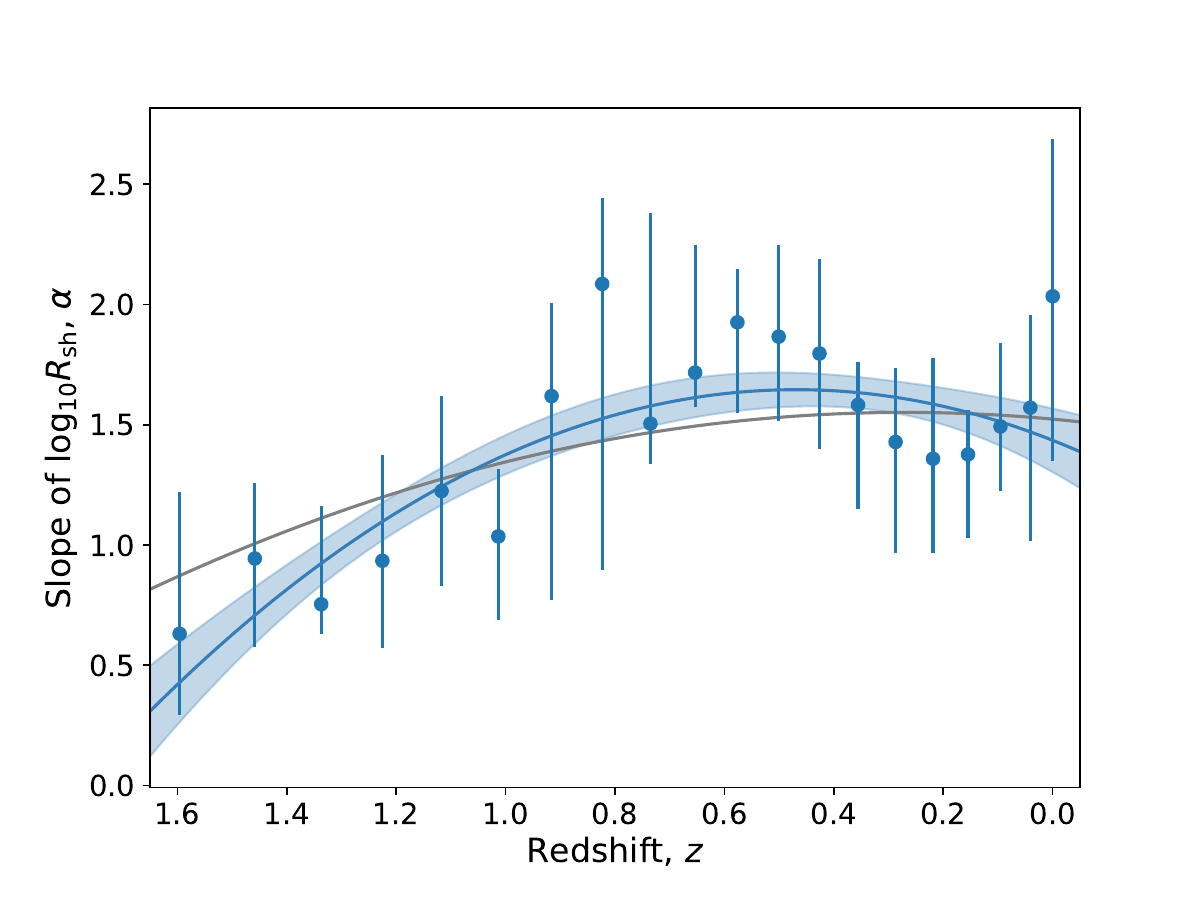}~
    \includegraphics[width=.5\textwidth]{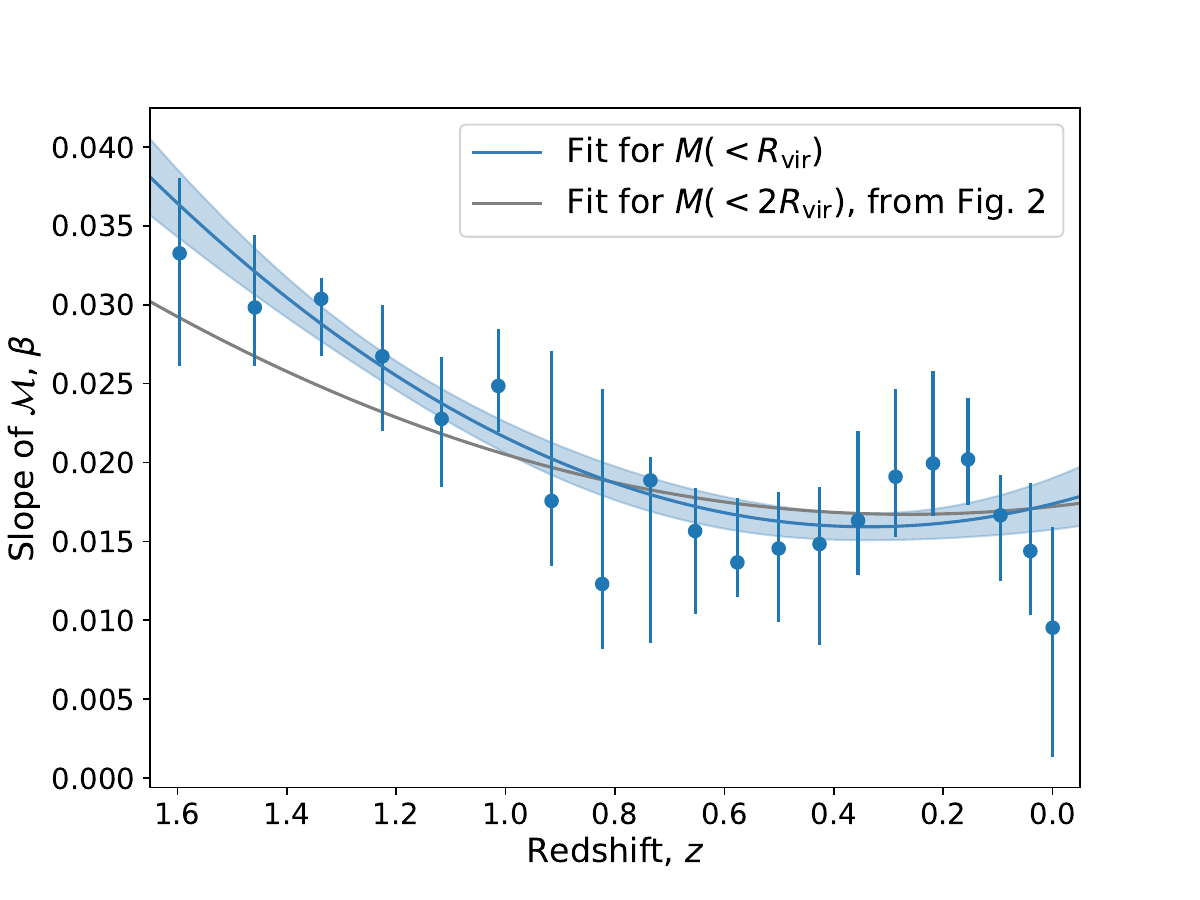}
    \includegraphics[width=.5\textwidth]{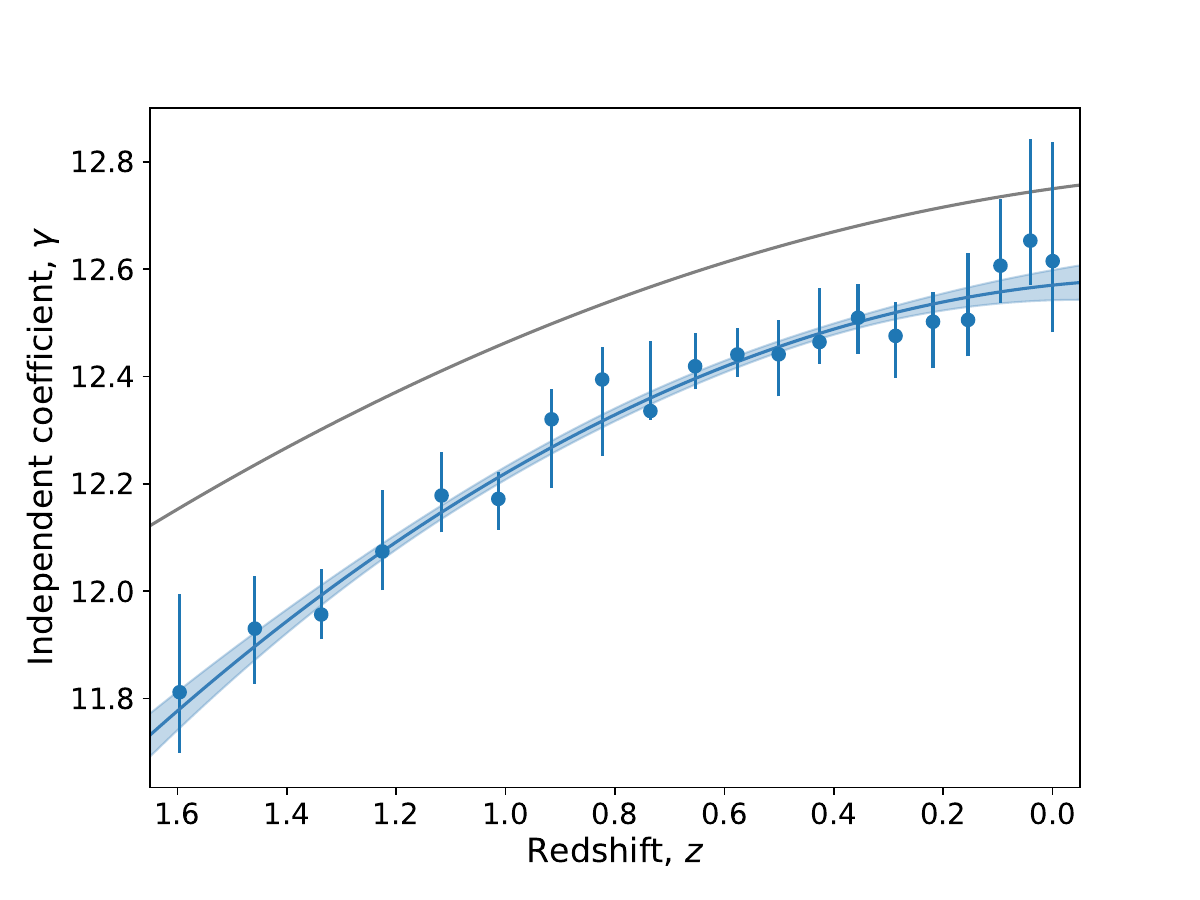}~
    \includegraphics[width=.5\textwidth]{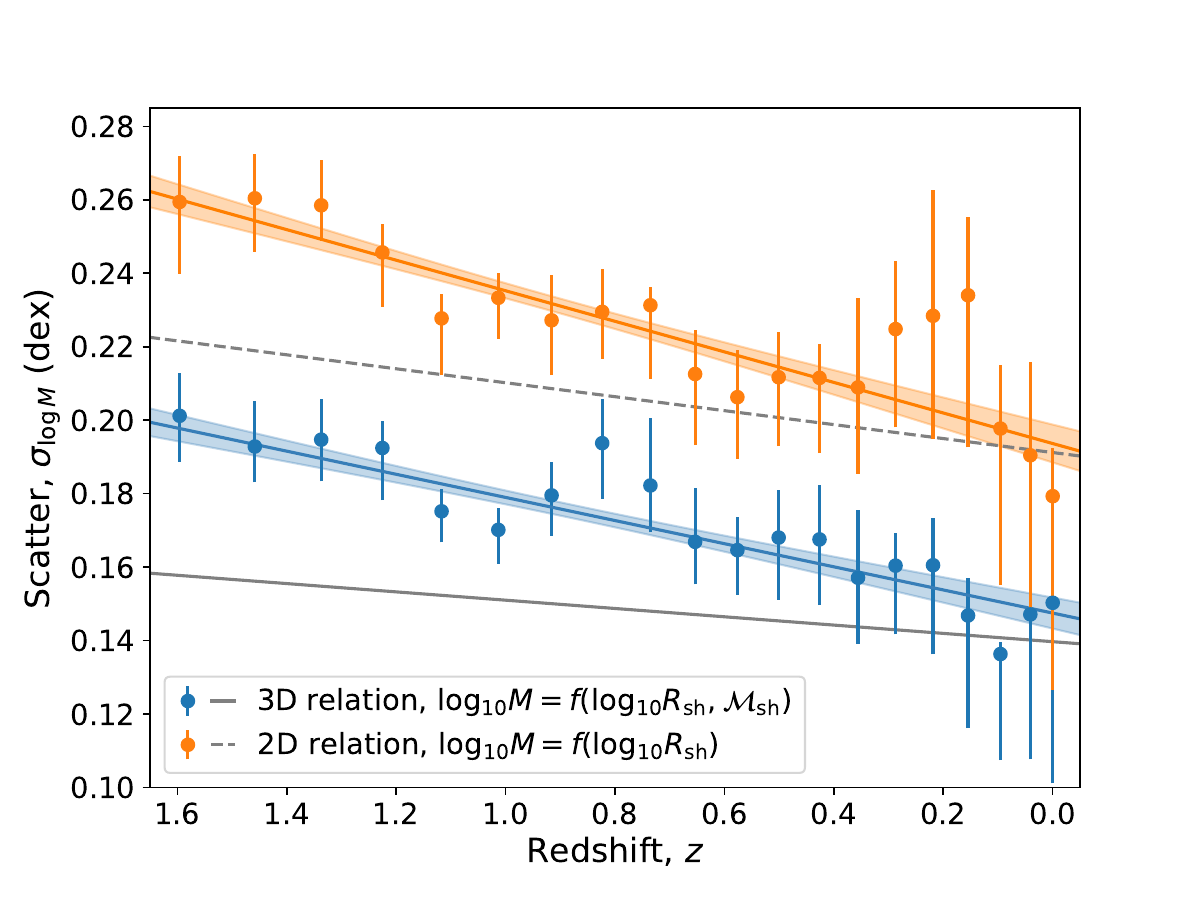}
    \caption{Scaling relation evolution summary, totally analogue to Fig. \ref{fig:evolution} in the main text, for the fit measuring the masses within $R_\mathrm{vir}$. Here, all plot elements correspond the fits within $R_\mathrm{vir}$, except for gray lines, which contain the fits for $2 R_\mathrm{vir}$ that were shown in Fig. \ref{fig:evolution}, for reference. \textit{Top left:} evolution of the coefficient of the shock radius. \textit{Top right:} evolution of the coefficient of Mach number. \textit{Bottom left:} evolution of the independent term. \textit{Bottom right:} evolution of the scatter (in dex) around the three-dimensional scaling relation (blue line), and around the bidimensional scaling relation (orange line).}
    \label{figS4}
\end{figure}

In Suppl. Fig. \ref{figS4}, we show this by presenting the results for the three-dimensional scaling relation (Eq. \ref{eq:fit_3d_evolution}), taking the masses within $R_\mathrm{vir}$, instead of $2R_\mathrm{vir}$. The four panels in the figure are completely analogous to Fig. \ref{fig:evolution} in the main text, with the addition of the gray lines, which contain the original fits within $2 R_\mathrm{vir}$ for a better visual comparison. Generally, the evolution of the fit coefficients $\alpha$ and $\beta$ (the coefficients of $\log_{10} R_\mathrm{sh}$ and $\mathcal{M}$, respectively) with redshift vary minimally, pointing at a consistent behaviour at different radial apertures as stated. Naturally, the normalisation decreases when shifting from $2 R_\mathrm{vir}$ to $R_\mathrm{vir}$. Regarding the scatter evolution, while at $z \sim 0$ masses can be determined with similar accuracy at both radial volumes, at high redshift choosing the larger aperture reduces very significantly the scatter around our relation. 

%---------------------
\subsection{Results marginalised over $\mathcal{M}_\mathrm{sh}$: the 2d relation}
\label{s:suppl.2d_relation}
%---------------------

\begin{figure}[h]
    \centering
    \includegraphics[width=.7\textwidth]{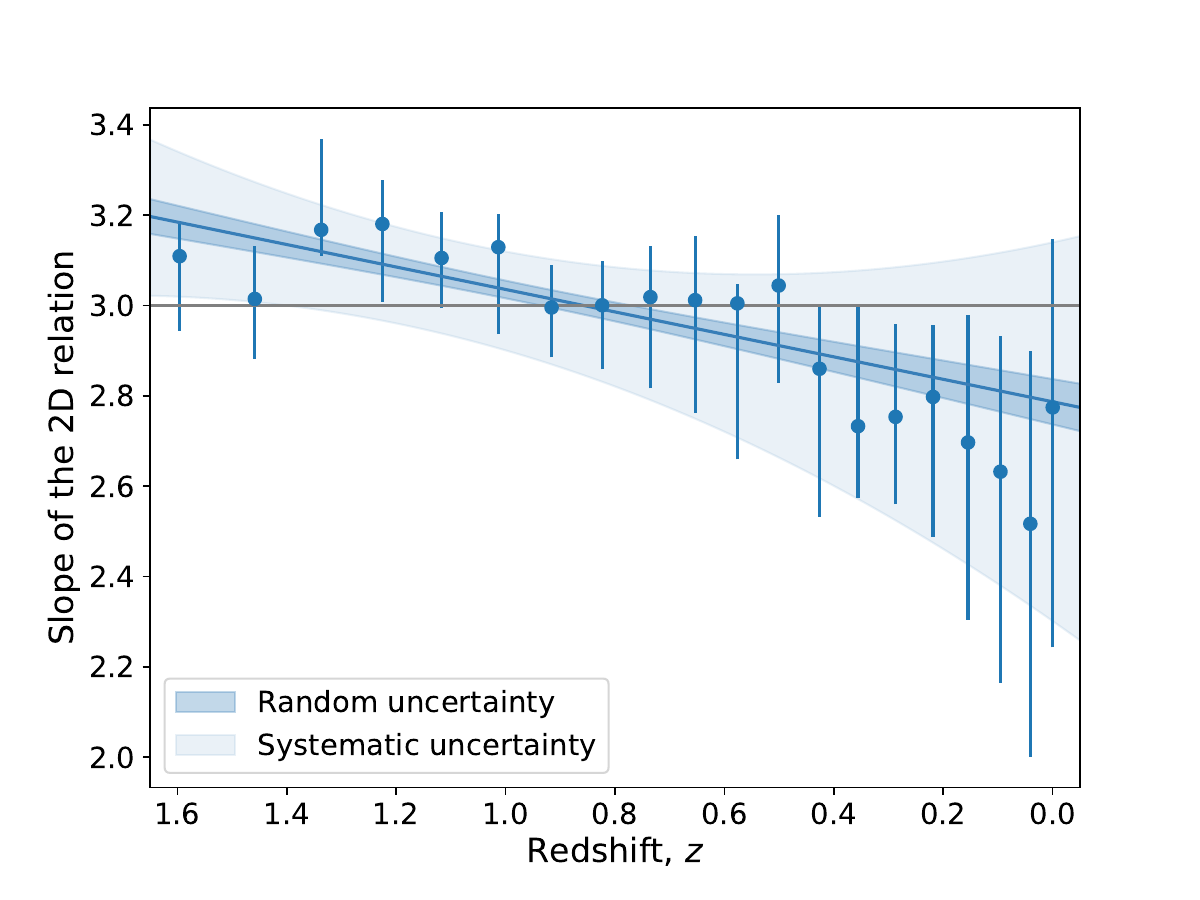}
    \caption{Evolution of the logarithmic slope of the bidimensional relation, $\log_{10} M(<2 R_\mathrm{vir}) = f(\log_{10} R_\mathrm{sh})$. Here, the dots correspond to the determinations of the slope at each snapshot, with the error bars obtained through bootstrap resampling. The dark contour indicates the $1\sigma$ statistical errors associated to the dispersion with respect to the fit, while the light countour contains, in addition to that, the \textit{systematic} uncertainties associated to the magnitude of the error bars.}
    \label{figS5}
\end{figure}

The bidimensional relation (i.e., marginalised over the shock intensity, $\log_{10} M(<2 R_\mathrm{vir}) = f(\log_{10} R_\mathrm{sh})$) is a useful baseline to compare our results with. While in Fig. \ref{fig:evolution} of the main text we have already compared the magnitude of the scatter around both relations, it is still interesting to study the logarithmic slope of the bidimensional relation. A logarithmic slope of 3 ($M \propto R_\mathrm{sh}^3$) would indicate a self-similar scaling of the marginalised data (even though, when shock intensity is considered, this additional information is capable of dramatically reducing the scatter). The evolution of this slope of the marginalised relation with cosmic time is shown in Suppl. Fig. \ref{figS5}, which is similar in its presentation to Fig. \ref{fig:evolution} of the main text. Even though this quantity exhibits a clear evolution, from $\simeq 3.2$ at $z\simeq 1.5$ to $\simeq 2.8$ at $z \simeq 0$, once the uncertainties associated to our limited statistics are considered, the slope of the marginalised relation is consistent with a self-similar behaviour (in the sense that it cannot be ruled out). Far from meaning that $10^{13} M_\odot$ groups and $5 \times 10^{14} M_\odot$ clusters behave likewise (the large scatter around the marginalised relation reflects that the behaviour is far from being that simple), this serves as a sanity check for our shock shell identification procedure.

%---------------------
\subsection{Further details on the fitted relations}
\label{s:suppl.scatter_relations}
%---------------------

\begin{figure}
    \centering
    \includegraphics[width=.5\textwidth]{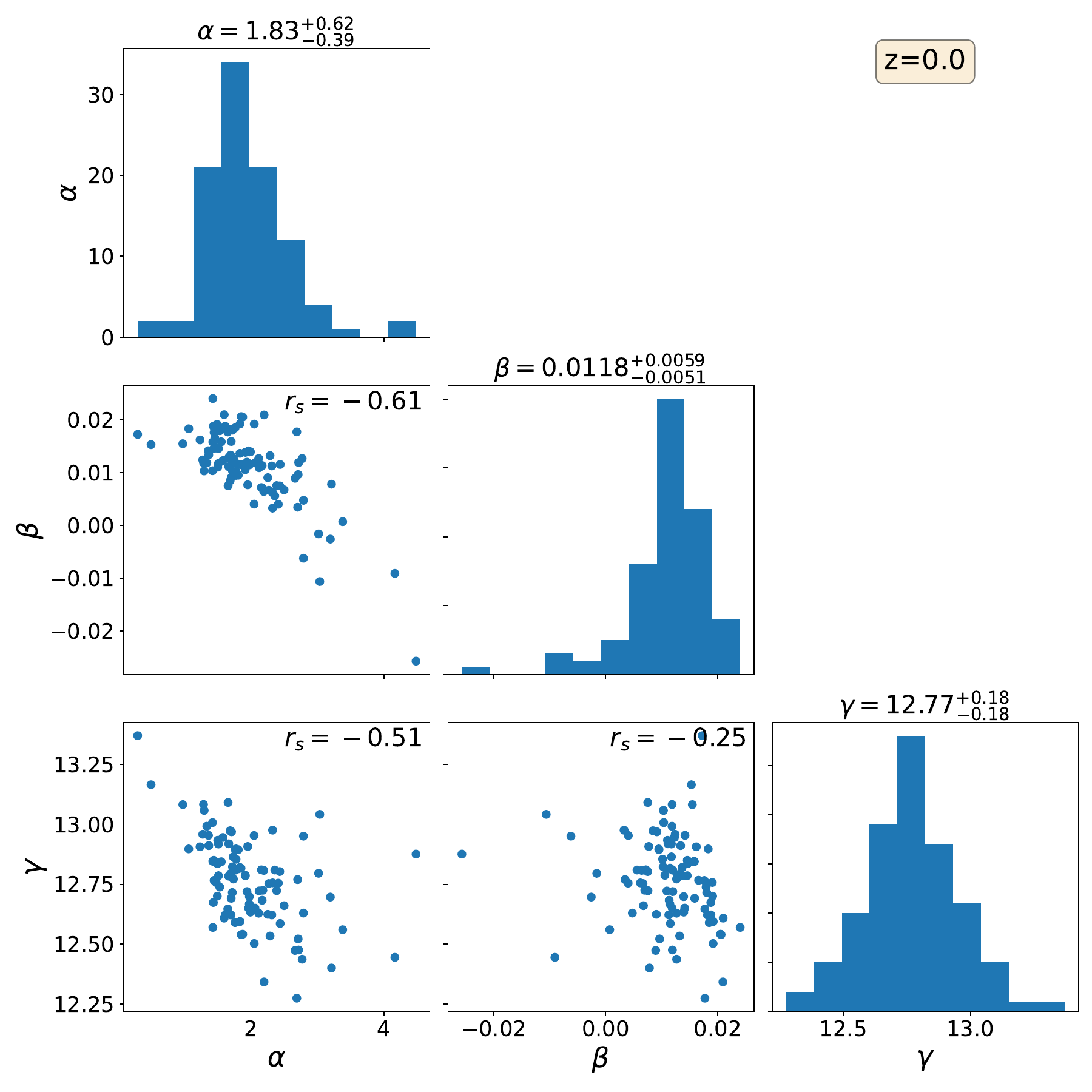}~
    \includegraphics[width=.5\textwidth]{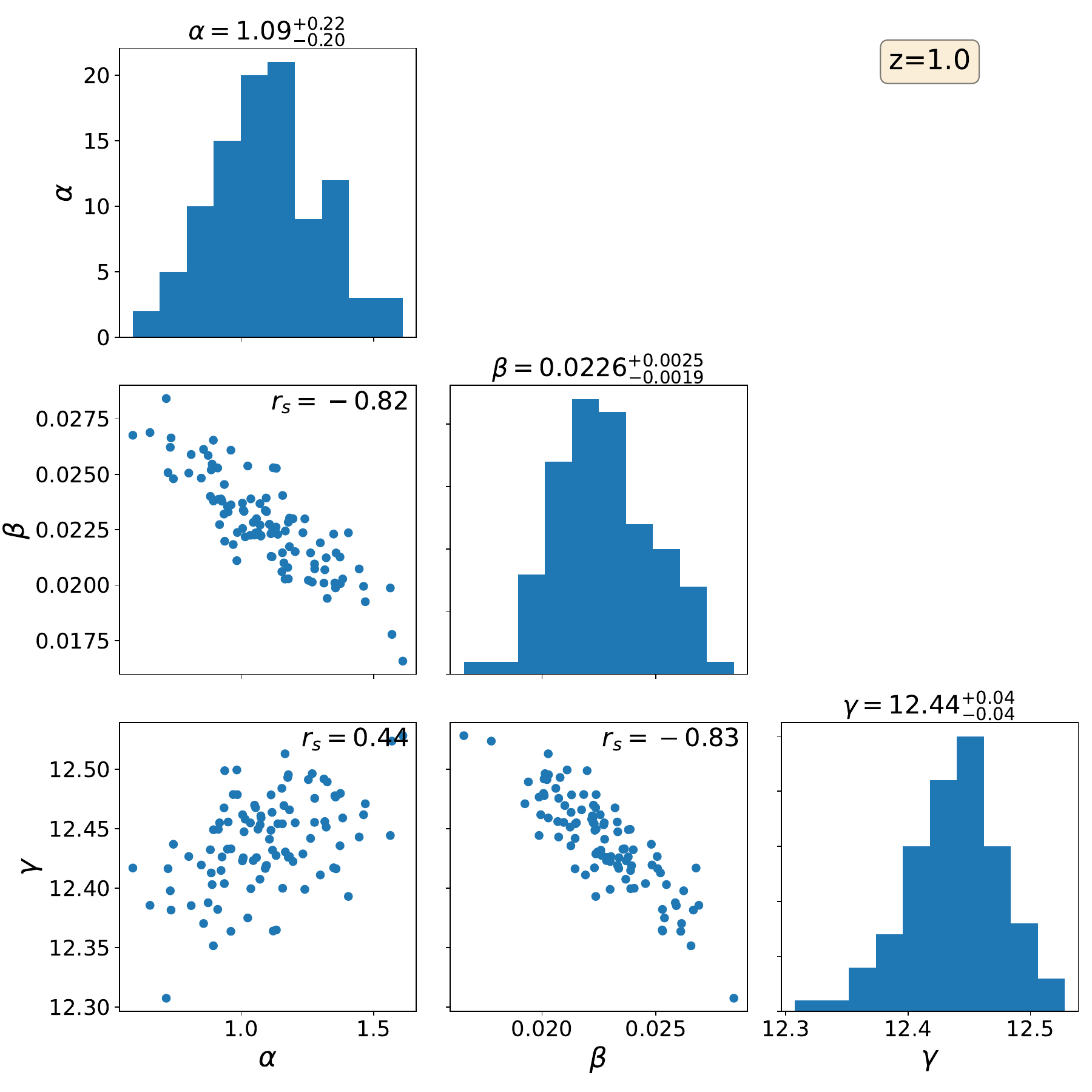}
    \caption{Distribution of the fit parameters in equation (\ref{eq:fit_3d_evolution}), at $z=0$ (left) and at $z=1$ (right). The panels along the diagonal contain the univariate distributions of each parameter ($\alpha$ being the coefficient of $\log_{10} R_\mathrm{sh}$, $\beta$ the cofficient of $\log_{10} M(<2R_\mathrm{vir})$, and $\gamma$ being the independent coefficient), with the value in the title being the median, together with the interquartilic ranges to measure the spread of the distribution. The off-diagonal panels represent the covariances between the fit parameters, with the Spearman rank correlation coefficient ($r_s$) written at the top of each panel.}
    \label{figS6}
\end{figure}

In equations \ref{eq:fit_3d_z0} and \ref{eq:fit_3d_z1} of the main text we have introduced our best fits for $z=0$ and $z=1$. Here, we provide some further information about the results of these fits. In particular, Suppl. Fig. \ref{figS6} presents graphically the uncertainties and correlations associated to the fit parameters $\alpha$ (the coefficient of $\log_{10} R_\mathrm{sh}$), $\beta$ (the cofficient of $\log_{10} M(<2R_\mathrm{vir})$), and $\gamma$ (the independent term), for $z=0$ (left) and $z=1$ (right). These plots are similar to those of Suppl. Fig. \ref{figS3}, but here each dot corresponds to a different estimation of the fit parameters (which we have produced through bootstrap resampling). Naturally, there is a reasonably high degree of uncertainty on each of the parameters, mostly associated to rather tight correlations amongst themselves. That is to say, upon a particular resampling, the fit might prefer a slightly higher value of the coefficient $\alpha$, at the expense of a lower value of the coefficient $\beta$, hence yielding a noticeable anticorrelation amongst these two variables. 

\begin{figure}
    \centering
    \includegraphics[width=.5\textwidth]{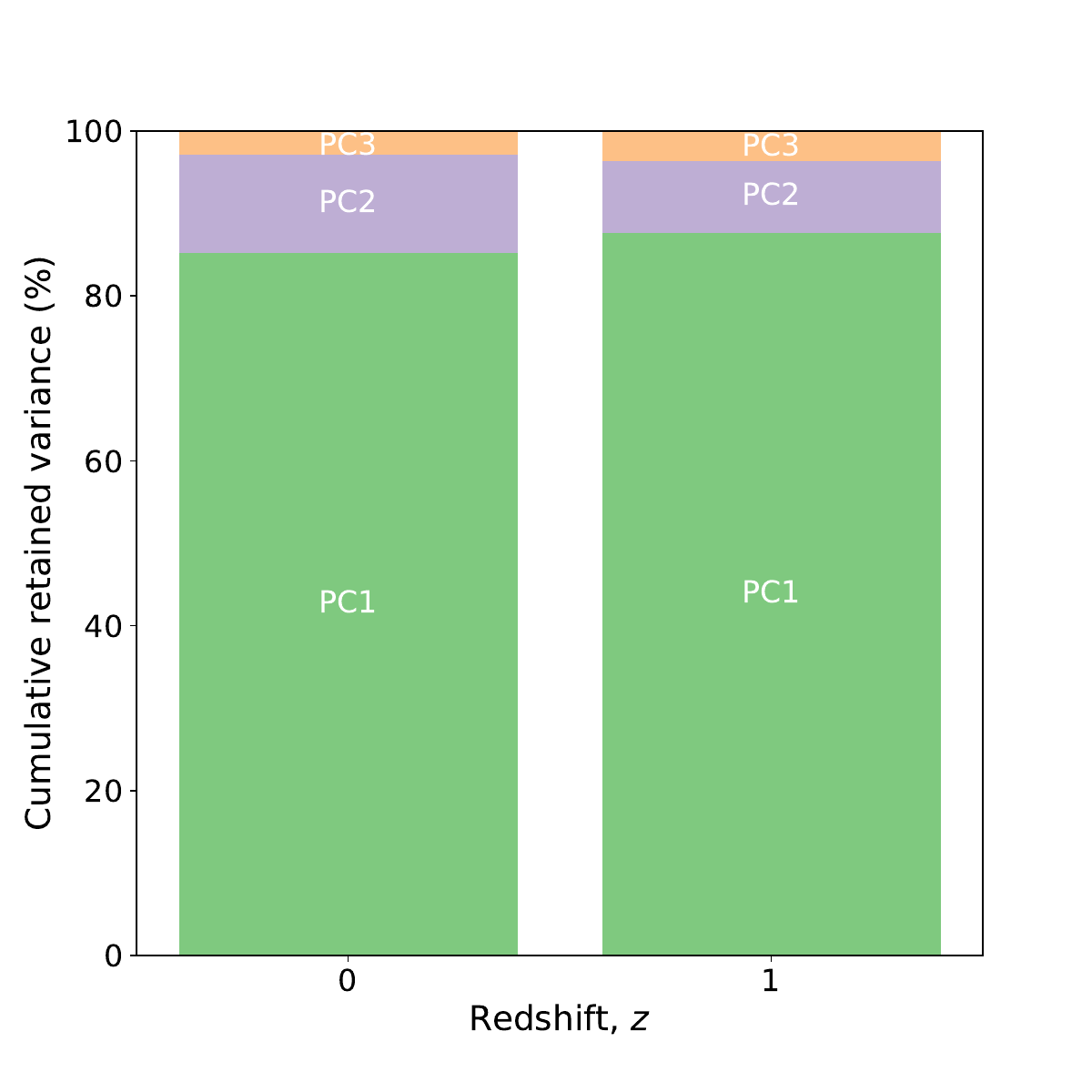}
    \caption{Fraction of variance explained by each of the principal components, at $z=0$ (left) and at $z=1$ (right).}
    \label{figS7}
\end{figure}

In Suppl. Fig. \ref{figS7} we show the fraction of variance retained by each of the principal components, again for $z=0$ (left) and $z=1$ (right). In both cases, a single principal component is capable of reproducing $\sim 85\%$ of the variance of the standardised variables (implying that an only variable would be enough to describe this data, losing only $15\%$ of the scatter). However, even though the distribution of points in this three-dimensional space is slightly prolate, the second principal component outbalances the third one by a factor of $3-4$ in both cases, implying that, rather than having a component retaining most of the variance and two uncorrelated, similarly important variables encapsulating the residuals, the addition of a second variable is capable of breaking most of the degeneracy and hence providing much more precise estimates.

%---------------------
\subsection{Relation of the accretion shock location to the virial accretion rate}
\label{s:suppl.accretion_rate}
%---------------------

\begin{figure}
    \centering
    \includegraphics[width=.5\textwidth]{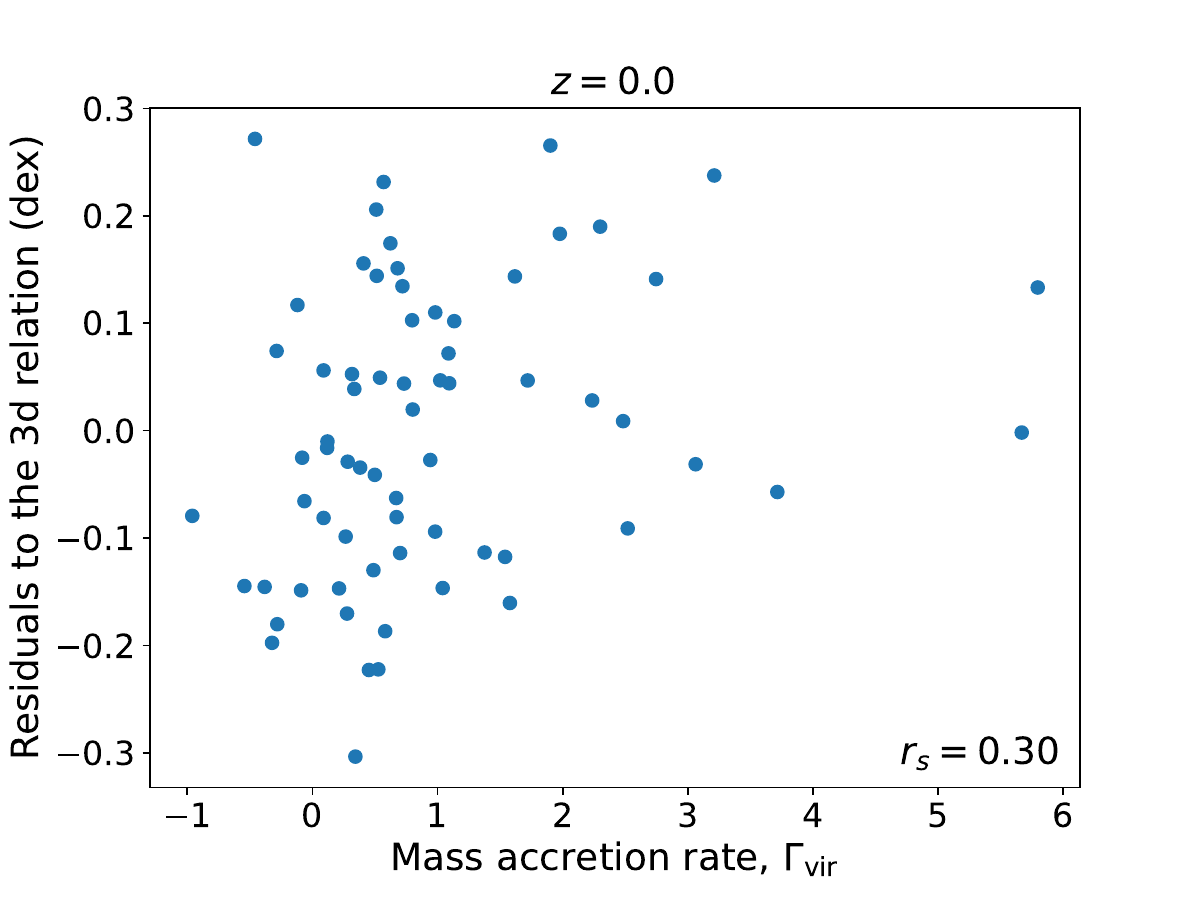}~
    \includegraphics[width=.5\textwidth]{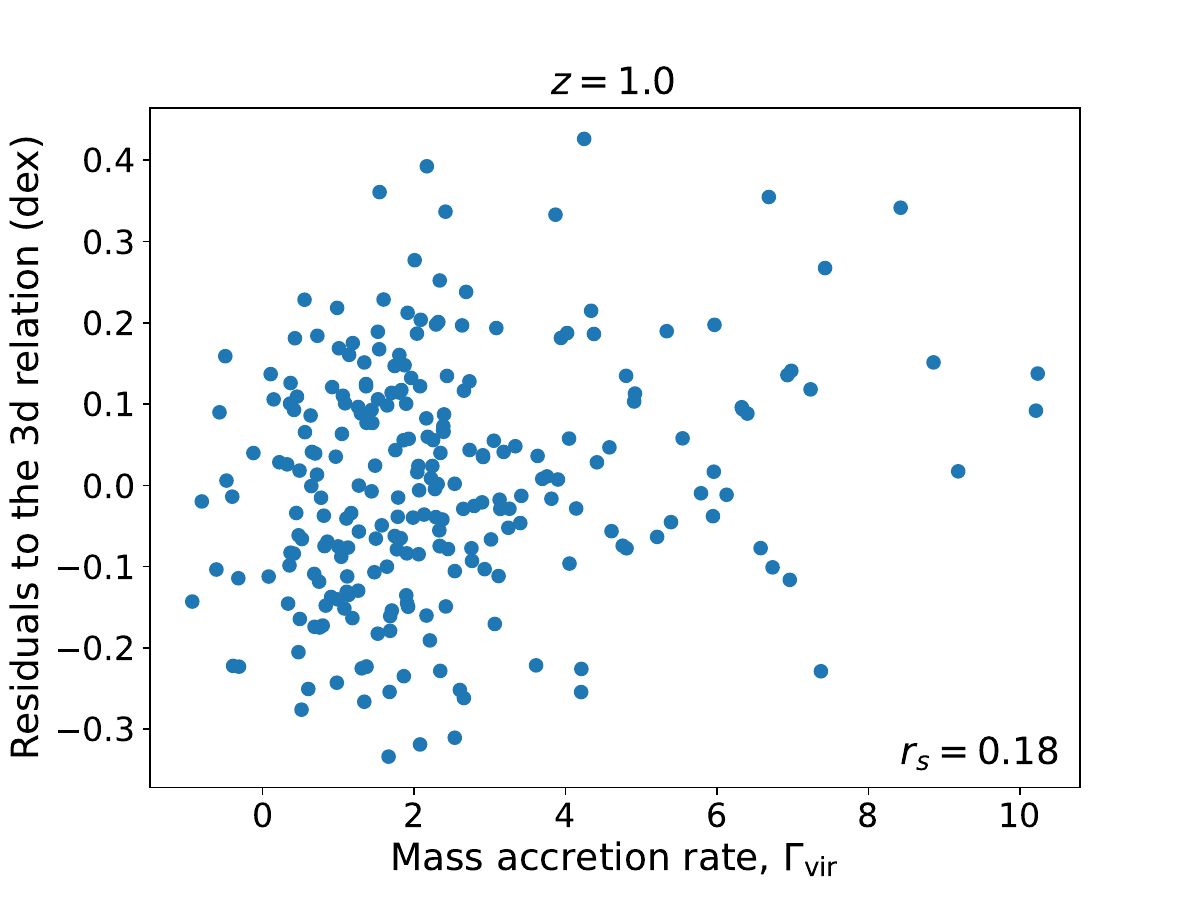}
    \caption{Correlation between the residuals with respect to the three-dimensional relation (equations \ref{eq:fit_3d_z0} and \ref{eq:fit_3d_z1}; the ones shown in the upper and lower panels of Fig. \ref{fig:3dscatter}) and the virial, total mass accretion rate, for the best-fits at $z=0$ and $z=1$, respectively, for the left-hand side and right-hand side panels. Written at the bottom right of each panel is the Spearman correlation coefficient between this pair of variables.}
    \label{figS8}
\end{figure}

While simple models for the collapse of an overdensity, assuming for example constant accretion rates, predict that the location of the accretion shock is determined by the accretion rate \citep{Bertschinger_1983}, the reality in a complex, cosmological context, where the growth of an object is the result of the integrated, time-dependent accretion rate, with anisotropic accretion and mergers playing a very significant role, makes these correlations much more insignificant \citep[e.g.,][their figure 6]{Aung_2021}. While in the panels of Suppl. Fig. \ref{figS3} it can already be seen that $R_\mathrm{sh}/R_\mathrm{vir}$ is only very loosely correlated to the accretion rate proxy $\Gamma_\mathrm{vir}$ (see its definition in the Methods section), in Suppl. Fig. \ref{figS8} we extend on this topic, by presenting the residuals of our three-dimensional relation (at $z=0$, left-hand side panel; and at $z=1$, right-hand side panel) as a function of the accretion rate. These two components are essentially uncorrelated, implying that the addition of information about the mass accretion rate over the last dynamical times brings very limited or no new information.

%---------------------
\subsection{Evolution of the explained variance fractions, uncertainties and correlations amongst the fit parameters}
\label{s:suppl.covariances_evolution}
%---------------------

\begin{figure}[h]
    \centering
    \includegraphics[width=.5\textwidth]{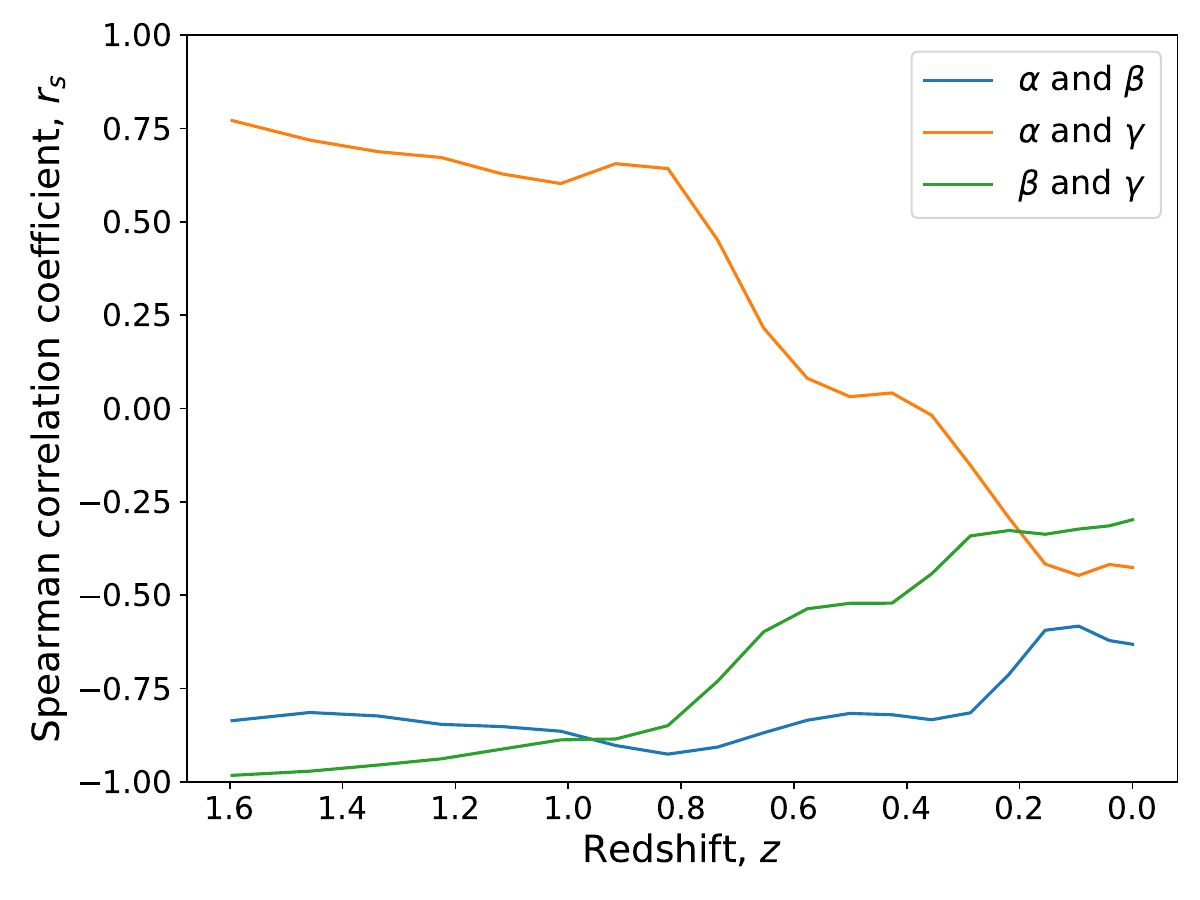}~
    \includegraphics[width=.5\textwidth]{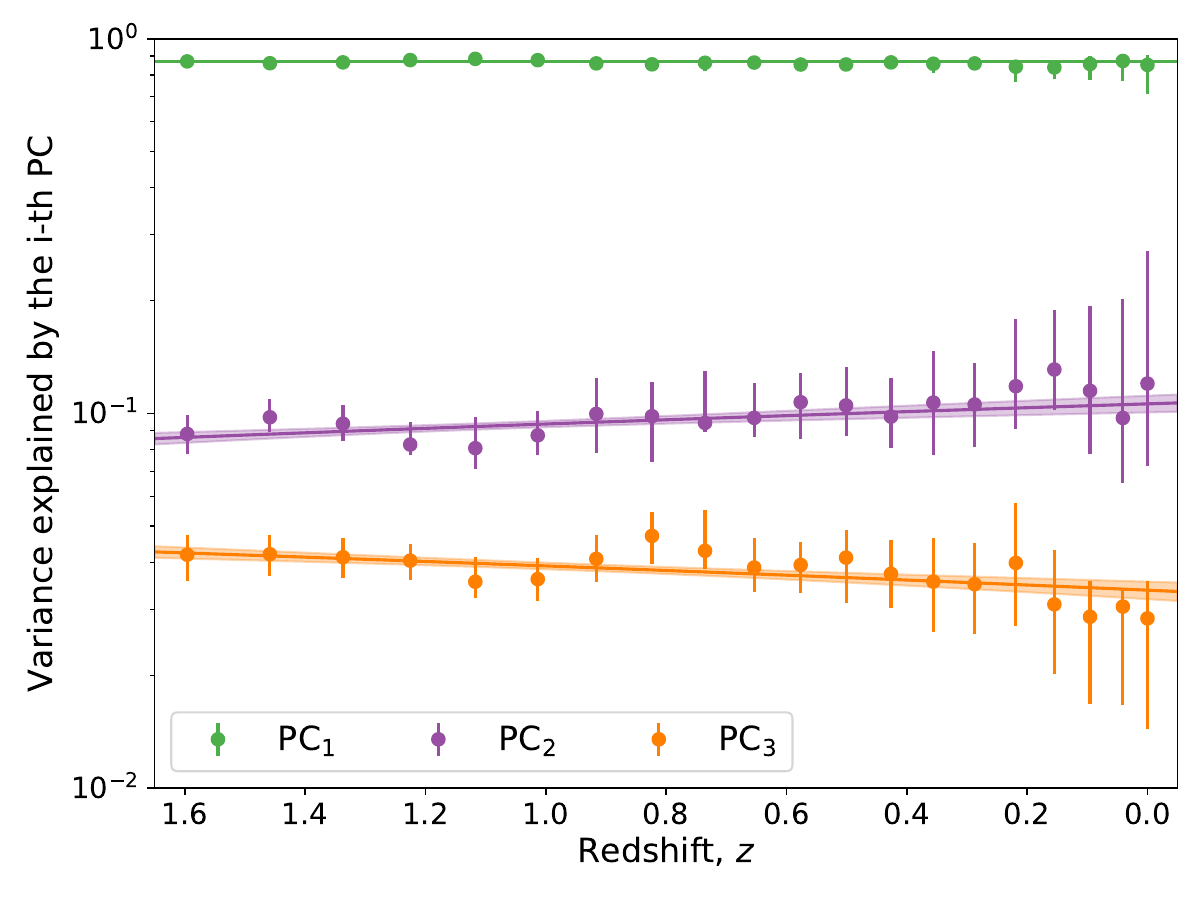}
    \caption{\textit{Left-hand side panel:} evolution of the Spearman rank correlation coefficients amongst the fit parameters $\alpha$, $\beta$ and $\gamma$ introduced in equation (\ref{eq:fit_3d_evolution}) The lines represent the data at each snapshot, smoothed with a Gaussian filter with standard deviation set to the $\Delta z$ between a pair of snapshots, to better visualise the trend. \textit{Right-hand side panel:} evolution of the fraction of variance explained by each of the three principal components. The data points are the values computed at each snapshot, with the $(16-84)\%$ confidence intervals obtained by bootstrap resampling. The lines contain the least-squares fits to this evolution.}
    \label{figS9}
\end{figure}

While the evolution of the intrinsic uncertainties in the determination of each of the coefficients can be seen in Fig. \ref{fig:evolution} of the main text, similarly to what we have shown in Suppl. Fig. \ref{figS6} for two specific snapshots of the simulation, in the left-hand side panel of Suppl. Fig. \ref{figS9} we present the evolution with decreasing redshift of the correlation amongst the fit parameters, obtained in an analogous manner to what we have described in Supplementary Section \ref{s:suppl.scatter_relations}. Generally speaking, most of the times the large uncertainties in the parameters seen in Fig. \ref{fig:evolution} of the main text are not directly relatable to uncertainty in the final solution, but rather to an important covariance or degeneracy amongst the parameters. Calibrating these relations on larger simulation volumes, both because they would host larger structures (therefore, increasing our range of values of radii, Mach numbers and masses) and because they would contain richer statistics, would help to bring down these correlations and therefore decrease the uncertainty in our parameters.

On the other hand, the right-hand side panel of Suppl. Fig. \ref{figS9} presents the evolution of the variance retained by each of the principal components. While the results do not change dramatically with cosmic time, and therefore the interpretation of the relation can be mantained, there is a clear trend for $\mathrm{PC}_2$ to become more relevant with respect to $\mathrm{PC}_3$ as the redshift decreases, meaning that the relation becomes more and more oblate as cosmic time progresses.

%---------------------
\subsection{Intrinsic scatter of the relation: redshift evolution of its mass dependence}
\label{s:supplementary_scattermass}
%---------------------

\begin{figure}
    \centering
    \includegraphics[width=.5\textwidth]{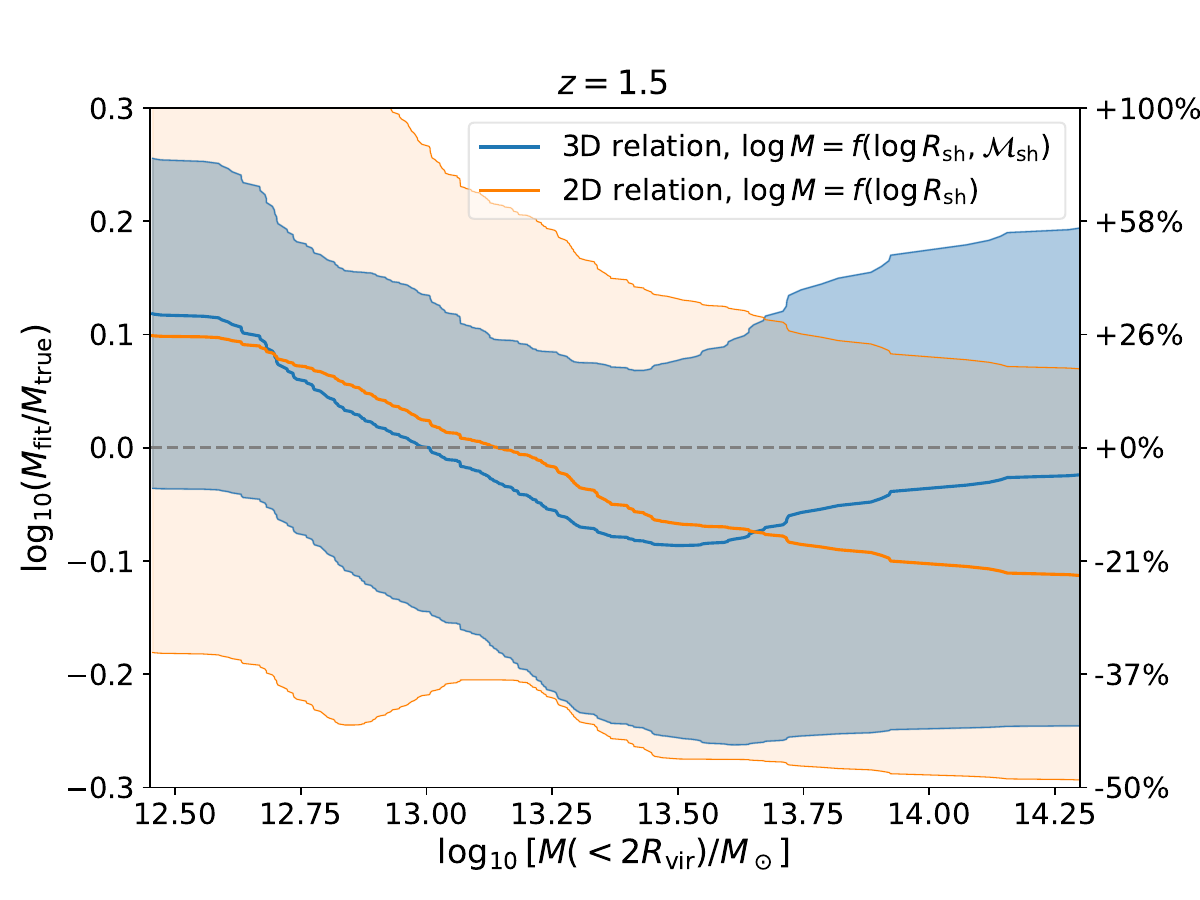}~
    \includegraphics[width=.5\textwidth]{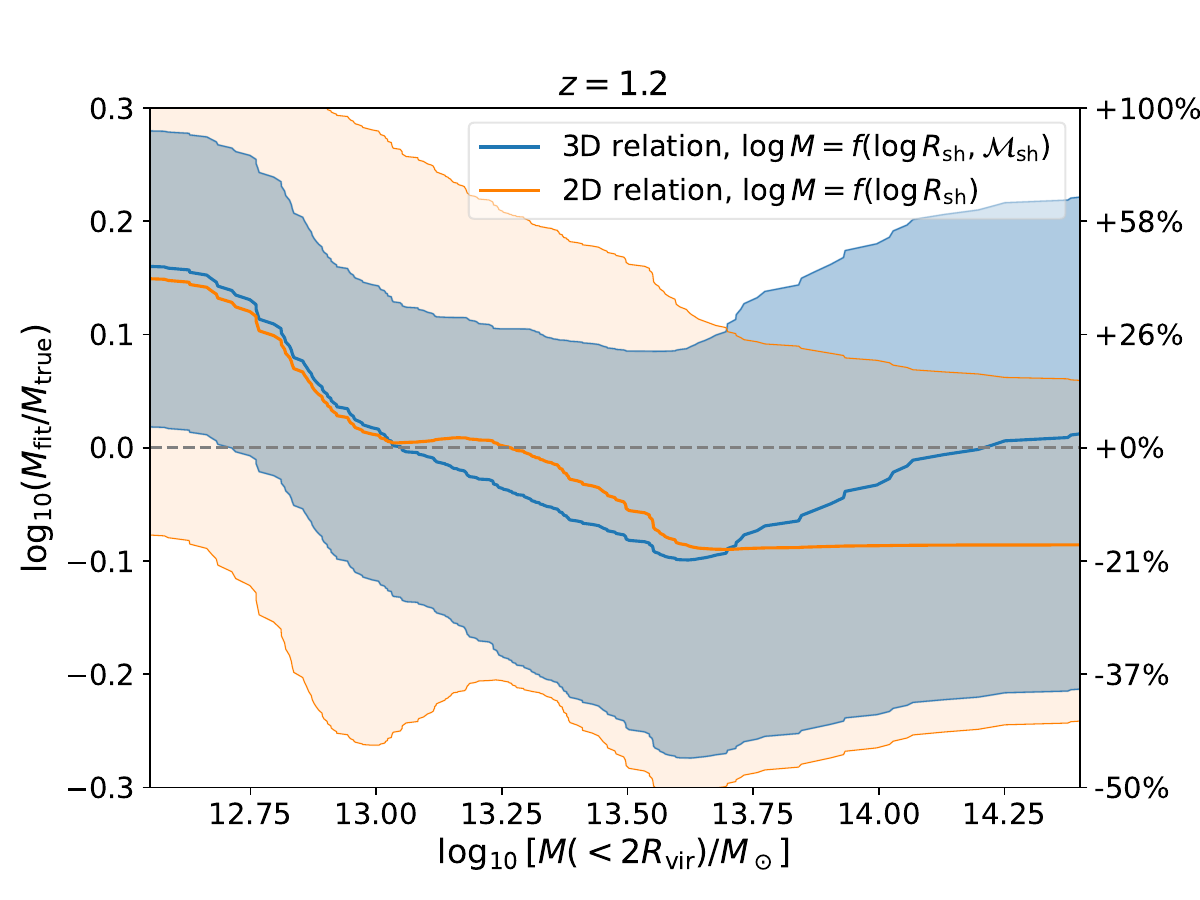}
    \includegraphics[width=.5\textwidth]{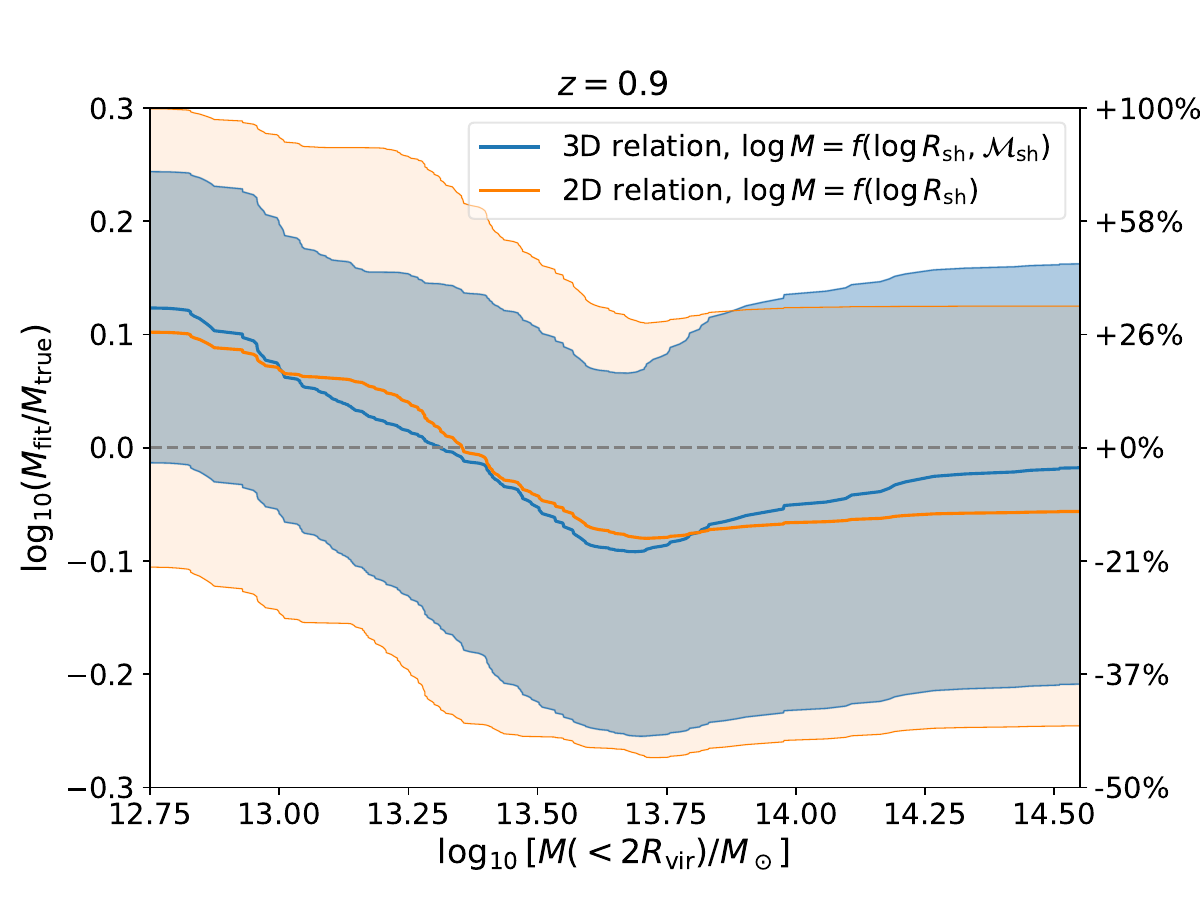}~
    \includegraphics[width=.5\textwidth]{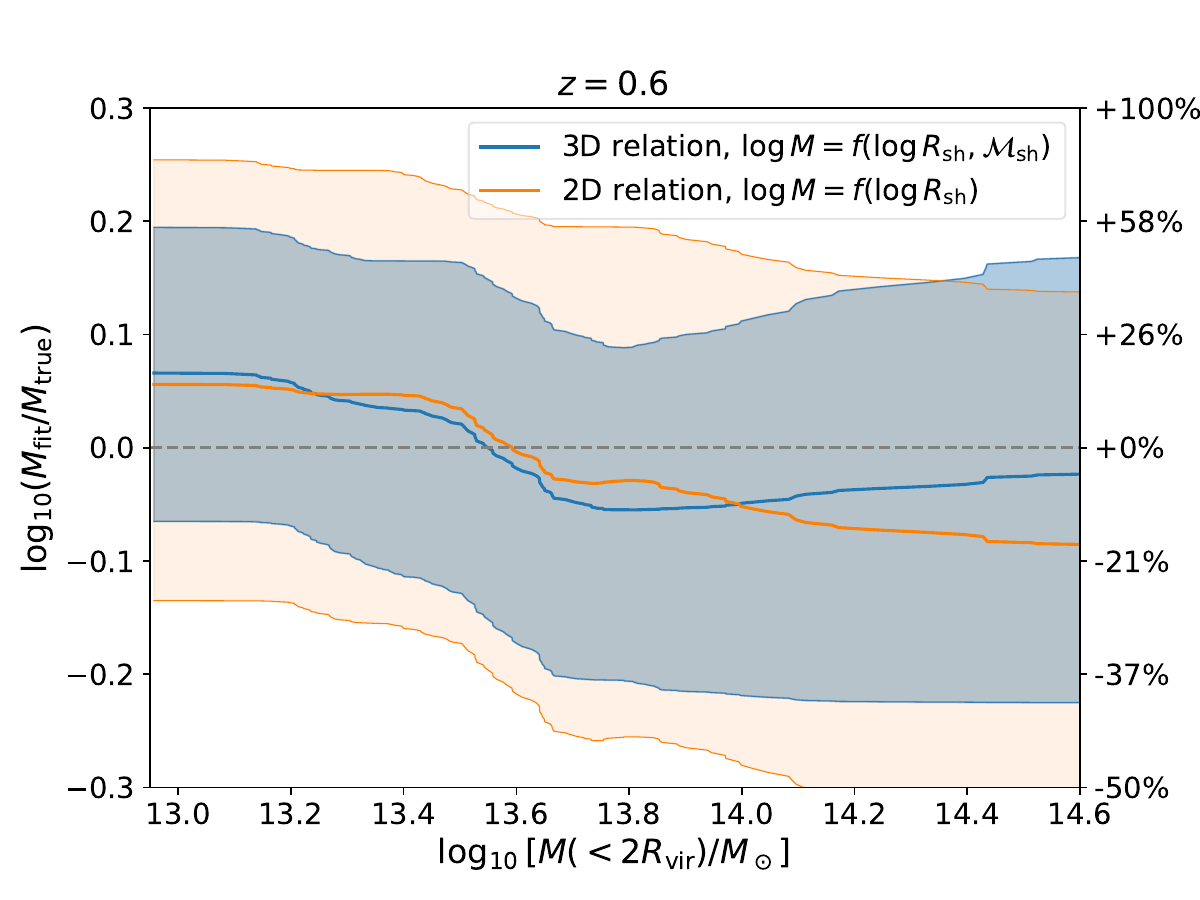}
    \includegraphics[width=.5\textwidth]{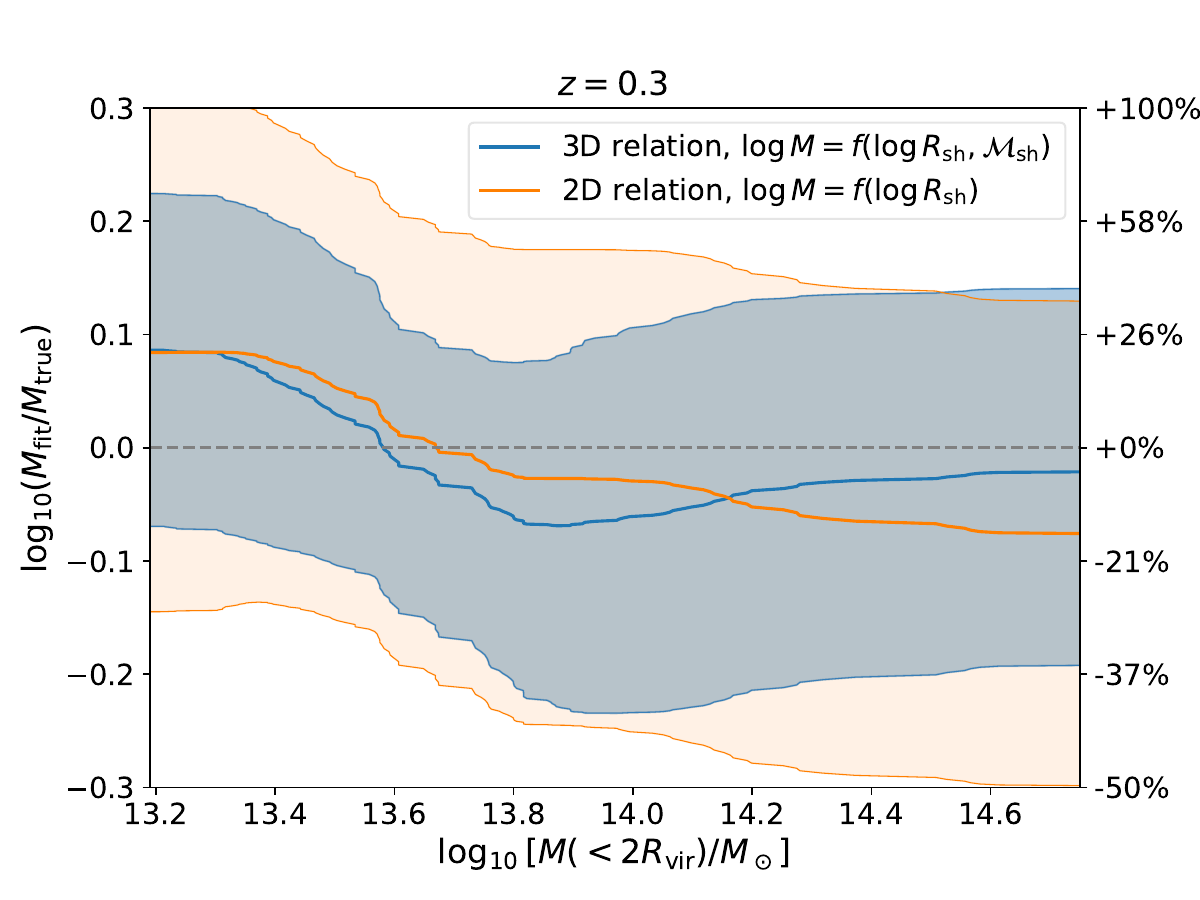}~
    \includegraphics[width=.5\textwidth]{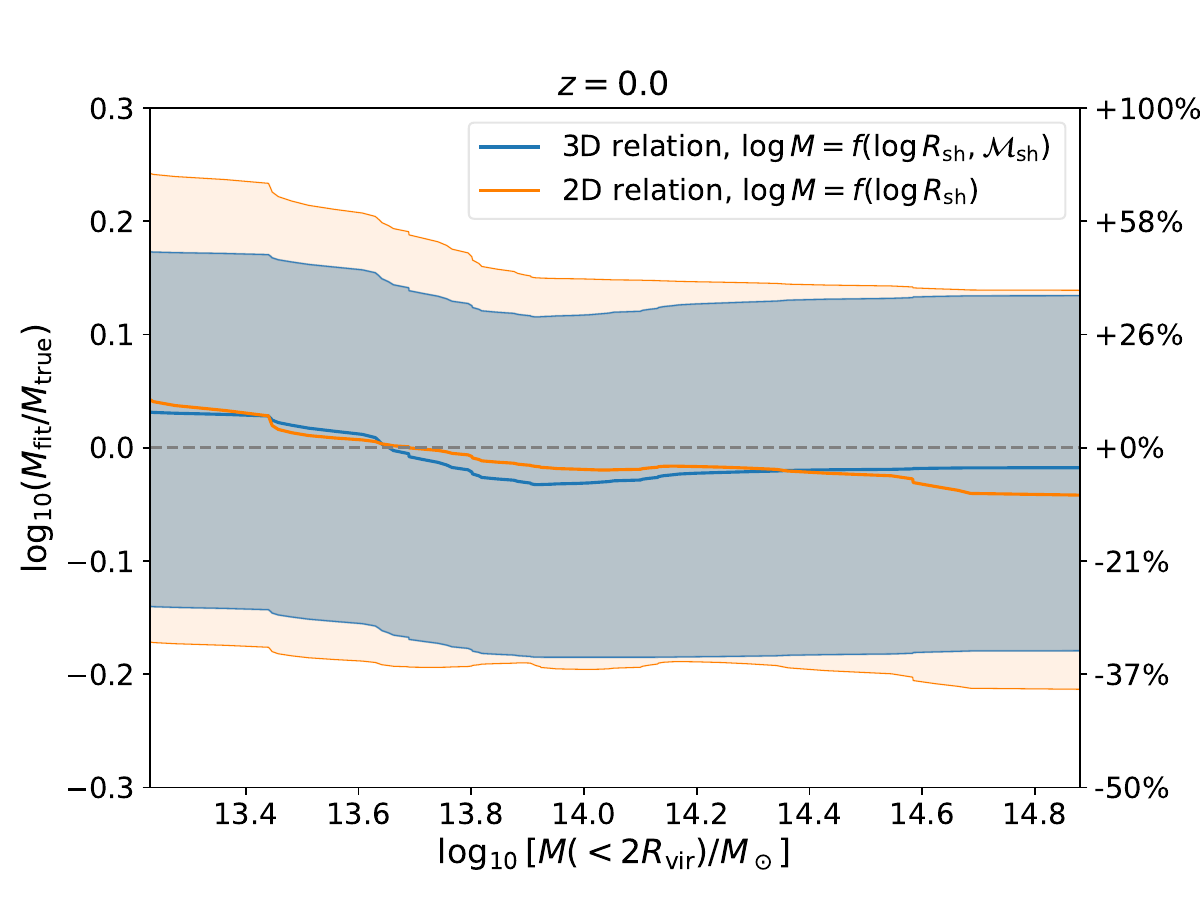}
    \caption{Distribution of the residuals ($\log_{10} M_\mathrm{fit} / M_\mathrm{true}$) with respect to our fitted three-dimensional scaling relation, at different cosmic times (from $z=1.5$ to $z=0$ in increments of $\Delta z = 0.3$). Each panel is similar to the right-hand side panel of Fig. \ref{fig:residuals} of the main text, but here we show only the contours enclosing the $(16-84)\%$ percentiles around the three-dimensional relation (blue) and the two-dimensional relation (ignoring the information about shock intensity; orange). The solid line corresponds to the mean value of the residuals.}
    \label{figS10}
\end{figure}

In the right-hand side panel of Fig. \ref{fig:residuals} of the main paper, we had presented the mass-dependence of the residuals with respect to our best-fit relation at $z=0$. This figure showed that there is a small bias with mass (of a hundredths of a dex, corresponding to a $\lesssim 7\%$ bias), as a consequence that the underlying relation might be non-linear. Interestingly, as we show in the different panels of Suppl. Fig. \ref{figS10}, the behaviour of this small bias with mass is consistent accross our whole redshift range. This hints at a behaviour that could be easily corrected, by calibrating this bias, and therefore would contribute to lower the scatter of our mass determinations. However, we leave this endeavour for future work with enhanced statistics.

\end{document}